\documentclass{aa}
\usepackage[utf8]{inputenc}
\usepackage{graphicx}
\usepackage{hyperref}
\usepackage{txfonts}
\usepackage{booktabs}
\usepackage{amsmath}
\usepackage{amssymb}
\usepackage{textcomp}
\usepackage{xcolor}
\usepackage{url}
\usepackage{lineno}
\usepackage{tabularx}
\usepackage{gensymb}
\usepackage{placeins}

\begin{document} 
\title{Caught in the act: interaction-driven evolution in the nearby compact galaxy group Roberts Quartet (SCG0018-4854)}

\titlerunning{A Compact Quartet in Turmoil}

   \author{Saili Keshri
          \inst{1}\fnmsep\inst{2} 
          \and
          Sudhanshu Barway
          \inst{1}\fnmsep\inst{2} 
          \and
          Mousumi Das
          \inst{1}
          \and
          Abhishek Paswan
          \inst{3}
        }

   \institute{Indian Institute of Astrophysics, Kormangala II Block, Bengaluru, India, 560034\\
            \email{saili.keshri@iiap.res.in}
         \and
             Department of Physics, Pondicherry University, R.V.Nagar, Kalapet, Puducherry, India, 605014\\
             \and
            Department of Physics, University of Allahabad, Prayagraj 211002, Uttar Pradesh, India
             }

\abstract {We present a spatially resolved multiwavelength study of the compact galaxy group Roberts Quartet (SCG0018-4854), aimed at understanding interaction-driven galaxy evolution in dense environments. The system comprises four galaxies (NGC~87, NGC~88, NGC~89, and NGC~92) that span a range of masses and evolutionary states. Using GALEX ultraviolet imaging, DECaLS optical data, VLT/MUSE integral-field spectroscopy, and infrared observations from VISTA/VIRCAM, 2MASS, and WISE, we investigate the interplay between kinematics, star formation, and stellar populations across the group. The spatially resolved analysis reveals disturbed stellar and gas kinematics, enhanced turbulence, and asymmetric structures in all members, consistent with repeated gravitational interactions. The most massive galaxy, NGC~92, exhibits prominent tidal features, a bar, and ring-like star-forming structures, indicative of interaction-driven gas inflows. Another massive member, NGC~89, shows suppressed star formation and signatures of AGN-driven feedback, while the lower-mass galaxies NGC~88 and the dwarf galaxy NGC~87 display enhanced star formation and, in one case, kinematic decoupling between stellar and gaseous components consistent with recent gas accretion. Combining ultraviolet-based age estimates with non-parametric star formation histories, we constrain the recent interaction timescale of the group to $\lesssim$500 Myr, whereas the crossing timescale is 424 Myr. These results indicate that Roberts Quartet is a dynamically young system undergoing ongoing assembly, where interactions, gas exchange, and feedback processes are actively shaping galaxy evolution. The dynamical complexity of the group further suggests that its present configuration may involve more than four progenitor components. In this context, the Roberts Quartet provides a valuable nearby analogue of compact, rapidly evolving groups observed at high redshift by recent JWST observations, offering a resolved view of the physical processes governing galaxy assembly in the early Universe.} 

   
\maketitle
\nolinenumbers 

\section{Introduction}\label{sec:Sec_1}

The processes of galaxy formation and transformation are crucial to their evolution. According to the hierarchical galaxy formation model, interactions between small galaxies at high redshift form the basis for the formation of present-day galaxies \citep{Toomre72}. Thus, detailed studies of local interacting and merging systems offer valuable insights into phenomena prevalent in the distant universe. Multiple studies have compared properties (for example, gas mass and star formation rate) of galaxies in the local and distant universe and found they evolved with redshift \citep{Mannucci10, Behroozi13, Madau14, Boyett22}. Thus, it is more relevant to compare the physical processes and phenomena than the comparison of the absolute values. Interacting and merging systems are unique laboratories to understand the galaxy transformation; however, to have a complete picture, it is imperative to combine observational results with simulations. \citet{Barnes1992} and subsequent numerical studies \citep[e.g.,][]{Springel2005, DiMatteo2007, DiMatteo2008, Bekki2011} have shown that repeated interactions and mergers in dense environments can trigger a chain of evolutionary transformations. Tidal forces redistribute angular momentum, drive gas inflows toward galactic centres, and compress the interstellar medium, inducing circumnuclear starbursts \citep{Scott2014, Bitsakis2011, Jyoti23}. These inflows can also feed central supermassive black holes, igniting active galactic nuclei (AGN) and potentially initiating feedback that quenches further star formation \citep{Weigel2018}. Over time, the depletion of cold gas and tidal heating transform gas-rich, late-type galaxies into red, quiescent early-type systems \citep{VerdesMontenegro2001}. Compact groups (CGs), therefore, serve as nearby analogues of the high-redshift universe, where interactions were more frequent \citep{RodriguezBaras2014}.

CGs of galaxies consist of a small number of galaxies, typically four to seven, separated by projected distances of only a few tens of kiloparsecs and have low velocity dispersions of $\approx$200 km/s \citep{Hickson1982, Hickson1992}. Such proximity leads to strong gravitational interactions, frequent encounters, and a high probability of mergers occurring over relatively short timescales. As a result, they play an important role in the environment-driven evolution of galaxies \citep{Mendes-de-Oliveira94, Coziol07}. Observationally, CGs show remarkable diversity in morphology and activity, from irregular and spiral systems with intense star formation to lenticular and elliptical galaxies that are dynamically relaxed and quiescent \citep[][and references therein]{Hickson97}. Multiwavelength studies reveal that many CGs contain both gas-rich and gas-poor members within the same environment \citep{Bitsakis2014, Walker2010}. Star formation frequently occurs in tidal bridges, producing patchy UV and H$\alpha$ emission \citep{deMello2008, Tzanavaris2010}. Some galaxies display rejuvenated star formation from fresh gas accretion or interaction-induced inflows \citep{Gallagher2008, TorresFlores2014}, while others exhibit AGN-driven suppression or morphological quenching \citep{Plana2020}. A higher incidence of AGN has been reported in CGs relative to field galaxies \citep{Coziol1998, Martinez2010, Sohn2013}, with the prevalence of nuclear activity often correlating with signs of tidal perturbation. This mixture of morphologies and activity levels reflects different evolutionary stages within a single gravitationally bound system, demonstrating how CGs evolve through cycles of star formation, merger-induced transformation, and feedback. 

Dynamical models predict that galaxies in groups undergo strong interactions and merge on a dynamical time scale \citep{Barnes85, Governato96}. However, major mergers are rare in CGs \citep{Mendes-de-Oliveira94}. CG galaxies preferentially merge in a dry condition, after losing most of their gas through interactions \citep{Coziol07}. This scenario is supported by HI observations showing disturbed distributions in CGs \citep{VerdesMontenegro2001}, implying that their evolution is driven primarily by galaxy interactions and tidal stripping. Thus, it is important to reconstruct the evolutionary and interaction history of CGs by examining the properties of their member galaxies and relating them to the global properties of the group. An integral-field spectroscopy study will help us to reveal complex velocity fields and ionisation structures consistent with recent interactions. Such evidence supports the view that compact configurations accelerate both star formation and black-hole growth.

In this work, we present a detailed multiwavelength analysis of the Southern Compact Group SCG0018-4854 \citep{Iovino02} or Robert’s Quartet (RQ hereafter), combining UV, optical (including imaging and integral-field spectroscopy), and IR data. RQ is a CG of galaxies located at a distance of $\sim$44.7 Mpc \citep{Mould00} (At the distance of RQ, 1\arcsec = 0.217 kpc). It comprises four morphologically diverse galaxies: NGC 87, NGC 88, NGC 89 and NGC 92 \citep{Presotto10} and details of each galaxy are given in Table~\ref{tab:morp}. Prominent tidal tails, disturbed morphologies, and extended UV and H$\alpha$ emission across the group indicate signatures of both ongoing and past interactions. The group is also found to have a common envelope of HI gas around the galaxies \citep{Pompei07}. All four galaxies are embedded in a hot intergalactic medium (KT $\approx$ 0.2 keV) as detected by X-ray observation \citep{Trinchieri08}. The diversity of morphologies and activity within this compact system provides an ideal opportunity to understand how tidal interactions, gas inflows, and AGN feedback collectively shape galaxy evolution in a dense, low-velocity dispersion environment. We aim to investigate the spatially resolved kinematics, star formation and stellar populations of its member galaxies and to explore how interaction-driven processes influence their current evolutionary states. 

The paper is organised as follows: Sect. ~\ref{sec:data} describes the datasets, Sect. ~\ref{sec:analysis} explains the methods used for our analysis. The results are presented in Sect. ~\ref{sec:result} followed by a discussion in Sect. \ref{sec:Disc}. Our main conclusions are presented in Sect. \ref{sec:Conc}. We adopt a flat $\Lambda$CDM cosmology with $\Omega_M=0.3$, $\Omega_\Lambda=0.7$, and $H_0=70$ km s$^{-1}$ Mpc$^{-1}$ throughout this study.

\begin{figure}
    \centering
    \includegraphics[width=0.8\linewidth]{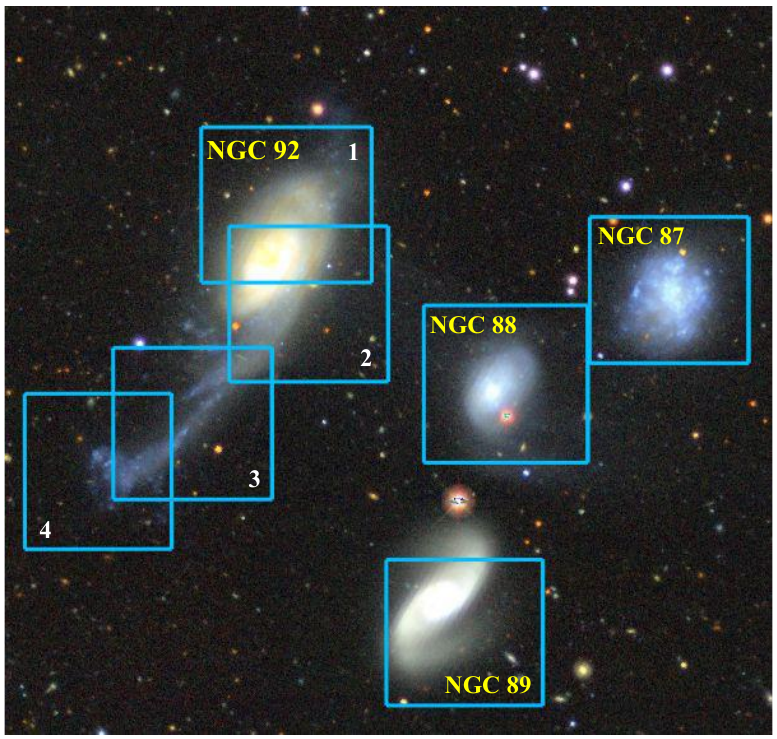}
    \caption{Optical image of RQ. The blue squares show the locations of the MUSE-DEEP IFU datacube footprints. The optical image is taken from the DECaLS \citep{Dey19}. North is up, and east is towards the left.}
    \label{fig:group}
\end{figure}

\begin{table}[h]
    \centering
     \caption{Details of the galaxies of RQ group.}
        \begin{tabular}{lcccr}
        \hline
        \textbf{Source} & \textbf{R.A}  & \textbf{Dec}  & \textbf{Velocity} \\
        & (hh mm ss) & (dd mm ss) & (km/s)\\
        \hline
        NGC 87 & 00 21 14.21 & -48 37 42.81 & 3426 $\pm$ 2.5 \\
        NGC 88 & 00 21 22.06 & -48 38 25.21 & 3409 $\pm$ 3.9 \\ 
        NGC 89 & 00 21 24.36 & -48 39 54.88 & 3295 $\pm$ 2.9 \\
        NGC 92 & 00 21 31.66 & -48 37 28.82 & 3362 $\pm$ 14 \\
        \hline
        \end{tabular}
    \begin{flushleft}
    \footnotesize
    \textbf{Description:} {Columns 2 and 3 are the coordinates of galaxies in J2000. The radial velocity is tabulated in column 4. The coordinates are listed from NED, and the velocity is estimated from MUSE data cubes.}
    \end{flushleft}
    \label{tab:morp}
\end{table}

\begin{table}[h]
\centering
\caption{Details of the archival imaging data used in this work.}
\resizebox{\columnwidth}{!}{%
\begin{tabular}{lcccr}
\hline
\textbf{Telescope/} & \textbf{Filter} & \textbf{Spatial sampling} & \textbf{FWHM} & \textbf{Point-source limit}\\
\textbf{Survey} & & (arcsec/pixel) & (arcsec) & ($5\sigma^{*}$/$10\sigma^{**}$, AB mag)\\
\hline
GALEX & FUV & 1.50 & 4.20 & 22.60$^{*}$ \\
GALEX & NUV & 1.50 & 5.30 & 22.70$^{*}$ \\
DECaLS & g & 0.26 & 0.97 & 23.95$^{*}$ \\
DECaLS & r & 0.26 & 0.83 & 23.54$^{*}$ \\
DECaLS & i & 0.26 & 0.77 & 23.80$^{*}$ \\
DECaLS & z & 0.26 & 0.87 & 22.50$^{*}$ \\
2MASS & J & 1.0 & 2.89 & 16.71$^{**}$ \\
2MASS & H & 1.0 & 2.83 & 16.49$^{**}$ \\
2MASS & Ks & 1.0 & 2.98 & 16.15$^{**}$ \\
VISTA & Ks & 0.34 & 1.12 & 20.36$^{*}$ \\
WISE & W1 & 1.38 & 6.08 & 19.20$^{*}$ \\
WISE & W2 & 1.38 & 6.84 & 18.84$^{*}$ \\
WISE & W3 & 1.38 & 7.36 & 16.37$^{*}$\\
WISE & W4 & 1.38 & 11.99 & 14.52$^{*}$ \\
\hline
\end{tabular}
}
\begin{flushleft}
\footnotesize
\textbf{References.} GALEX: \citet{Morrissey07};
DECaLS: \citet{Dey19};
2MASS: \citet{Skrutskie06};
VISTA: \citet{Sutherland15};
WISE: \citet{Wright10}.
\end{flushleft}
\label{tab:data}
\end{table}

\begin{figure*}
\centering
  \includegraphics[width=12cm]{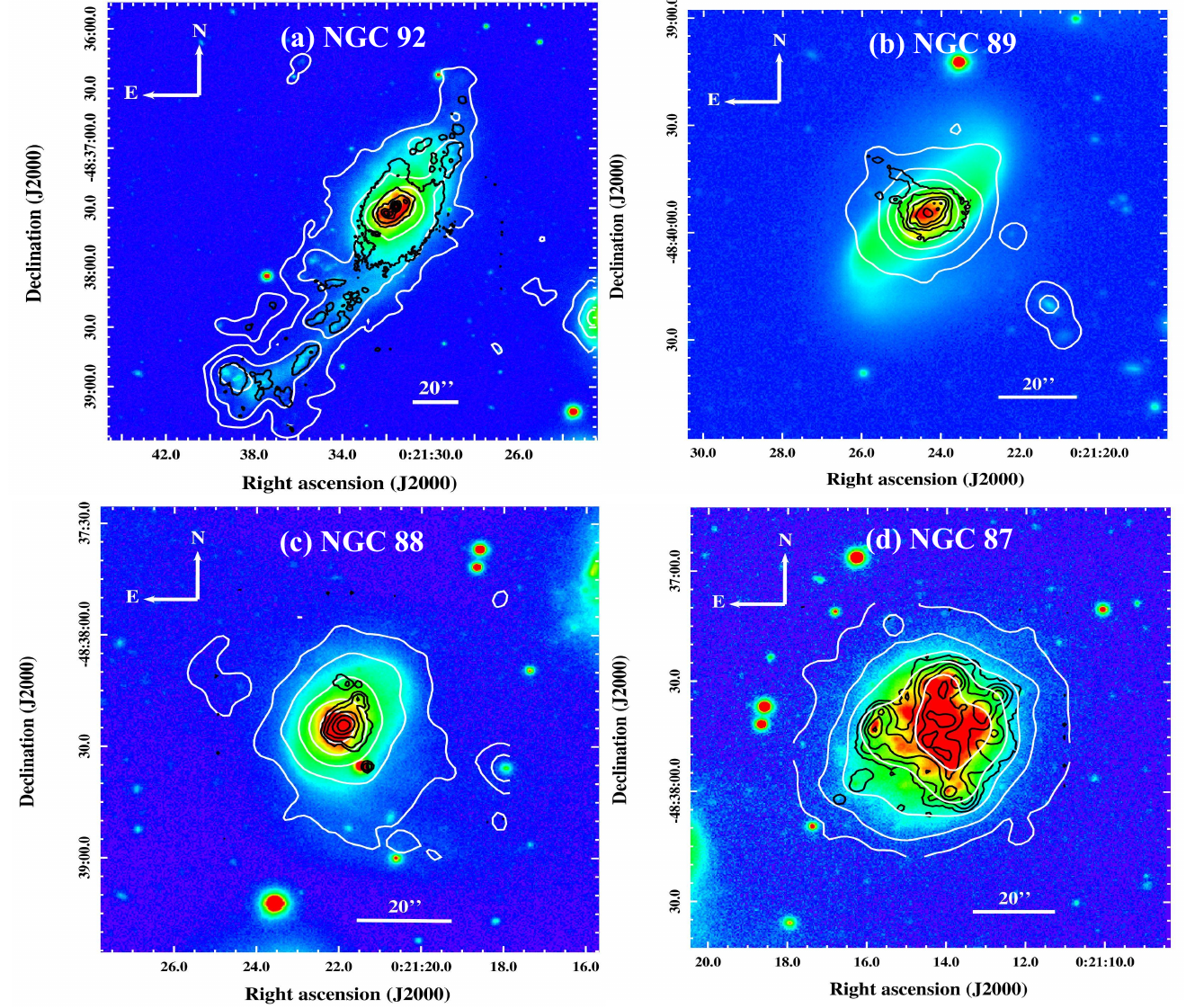}
     \caption{Multi-wavelength view of the RQ galaxies. DECaLS g-band images of (a) NGC 92, (b) NGC 89, (c) NGC 88, and (d) NGC 87 are presented with MUSE H$\alpha$ (black) and GALEX FUV (white) contours overlaid.}
     \label{fig:group_ind}
\end{figure*}

\section{Data} \label{sec:data}
To investigate the stellar and ionised gas properties of the galaxies in RQ, we combined ultraviolet, optical, and infrared data from multiple archives, each probing a complementary aspect of the system’s evolution. The datasets are briefly described below.
\paragraph{Ultraviolet imaging:}
The far- and near-ultraviolet (FUV, NUV) images were obtained from the \textit{Galaxy Evolution Explorer} (\textit{GALEX}) archive, which is a NASA UV space telescope that works in imaging and spectroscopy modes. It has two imaging filters, FUV (1344-1786 Å) and NUV (1771-2831 Å) \citep{Morrissey07}. GALEX observed RQ on 2005-09-09 and 2005-09-25 in NUV and FUV filters, with exposure times of 7419 and 4060 seconds, respectively. We accessed the science-ready data through the MAST portal \footnote{https://mast.stsci.edu/portal/Mashup/Clients/Mast/Portal.html}. The UV emission traces recent star formation (up to 300 Myr), enabling us to identify young stellar complexes and tidal star-forming regions within the group. 
\paragraph{Optical imaging:}
Deep $g$, $r$, $i$, and $z$-band optical images were taken from the Dark Energy Camera Legacy Survey (DECaLS; \citealt{Dey19}) Data Release 10 (DR10). The data reach an approximate 5$\sigma$ depth of $g = 23.72$, $r = 23.27$, $i = 23.50$ and $z = 22.22$ AB mag for a galaxy with an exponential disc profile and a half-light radius of 0.45 arcsec. The high surface-brightness ($\mu_{r} \approx 29$ $mag/arcse^{2}$) sensitivity of DECaLS enables the detection of faint tidal features and low-surface-brightness structures, providing constraints on morphological disturbances and stellar mass distributions.
\paragraph{Optical integral-field spectroscopy:}
Optical integral-field spectroscopic data were obtained from the Multi Unit Spectroscopic Explorer (MUSE; \citealt{Bacon2010}) at the ESO Very Large Telescope (VLT). MUSE datacube covers 4750–9350 \AA~range with a mean spectral resolution of R$\sim3000$ and spatial sampling of 0.2\arcsec/pixel over a 1\arcmin$\times$1\arcmin field of view. The science-ready MUSE-DEEP data cube of the galaxy group was retrieved from the ESO science archive portal \footnote{https://archive.eso.org/scienceportal/home}. The galaxy group was observed from 2018-05-09 to 2018-07-20, and the program ID is 0101.C-0329. The MUSE-DEEP datacube footprint of the galaxy group is shown in Fig.~\ref{fig:group}, where the optical image is taken from DECaLS \citep{Dey19}. In this work, we utilise datacube 2 of the four available datacubes covering NGC 92 to investigate the stellar and gas kinematics along with stellar population properties of the central region of NGC 92. Additionally, we have utilised datacube 3 to constrain the gas-phase metallicity of the tail. We have also used all four datacubes to produce an H$\alpha$ image of NGC 92.
\paragraph{Infrared imaging:}
Near-infrared ($J$, $H$, $Ks$) images were retrieved from the Two Micron All Sky Survey (2MASS, \citet{Skrutskie06}) Extended Source Catalogue and mid-infrared (W1 (3.4 $\mu$m), W2 (4.6 $\mu$m), W3 (12 $\mu$m), and W4 (24 $\mu$m)) data were taken from the Wide-field Infrared Survey Explorer (WISE, \citet{Wright10}). The measurements obtained from the 2MASS images were converted to flux units using scaling factors given in the respective image headers. For WISE, we used the scaling factors available in its documentation. We have also utilised high resolution $Ks$-band images from VISTA InfraRed CAMera (VIRCAM) with a 1.65$\degree$ field of view at Visible and Infrared Survey Telescope for Astronomy (VISTA) \citep{Sutherland15}. Detailed information on the data used in this work is given in Table~\ref{tab:data}.

\section{Analysis} \label{sec:analysis}
The analysis combines morphological, photometric, and spectroscopic diagnostics to characterise the stellar and gas kinematics and stellar populations of the RQ member galaxies. It allows us to construct a coherent view of how tidal interactions, gas flows, and feedback shape their current evolutionary state.

\begin{figure}
    \centering
    \includegraphics[width=9cm, height = 7.5cm]{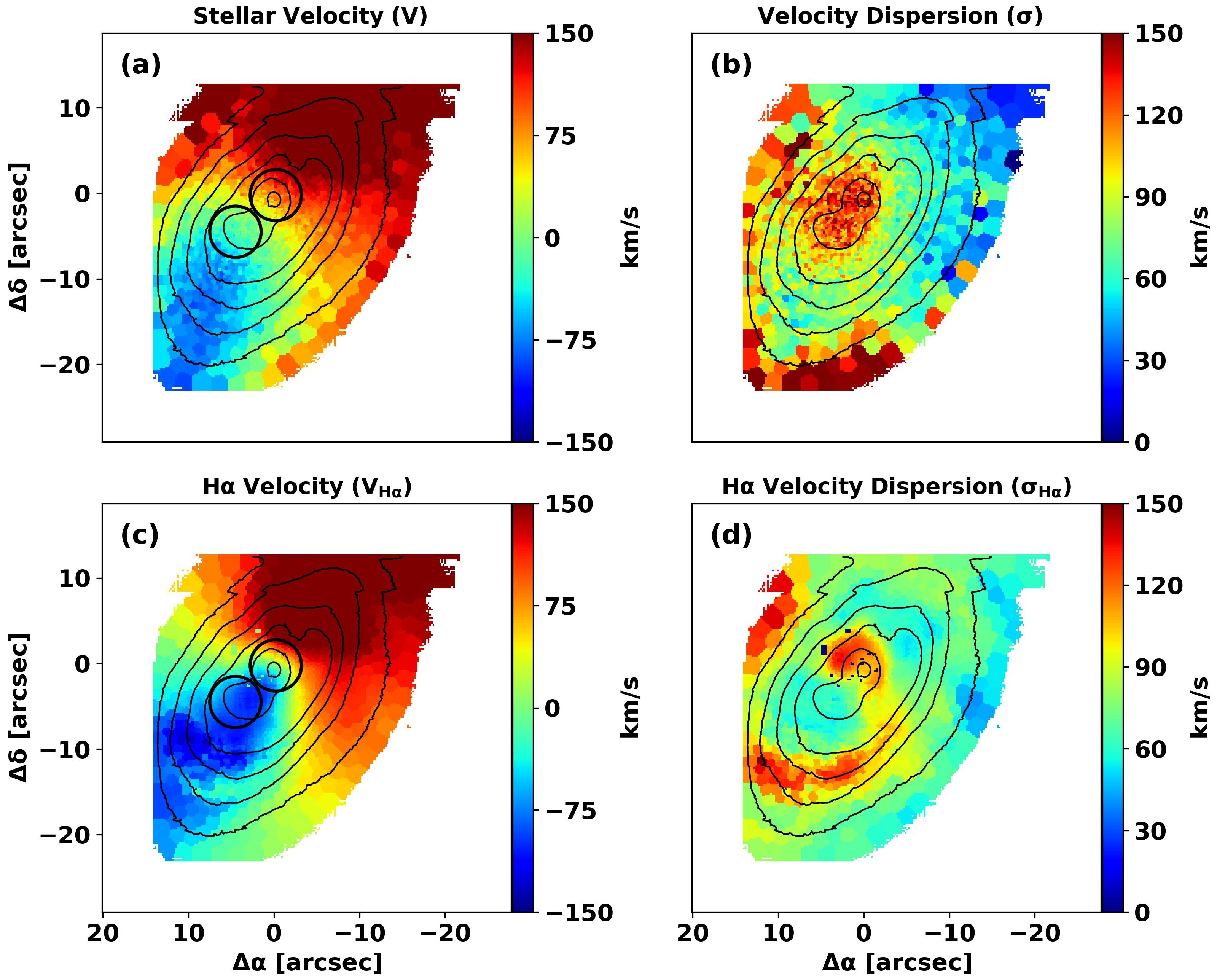}
   \caption{Spatially resolved maps of the stellar and gas kinematics of NGC 92 derived from MUSE observations. Top row: (a) Stellar velocity, (b) Velocity dispersion ($\sigma_{*}$), Bottom row: (c) Ionised gas (H$\alpha$) velocity, (d) Gas velocity dispersion ($\sigma_{H\alpha}$). In each panel, north is up and east is towards the left.}
    \label{fig:NGC92_kin}
\end{figure}

\begin{figure*}
    \centering
    \includegraphics[width=18cm, height =8cm]{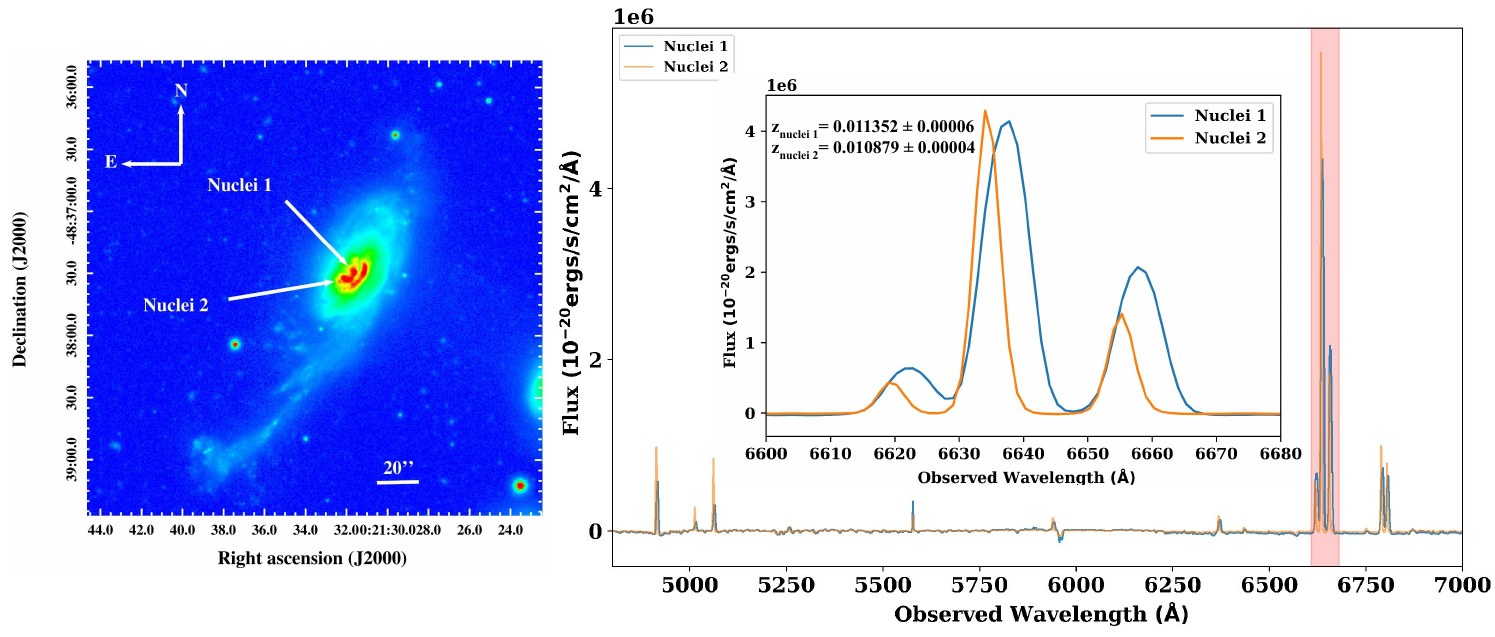}
    \caption{Left: DECaLS g-band image of NGC 92. Right: The MUSE spectra for nuclei 1 (blue) and nuclei 2 (orange) from the left panel, with zoomed-in views highlighting the H$\alpha$ and [NII] emission lines. Gaussian profiles are used to model these lines to calculate the redshifts for both nuclei.}
    \label{fig:92_spectra}
\end{figure*}

\subsection{Global Properties using SED modelling}\label{sec:sed}
To characterise the global properties of NGC 87, NGC 88, NGC 89, and NGC 92, we performed broadband spectral energy distribution (SED) modelling. The analysis required multi-band photometry, which was obtained by convolving all images to match the \textit{WISE} W4 point-spread function (PSF). Source detection and segmentation maps were generated using the \texttt{SExtractor} software \citep{Bertin96}, adopting a detection threshold of 3$\sigma$ on the \textit{DECaLS} $g$-band image of each galaxy. The resulting segmentation maps were applied uniformly across all other bands to extract consistent flux measurements. Photometry was obtained in 13 bands spanning from the UV to the mid-IR: \textit{GALEX} (FUV, NUV), \textit{DECaLS} ($g$, $r$, $i$, $z$), 2MASS ($J$, $H$, $Ks$), and \textit{WISE} (W1, W2, W3, W4). In addition, VISTA $Ks$-band imaging was only used for the structural decomposition and GALFIT modelling owing to its higher spatial resolution.

The observed UV–optical–IR SEDs were modelled using the \texttt{Code Investigating GALaxy Emission} (\texttt{CIGALE}; \citealt{Boquien19}), which implements the energy-balance principle for SED fitting to derive key physical properties such as the star formation rate (SFR), stellar mass ($M_{\star}$), stellar population age, dust attenuation, and star formation history (SFH) parameters. \texttt{CIGALE} constructs model spectra through a Bayesian analysis framework by combining multiple physical modules. We adopted a delayed SFH with an optional burst/quench component (\texttt{sfhdelayeddbq}), coupled with \citet{Bruzual03} single stellar population (SSP) models assuming a Salpeter initial mass function (IMF; \citealt{Salpeter55}) with mass limits of 0.1–100 $M_{\odot}$. The nebular emission from the interstellar medium was modelled using the \texttt{nebular} module, and dust attenuation was treated using the modified starburst law of \citet{Calzetti00} with zero UV bump amplitude. Infrared emission from dust heated by stellar radiation was modelled using the dust emission models of \citet{Draine07, Draine14}. To account for the contribution from an active galactic nucleus (AGN), we included the \texttt{SKIRTOR} models of \citet{Stalevski16}. A summary of the \texttt{CIGALE} modules and their corresponding input parameters is presented in Table~\ref{tab:sed}. The derived stellar masses and global star formation rates for all four member galaxies are summarised in Table~\ref{tab:sed_out}.

\begin{table}[h]
    \centering
        \caption{Stellar mass (M$_{*}$) and SFR derived from broad-band SED analysis.}
        \begin{tabular}{lcr}
        \hline
        Source & Stellar mass & SFR \\
        & ($M_{\odot}$) & ($M_{\odot}$yr$^{-1}$)\\
        \hline
        NGC 87 & 1.87 $\times$ 10$^{9}$ & 1.57 $\pm$ 0.19 \\
        NGC 88 & 6.26 $\times$ 10$^{9}$ & 0.39 $\pm$ 0.05 \\ 
        NGC 89 & 2.70 $\times$ 10$^{10}$ & 0.16 $\pm$ 0.03 \\
        NGC 92 & 5.61 $\times$ 10$^{10}$ & 1.08 $\pm$ 0.14\\
        \hline
        \end{tabular}
    \label{tab:sed_out}
\end{table}

\subsection{MUSE data cube analysis}\label{sec:MUSE}
We utilised the MUSE WFM-NOAO (Wide Field Mode, no adaptive optics) science-ready data cube of the RQ for the spectroscopic analysis. The data were processed using version 3.1.0 of the \texttt{Galaxy IFU Spectroscopy Tool} (\texttt{GIST}; \citealt{Bittner19}), a comprehensive framework designed for analysing fully reduced integral-field spectroscopic observations. The \texttt{GIST} pipeline employs the penalised pixel-fitting code (\texttt{pPXF}; \citealt{Cappellari04, Cappellari17, Cappellari2022}) to derive the stellar line-of-sight velocity distribution (LOSVD) and associated stellar kinematics, while the Gas and Absorption Line Fitting (\texttt{GandALF}) module \citep{Sarzi06, Falcon06} is used to fit emission lines and extract gaseous kinematics and line fluxes. Spatial binning was performed using the Voronoi tessellation method \citep{Cappellari03} within the wavelength range 480–700 nm to achieve a target signal-to-noise ratio (S/N) of 40 per bin. The S/N is computed for each unbinned spaxel per spectral pixel ($\approx$ per \AA) within the selected wavelength range. Spaxels with S/N $<$ 3 are excluded from the binning process to minimise the contribution of noisy data before binning. The remaining spaxels are then adaptively combined until the target S/N of 40 per bin is reached, whereas individual spaxels with S/N $>$ 40 were left unbinned. All spectra were subsequently corrected for galactic extinction in the direction of the target. A detailed discussion on MUSE analysis is presented in \citet{Keshri25b, Keshri25a}. 

\subsubsection{Non-parametric Star Formation History}
We derived global properties for each member of the group using SED modelling of multi-band images, which is discussed in detail in Sec~\ref{sec:sed}. Here we utilised MUSE cube to derive spatially resolved mass-weighted stellar populations and star formation histories (SFHs) of each galaxy in RQ using the SFH module of the \texttt{GIST} pipeline. This module employs the \texttt{pPXF} algorithm to perform a regularised full spectral fit, enabling the recovery of non-parametric SFHs and stellar population properties. The analysis was carried out on emission-line-subtracted spectra to ensure a robust continuum fit. Fitting was performed over the wavelength range 480–550 nm, incorporating an eighth-order multiplicative Legendre polynomial to correct for residual continuum mismatches between the observed and template spectra. The spectral modelling used a linear combination of templates from the MILES library \citep{Vazdekis10}, which spans stellar ages from 0.03 to 14 Gyr and metallicities ([M/H]) in the range of -2.27 to +0.40. The models are based on the BaSTI isochrones \citep{Pietrinferni04, Pietrinferni06, Pietrinferni09, Pietrinferni13} and assume a revised Kroupa initial mass function (IMF; \citealt{Kroupa01}).

\subsubsection{Gas-Phase Metallicity Analysis}
To estimate the spatially resolved oxygen abundance across all galaxies, we derived the gas-phase metallicity ($Z_{\mathrm{gas}}$) using the N2 diagnostic of \citet{Pettini04}, expressed as

\begin{equation}
12 + \log(\mathrm{O/H}) = 8.9 + 0.57 \times \log(\mathrm{N2}),
\label{eq:gas-phase}
\end{equation}

where N2 represents the flux ratio of the [NII] $\lambda6583$ to H$\alpha$ emission lines. For reference, the solar metallicity corresponds to $12 + \log(\mathrm{O/H}) = 8.66$ \citep{Asplund06}.

The N2 index is highly sensitive to the gas-phase oxygen abundance, responding to both ionisation and chemical effects. When (O/H) falls below the solar value, the ionisation parameter and/or hardness of the ionising radiation typically increase, leading to a decrease in the [NII]/[NIII] ratio. Conversely, at higher abundances, the secondary nature of nitrogen causes the (N/O) ratio to decline. These dependencies make N2 an efficient proxy for tracing metallicity variations. Using the resulting $Z_{\mathrm{gas}}$ maps, we investigated the radial and spatial metallicity gradients across each member of RQ, as discussed in the following sections.

\begin{figure*}[!t]
    \centering
    \includegraphics[width=18cm, height = 4.5cm]{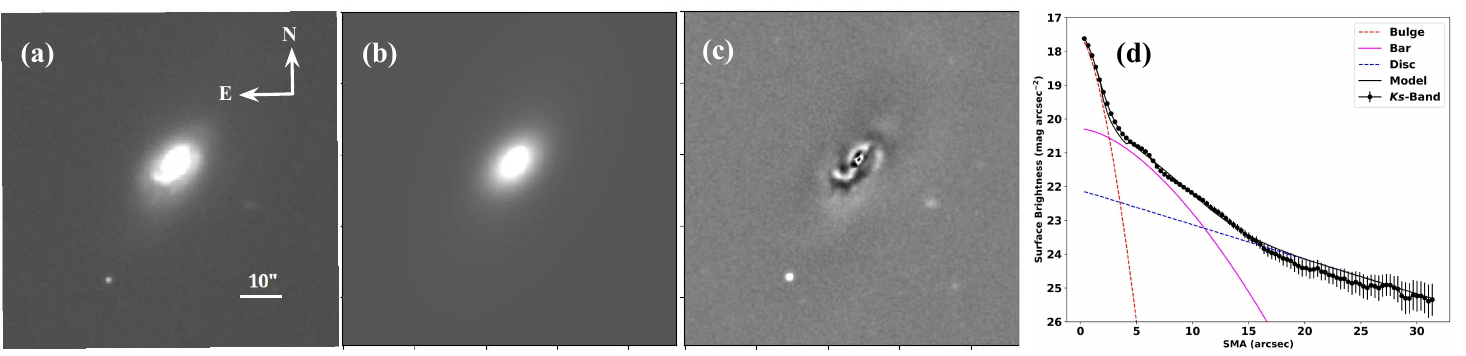}
    \caption{Output of the GALFIT model along with the azimuthally averaged surface brightness profile of NGC 92. (a) $Ks$-band image of the galaxy, (b) corresponding best-fit model, (c) residual image and (d) azimuthally averaged surface brightness of the best-fitting ellipses are plotted as a function of the semi-major axis (solid points), with the error bars representing the rms uncertainty of the intensity measured along each fitted isophote. The surface brightness profiles are decomposed into two Sérsic components and one exponential component, corresponding to a bulge (red dashed line), a bar (magenta dashed line), and a disc (blue dashed line). The black solid line shows the combined contribution of all three components.}
    \label{fig:SB}
\end{figure*}

\begin{figure}
    \centering
    \includegraphics[width=9cm, height = 8cm]{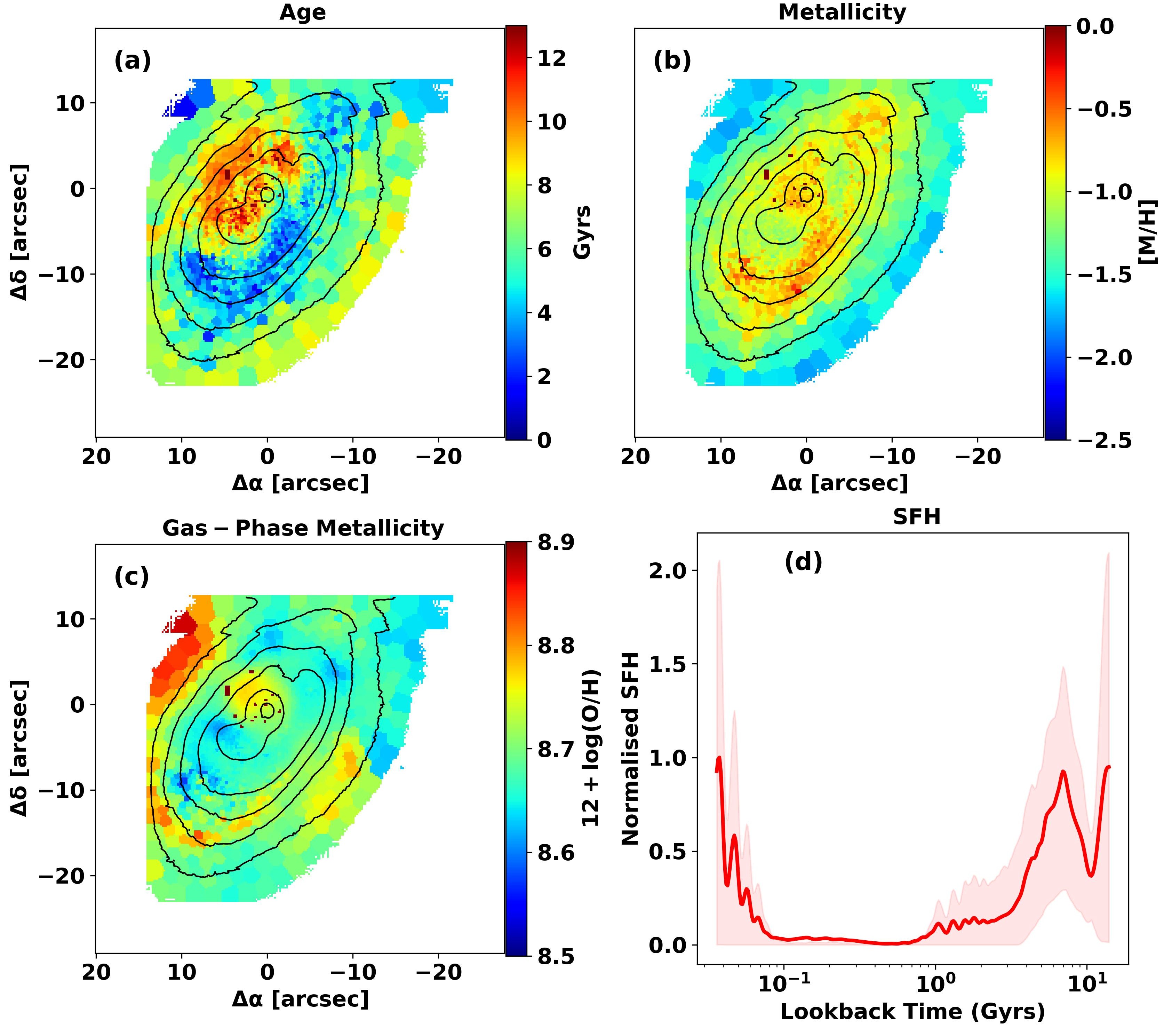}
    \caption{Spatially resolved maps of stellar population properties of NGC 92 derived from MUSE observations. (a) Mass-weighted stellar age, (b) Stellar metallicity ([M/H]), (c) Gas-phase metallicity (12 + log(O/H)), (d) Average mass-weighted star formation history (SFH) of the galaxy as a function of lookback time is presented. The red curve shows the mean SFH, while the shaded region represents the 16th–84th percentile range, indicating the spatial variation in SFH across the galaxy. The SFH is normalised to its peak value.}
\label{fig:NGC92_sfh}
\end{figure}

\subsubsection{BPT and WHAN classification}
Understanding the relative contributions of star formation and AGN activity in interacting systems is essential for disentangling the physical processes governing galaxy evolution. Interactions can compress the interstellar medium, triggering enhanced star formation, while also driving gas inflows that feed central supermassive black holes. To examine the dominant ionisation mechanisms within the galaxies of RQ, we employed spatially resolved emission-line diagnostics derived from the MUSE IFU data. Emission-line fluxes extracted from the MUSE cube were used to construct Baldwin–Phillips–Terlevich (BPT; \citealt{Baldwin81}) diagrams, which classify the ionisation sources based on characteristic nebular emission-line ratios. We used [OIII]$\lambda5007$/H$\beta$ versus [NII] $\lambda6583$/H$\alpha$ flux ratios, measured for each Voronoi bin across the MUSE field of view. These ratios separate regions ionised by hot, young stars in HII regions from those excited by harder radiation fields typically associated with AGN. The theoretical upper limit for pure star-forming regions was defined by \citet{Kewley01} using stellar population synthesis and photoionisation models. Galaxies located above this curve cannot be explained by star formation alone and are attributed to AGN-dominated ionisation. Large spectroscopic surveys, such as the SDSS, have shown that galaxies populate the BPT plane in a characteristic “seagull-shaped” distribution \citep{Kauffmann03}. The left branch corresponds to ionisation due to star formation, the right to AGN-dominated systems, and the intermediate “composite” region represents mixed ionisation from both mechanisms. The AGN branch itself bifurcates into a high-ionisation Seyfert sequence and a low-ionisation LINER sequence \citep{Schawinski07}, reflecting different physical excitation processes such as low-luminosity AGN, shock heating, or ionisation by evolved stars. We derived spatially resolved BPT classifications for the RQ galaxies. This approach enables us to map the spatial variation of ionisation mechanisms, identifying regions dominated by star formation, AGN activity, or a composite of both across the group members.

An important caveat in interpreting BPT diagnostics concerns the origin of emission-line ratios traditionally classified as LINER-like. \citet{Stasinska08} demonstrated that “retired galaxies (RG)”, which have ceased star formation, can occupy the same region of the BPT diagram as LINERs. In these systems, the ionising photons are not necessarily produced by accretion onto a supermassive black hole but may instead arise from hot, low-mass evolved stars, such as post-asymptotic giant branch (post-AGB) stars or white dwarfs, that can ionise the surrounding interstellar medium and produce optical line ratios similar to those of LINERs. Consequently, a significant fraction of galaxies located along the LINER branch of the BPT “right wing” may represent retired systems rather than genuine AGN. To mitigate this ambiguity, \citet{Cid11} introduced the WHAN diagram, which combines the [NII]/H$\alpha$ flux ratio with the equivalent width (EW) of H$\alpha$. This approach provides a more robust separation between true AGN and ionisation from evolved stellar populations. The inclusion of H$\alpha$ EW effectively distinguishes weak AGN (wAGN) from RGs (galaxies that have stopped forming stars and whose residual gas is ionised by their low-mass hot evolved stars) and passive galaxies (galaxies with no emission lines), thereby refining the classification of low-ionisation spectra beyond the limits of the traditional BPT framework.

\subsection{Dynamical state of RQ}\label{sec:dynamics}
The convenient way of estimating the dynamical state of a group is to measure the dimensionless crossing time, $H_{0}t_{c}$, where $H_{0}$ is the Hubble-Lemaitre constant. It measures the time taken by a galaxy to pass through the group and is estimated using the relation given by \citet{Hickson1992} as follows:

\begin{equation}
    t_{c} = \frac{4R}{\pi \sigma_{3D}}
    \label{eq:time}
\end{equation}

where R represents the median of the 2D galaxy-galaxy separation vector in kpc. We measured it by estimating the separation between each galaxy in arcsec, then determined the median of these separations in arcsec and converted it to the linear distance in kpc using the distance of RQ. $\sigma_{3D}$ is the 3D velocity dispersion and is defined as $\sigma_{3D} = \left[3\left(\langle v^2 \rangle - \langle v \rangle^2 - \langle \delta v^2 \rangle\right)\right]^{1/2}$, where $<v>$ is the average of the radial velocities of each galaxy in the group and $<\delta v^{2}>$ is the average of velocity errors squared. We utilised velocity and error as given in Table~\ref{tab:morp}.

We also estimated the LOS velocity dispersion ($\sigma_{G}$) of the group using a variant of the gapper estimator as described by \citet{Beers90}, since it is less biased for groups with a small number of members compared to the standard deviation. The method involves sorting the velocities ($v_{i}$) of each galaxy in ascending order and defining the gap as
\begin{equation}
    g_{i} = v_{i+1} - v{i},\qquad i = 1,2,...N-1
    \label{eq:gap}
\end{equation}
Then, the rest-frame LOS $\sigma_{G}$ is estimated using the relation given by \citet{yang05, Zheng21}
\begin{equation}
    \sigma_{G} = \frac{\sqrt{\pi}}{(1+z)N(N-1)}\sum_{i=1}^{N-1} w_{i}g_{i}
    \label{eq:dispersion}
\end{equation}
Here $z$ is the CG redshift, and $w_{i}$ is the weight defined as $w{i}$ = $i(N-i)$, where N represents the number of members in CG.

Under the assumption that one of the galaxies is moving at the centre-of-mass velocity of the group, $\sigma_{G}$ should be corrected by a factor of $N(N-1)$ \citep{Eke04}. Also, redshift measurement errors increase the estimate of $\sigma_{LOS}$; thus, it should be removed in quadrature. The $\sigma_{LOS}$ of each group is given by
\begin{equation}
    \sigma_{LOS} = \sqrt{max \left( 0, \frac{N\sigma_{G}^{2}}{N-1} - <\delta v^{2}> \right)}
    \label{eq:dispersion_los}
\end{equation}
Our estimates of group crossing time and velocity dispersion suggest that RQ is a dynamically young CG with a crossing time ($t_{c}$) of $\sim$ 424 Myr ($log(H_{0}t_{c}) \sim -1.52$) and a velocity dispersion of $\sim$74 km/s.

\section{Results} \label{sec:result}

RQ provides an excellent nearby laboratory for investigating the impact of galaxy interactions on star formation, stellar populations, and gas dynamics in compact-group environments. Although the member galaxies reside within the same gravitational potential, they exhibit a wide diversity in morphology, stellar mass, star formation activity, and evolutionary state. This diversity highlights the complex and asynchronous nature of galaxy evolution in compact groups. In the following sections, we analyse each galaxy individually, focusing on its structural properties, stellar populations, ionised-gas distribution, and kinematic signatures. Together, these properties trace different stages of interaction-driven evolution and transformation within the group environment.

\begin{table}
	\centering
        \caption{Output of three-component (bulge, bar and disc) GALFIT modelling of NGC 92 in VIRCAM/VISTA $Ks$-band image.}
	\begin{tabular}{lccccr}
		\hline
        \hline
        Bulge & R$_{e}$ (kpc) & 0.28 \\
              & n  & 0.65\\
        \hline
        Bar & R$_{e}$ (kpc) & 1.23 \\
            & n & 0.62 \\
        \hline
        Disc & R$_{s}$ (kpc) & 2.31 \\          
        \hline		
	\end{tabular}
    \begin{flushleft}
    \footnotesize
    \textbf{Description:} {R$_e$ is the effective radius in kpc for the bulge and bar, R$_s$ is the disc scale length in kpc, and n is the Sersic index.}
    \end{flushleft}
	\label{tab:galfit}
\end{table}

\begin{figure}
    \centering
    \includegraphics[width=9cm, height = 8cm]{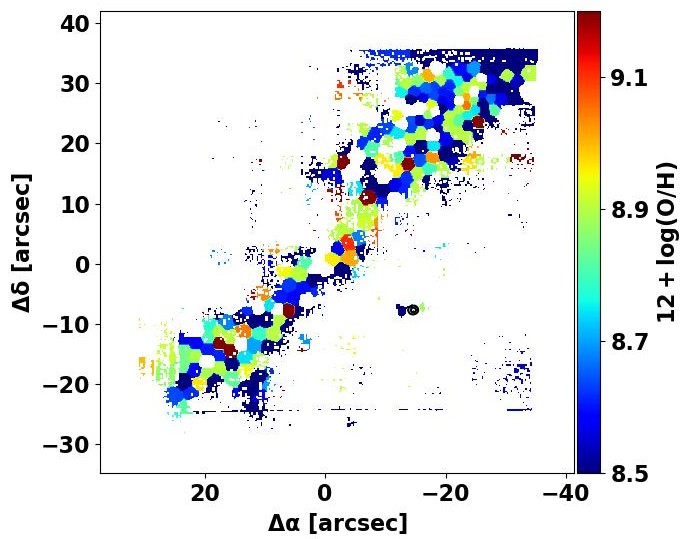}
    \caption{Spatially resolved gas-phase metallicity (12 + log(O/H)) of NGC 92 tail (cube 3 of Fig.~\ref{fig:group}).}
\label{fig:NGC92_tail}
\end{figure}

\subsection{NGC 92}
NGC~92 is the most massive and actively star-forming member of RQ, with a stellar mass of $\sim5.6\times10^{10}~M_{\odot}$ and a total SFR of $\sim$1.1 $M_{\odot}~{\rm yr^{-1}}$, as derived from  SED modelling. The \textit{GALEX} FUV imaging reveals six prominent star-forming clumps in the galaxy, including the tidal tail, which is extending up to $\sim$30 kpc \citep{Temporin05}. The derived ages of these clumps range from 27 to 204 Myr, with the central clump having an age of $\sim$204 Myr (Fig.~\ref{fig:age_NGC92}). The tidal tail has also been detected in X-ray observations of the group \citep{Trinchieri08}.

In optical images, NGC~92 exhibits a morphology reminiscent of a ring structure. Interactions are known to produce ring-like morphologies when an intruder passes close to a massive gas-rich disc galaxy \citep{Lynds76, Theys76}. Such encounters can additionally generate one-armed or asymmetric tidal structures, as demonstrated in the simulations of \citet{Toomre78}, whose final configurations closely resemble the morphology of NGC~92. The galaxy also hosts a candidate tidal dwarf galaxy within the tidal tail \citep{Torres-Flores09}, a double nucleus \citep{Danks81}, and a prominent dust lane. These features collectively indicate that NGC~92 is undergoing strong interaction-driven evolution and likely plays a central role in shaping the dynamical state of the group.

To investigate the nature of the central structure in NGC~92, we examine the stellar and ionised-gas kinematics derived from the MUSE data. The stellar line-of-sight velocity field within the central $36\arcsec \times 36\arcsec$ region (Fig.~\ref{fig:NGC92_kin}a) exhibits a well-ordered rotational pattern with amplitudes reaching $\pm$150 km/s. While the velocity field is broadly regular at intermediate radii, the innermost region shows subtle twists in the zero-velocity curve and isophotal asymmetries, suggesting the presence of non-axisymmetric structures such as a bar or perturbations induced by recent tidal interactions.

The H$\alpha$ velocity field (Fig.~\ref{fig:NGC92_kin}c) broadly follows the stellar rotation pattern, although stronger asymmetries are observed in the central region. Within this region, we identify two high surface-brightness components (hereafter nuclei~1 and nuclei~2), marked in the left panel of Fig.~\ref{fig:92_spectra}. The right panel of Fig.~\ref{fig:92_spectra} shows the integrated spectra extracted from these nuclei, together with a zoomed-in view of the H$\alpha$ $\lambda6563$\AA\ and [NII] $\lambda\lambda6548,6583$\AA\ emission lines used for the redshift measurements. Gaussian fitting yields systemic redshifts of $z_{\rm nuclei1}=0.011352\pm0.00006$ and $z_{\rm nuclei2}=0.010879\pm0.00004$, corresponding to a velocity offset of $\sim$140 km/s. Although this offset could initially suggest dynamically distinct components, both nuclei lie along the same large-scale rotational gradient in the stellar and gas velocity fields (Fig.~\ref{fig:NGC92_kin}a,c), with no clear evidence for kinematic decoupling. This indicates that the observed velocity difference is consistent with internal motions within the rotating disc rather than two separate galaxies.

To further investigate the central structure of NGC~92, we analysed the near-infrared morphology using the VIRCAM/VISTA $Ks$-band image. We performed a two-dimensional photometric decomposition with GALFIT \citep{Peng02}. A detailed description of the decomposition methodology is provided in \citet{Keshri25a}, and the same procedure was adopted here to model the bulge, bar, and disc components. The bulge and bar were fitted with Sérsic profiles, while the disc was modelled with an exponential profile. The resulting structural parameters are listed in Table~\ref{tab:galfit}. The observed $Ks$-band image, best-fit model, residual map, and one-dimensional surface-brightness profiles are presented in Fig.~\ref{fig:SB}.

\begin{figure}
    \centering
    \includegraphics[width=9cm, height = 7.2cm]{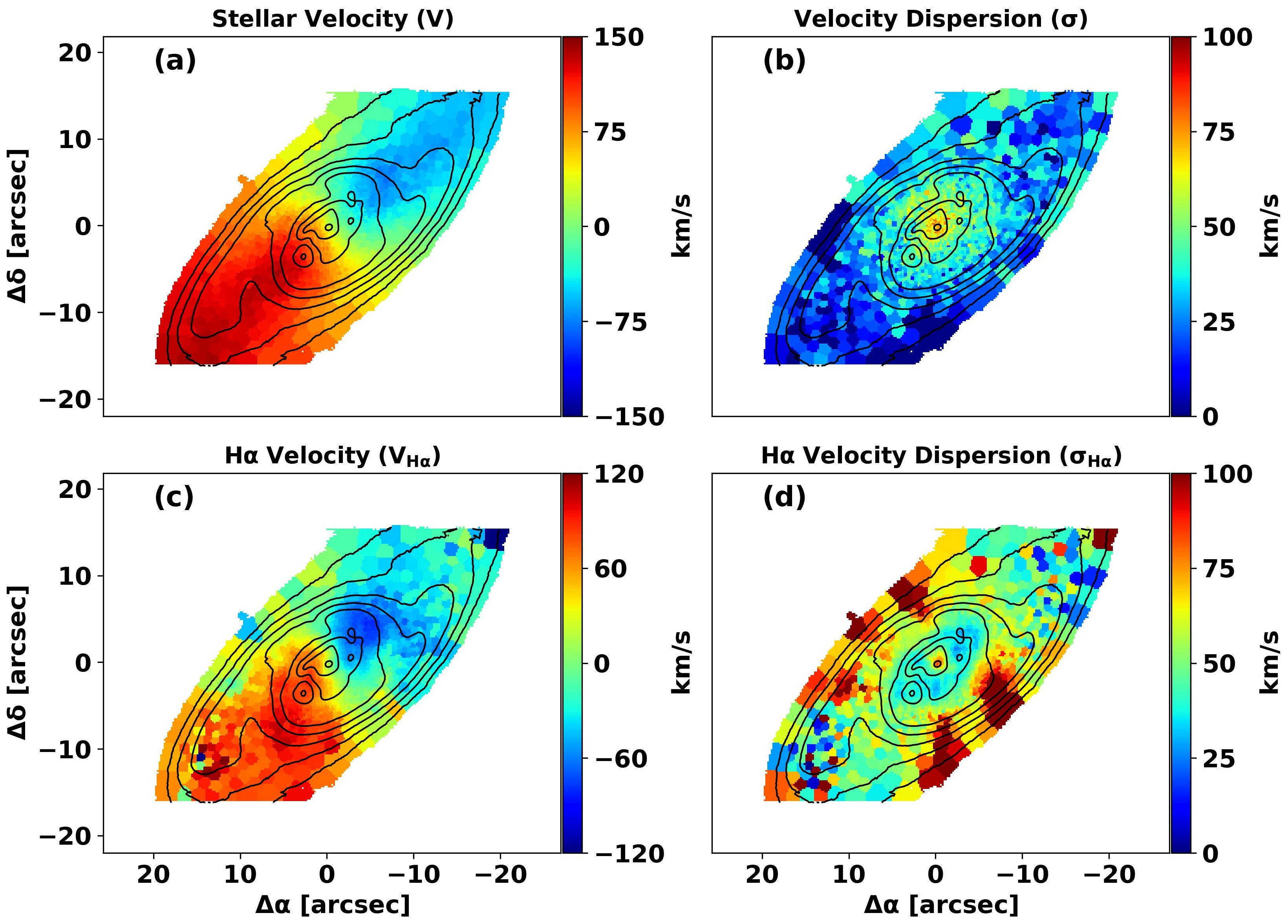}
    \caption{Spatially resolved maps of stellar and gas kinematics of NGC 89 derived from MUSE observations. All the panels are the same as in Figure \ref{fig:NGC92_kin}.}
    \label{fig:NGC89_kin}
\end{figure}

Previous work by \citet{Presotto10} inferred the presence of a bar in NGC~92 based on long-slit kinematic analysis and visual inspection of $Ks$-band NIR image, although no photometric decomposition was performed. Our two-dimensional decomposition clearly reveals a prominent stellar bar with a length of 1.232 kpc. Combining the photometric and kinematic results, we find that the second "nucleus" does not correspond to a dynamically distinct component. Instead, it is associated with the inner spiral-arm structure, where enhanced surface brightness and projection effects likely mimic a secondary nucleus. This interpretation removes the need for a recent merger and suggests that the central morphology and kinematics of NGC~92 are primarily shaped by bar-driven dynamics and interaction-induced perturbations.

\begin{figure}
    \centering
    \includegraphics[width=9cm, height = 7.5cm]{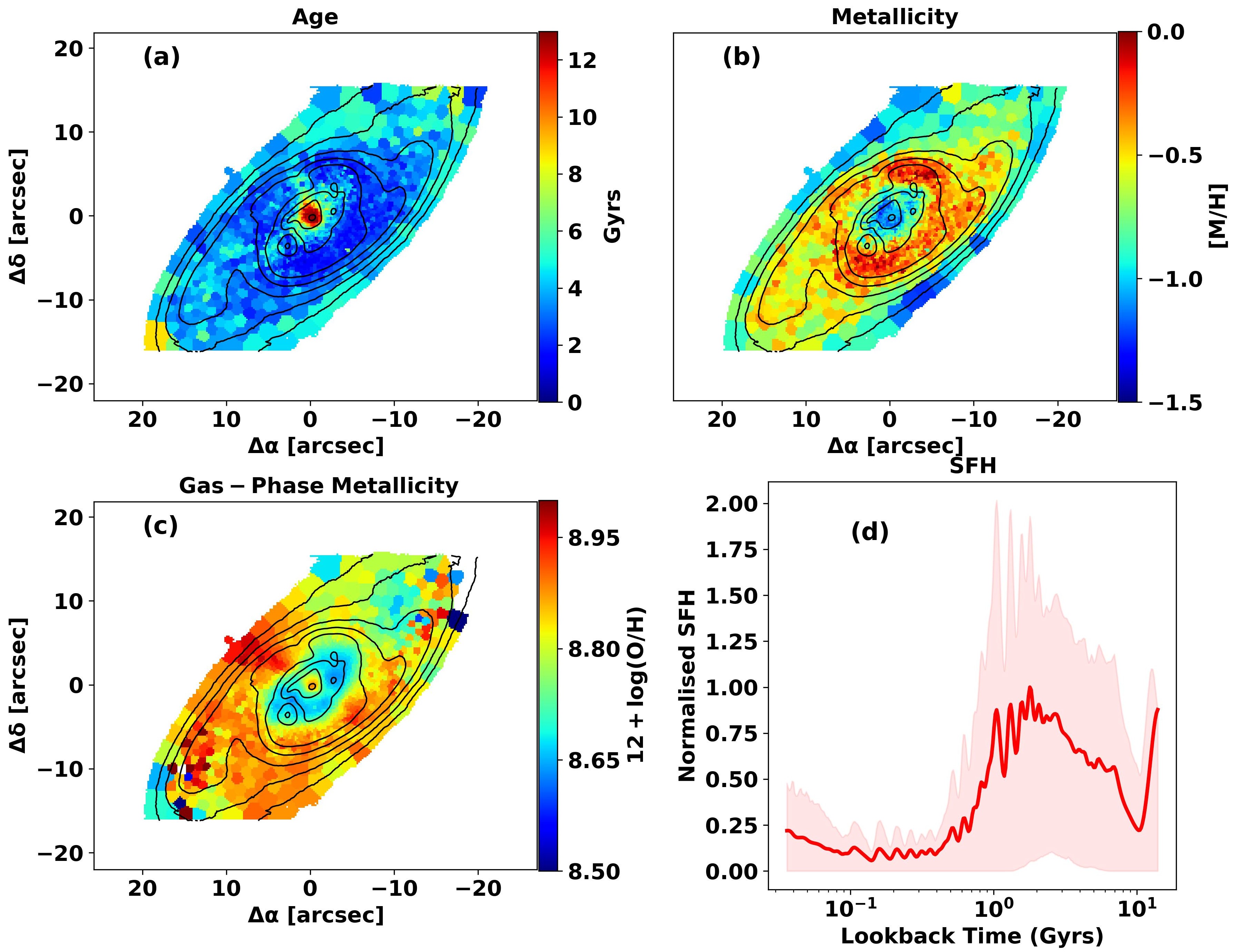}
    \caption{Spatially resolved maps of stellar population properties of NGC 89 derived from MUSE observations. All the panels are the same as in Figure \ref{fig:NGC92_sfh}.}
    \label{fig:NGC89_sfh}
\end{figure}

\begin{figure*}[h]
\centering
    \includegraphics[width=12cm]{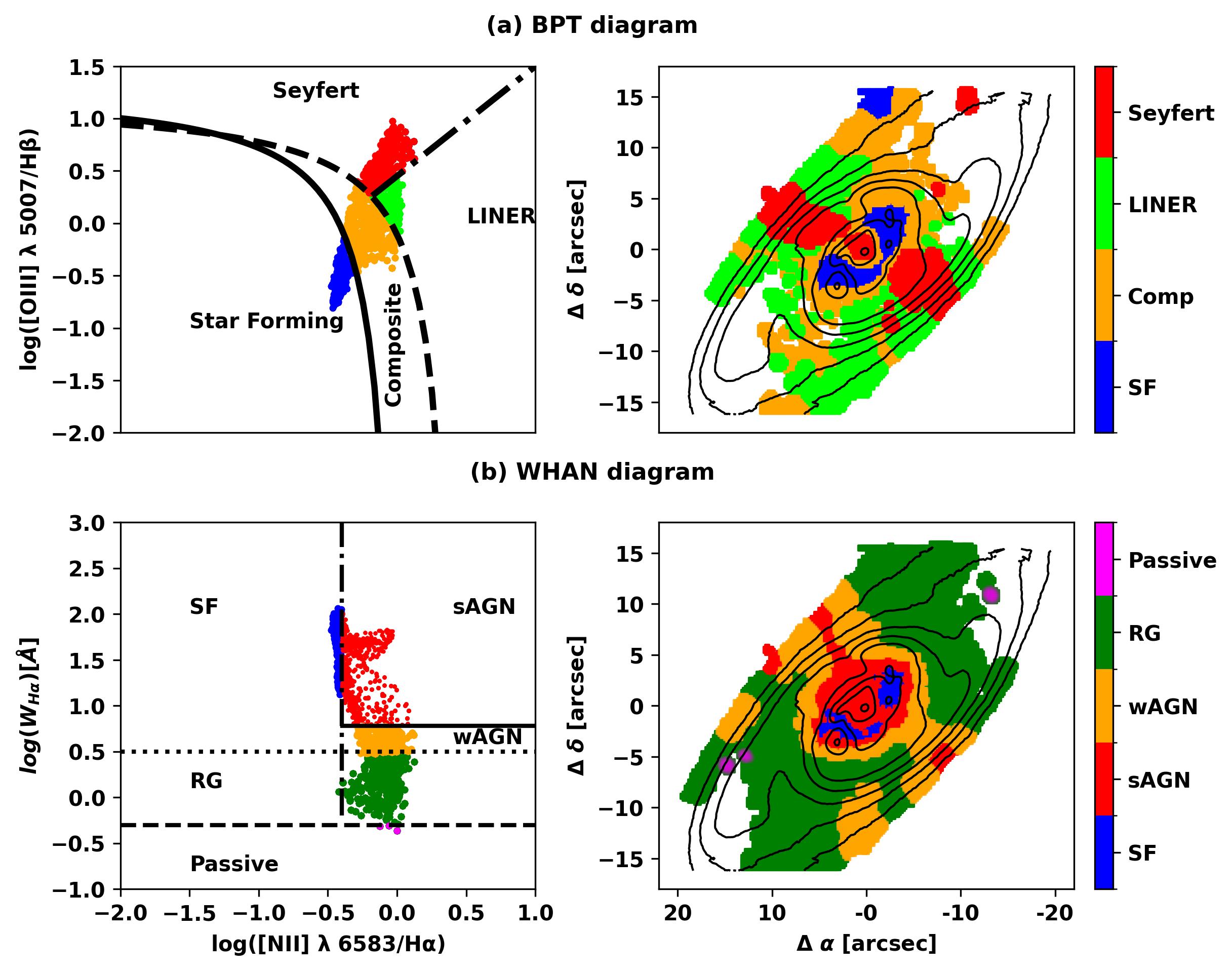}
    \caption{Ionisation diagnostic diagram of NGC 89. Top row: (left) BPT diagnostic diagram, classifying emission-line spaxels into star-forming (blue), composite (orange), LINER (green), and Seyfert (red) regions. (right) Spatial BPT map of the galaxy, showing a Seyfert-like nucleus surrounded by composite and LINER regions. Bottom row: (left) WHAN diagram plotting log([NII]/H$\alpha$) versus H$\alpha$ equivalent width, separating AGN, retired galaxies (RG), and star-forming regions. (right) Spatial WHAN map confirming a central strong AGN (sAGN) region surrounded by weak AGN (wAGN), RG, and SF ring-like ionisation in the disc. Contours in all spatial panels trace the white light continuum isophotes.}
    \label{fig:BPT_WHAN_NGC89}
\end{figure*}

The stellar velocity dispersion map ($\sigma_{*}$, Fig.~\ref{fig:NGC92_kin}b) further characterises the internal structure of NGC~92. Values exceeding $\sim$120 km/s are confined to the central kiloparsec, indicating the presence of a dynamically hot central component or enhanced mass concentration. The pronounced central increase in $\sigma_{*}$, together with the twisting of the inner velocity field, is consistent with bar-driven secular evolution that has funnelled gas toward the centre and contributed to the growth of a disc-like pseudobulge. The H$\alpha$ velocity dispersion map ($\sigma_{H\alpha}$, Fig.~\ref{fig:NGC92_kin}d) reaches values above 120 km/s in the circumnuclear region, indicating strong gas turbulence. The enhanced $\sigma_{H\alpha}$ also traces the prominent dust lane visible in the optical image. Such dynamical disturbances are likely associated with ongoing star formation and gas inflow driven by interactions within the group.

Stellar population diagnostics provide further evidence for the complex evolutionary history of NGC~92. The mass-weighted age map (Fig.~\ref{fig:NGC92_sfh}a) reveals an old stellar population (10-12 Gyr) concentrated in the central region, while significantly younger populations dominate the outer disc. These younger outer-disc populations are likely associated with tidally induced star formation, consistent with UV and optical imaging studies that show extended tidal features, disturbed spiral arms, and dust lanes \citep{Torres-Flores09, Torres-Flores14}.

The stellar metallicity map (Fig.~\ref{fig:NGC92_sfh}b) exhibits a pronounced radial gradient, indicating more efficient chemical enrichment in the central regions and comparatively lower metallicities in the outer disc. The gas-phase metallicity map (Fig.~\ref{fig:NGC92_sfh}c) shows oxygen abundances spanning $\sim$8.5-8.9, with localised enhancements across the disc. In contrast to the stellar component, the gas-phase metallicity distribution does not exhibit a radial gradient, suggesting efficient mixing of the ionised gas or recent inflows of lower-metallicity material that have flattened the abundance profile. The elevated gas-phase metallicities in the circumnuclear region ($\sim$8.8--8.9) coincide with regions of enhanced $\sigma_{H\alpha}$, supporting a scenario in which bar-driven inflows or interaction-induced shocks concentrate enriched gas near the centre, potentially sustaining recurrent circumnuclear star formation.

We further investigated the gas-phase metallicity along the tidal tail using a modified analysis of cube~3 of NGC~92, as indicated in Fig.~\ref{fig:group}. Voronoi binning was performed over the wavelength range 650-660 nm to achieve a target S/N of 20, while the remainder of the analysis followed the procedure described in Sec.~\ref{sec:MUSE}. The resulting spatially resolved metallicity map is shown in Fig.~\ref{fig:NGC92_tail}. The derived oxygen abundances span $\sim$8.5-9.1 dex, comparable to those measured in the galaxy's central regions.

Previous slit-spectroscopy measurements of star-forming regions along the tail by \citet{Torres-Flores14} reported slightly lower metallicities (8.45-8.67 dex). Nevertheless, both studies indicate that the disc and tidal tail share similar metallicities, implying a relatively flat abundance profile with no strong radial gradient. This suggests that enriched gas has been redistributed from the inner regions toward the outskirts, likely through recent interactions. The presence of a common HI envelope \citep{Pompei07}, together with the stellar mass ratios of NGC~88 to NGC~92 (1:8.8) and NGC~89 to NGC~92 (1:2.2), suggests that NGC~88 is the primary interacting companion, although the possibility of contribution from NGC~89 cannot be excluded.

These interactions appear to have strongly perturbed the outer disc while largely preserving the inner structure. The presence of a stellar bar in NGC~92 is consistent with this picture and resembles systems such as the Cartwheel galaxy, where head-on collisions disrupt the outer regions while leaving the inner structure comparatively intact \citep{Chayan24}. The combination of a stellar bar, enhanced central velocity dispersion, flattened gas-phase metallicity gradients, and young stellar populations indicates efficient gas mixing and interaction-driven gas redistribution sustaining recent star formation activity.

We also reconstructed the median SFH (Fig.~\ref{fig:NGC92_sfh}d) from the full spectral fitting, providing an understanding of the mass assembly history of NGC~92. The SFH reveals two dominant phases. The first corresponds to a major early episode around $\sim$10 Gyr ago, during which most of the stellar mass was assembled, consistent with the presence of an old population in the central region. The second component is significantly younger ($\sim$500 Myr) and is primarily associated with the disc and ring. This recent star formation episode is likely linked to tidally induced star formation triggered by the ongoing interaction with NGC~88.

\subsection{NGC 89}
NGC~89 is an SB0/a pec galaxy characterised by a relatively low SFR ($\sim$0.16 $M_{\odot}~{\rm yr^{-1}}$) and a high stellar mass ($M_{*}\sim2.7\times10^{10}~M_{\odot}$), as derived from the SED modelling. Its low SFR suggests that the galaxy is undergoing, or has recently undergone quenching, possibly driven by environmental processes such as gas stripping or internal feedback mechanisms. NGC~89 hosts a Seyfert~2 nucleus and H$\alpha$ emitting extraplanar filamentary structures on both sides of the disc, including a $\sim$4 kpc jet-like filament extending toward the NE \citep{Temporin05}. The SED fitting with \texttt{CIGALE} additionally requires an AGN component modelled using the \texttt{SKIRTOR} template, indicating a significant contribution of nuclear activity to the infrared emission. These properties support a scenario in which AGN feedback suppresses star formation in the galaxy. Gas inflow triggered during a previous interaction with other group members may have fuelled the AGN, while the resulting feedback contributed to the decline of star formation activity. \citet{Presotto10} estimated an interaction timescale of $\sim$200-700 Myr between NGC~89 and NGC~92, suggesting that gas exchange between the two galaxies may be linked to both the AGN activity and the extraplanar emission. In the sSFR-M$_{*}$ plane (Fig.~\ref{fig:ssfr-mass}), NGC~89 lies in the green-valley region, consistent with a transition or post-starburst evolutionary phase. It is also the only galaxy in the group with no HI detection \citep{Pompei07}.

The stellar velocity field (Fig.~\ref{fig:NGC89_kin}a) exhibits a regular and symmetric rotation pattern with amplitudes reaching $\pm$150 km/s. The stellar velocity dispersion ($\sigma_{*}$, Fig.~\ref{fig:NGC89_kin}b), shows a pronounced central peak, with values of $\sim$80-100 km/s in the inner few arcseconds, declining to $\sim$20-30 km/s toward the outskirts. Such elevated central $\sigma_{*}$ values are characteristic of a dynamically hot bulge or inner spheroidal component commonly associated with AGN-host galaxies. The H$\alpha$ velocity field (Fig.~\ref{fig:NGC89_kin}c) broadly follows the stellar rotation pattern. In contrast, the gas velocity dispersion ($\sigma_{H\alpha}$, Fig.~\ref{fig:NGC89_kin}d), exhibits widespread enhancements, particularly in the NE and SW regions, where values rise to $\sim$80-100 km/s. These elevated dispersions are indicative of turbulent gas motions, possibly associated with shocks or AGN-driven outflows.

The mass-weighted stellar age distribution (Fig.~\ref{fig:NGC89_sfh}a) reveals an old stellar population (10-12 Gyr) concentrated in the central region, while the disc is dominated by intermediate-age stellar populations($\lesssim$4 Gyr). This suggests that the central mass concentration was assembled at early epochs, while the disc experienced more extended or episodic star formation. The stellar metallicity distribution ([M/H]; Fig.~\ref{fig:NGC89_sfh}b) exhibits a pronounced radial gradient. The central few arcseconds are comparatively metal-poor, while metallicity reaches nearly solar values in a ring-like structure located at $\sim$8-10 arcsec from the centre, followed by a negative gradient at larger radii.

The gas-phase metallicity map (Fig.~\ref{fig:NGC89_sfh}c) spans $\sim$8.5-8.95 dex, with the highest values observed in the NE, south, and SW regions of the disc, possibly tracing localised enrichment or outflows of metal-rich gas. The SFH (Fig.~\ref{fig:NGC89_sfh}d) indicates an extended period of star formation over several Gyr, with comparatively little evidence for very recent star formation (within the last 1 Gyr) relative to the other galaxies in our sample. The lack of significant contributions from young stellar populations ($<100$ Myr) further supports the interpretation of NGC~89 as a quenched or post-starburst galaxy.

The ionisation diagnostics provide the clearest evidence for nuclear activity in NGC~89. The BPT diagram (Fig.~\ref{fig:BPT_WHAN_NGC89}, top left) classifies the emission-line spaxels using the [OIII]/H$\beta$ and [NII]/H$\alpha$ line ratios. The spaxels populate all four classical regimes: star-forming (blue), composite (yellow), LINER (lime), and Seyfert (red). The spatially resolved BPT map (Fig.~\ref{fig:BPT_WHAN_NGC89}, top right) confirms the presence of a Seyfert~2 nucleus surrounded by a star-forming ring, together with extended Seyfert-like emission.

The WHAN diagram (Fig.~\ref{fig:BPT_WHAN_NGC89}, bottom left) provides complementary constraints on the dominant ionisation sources by separating strong and weak AGN (sAGN and wAGN) from retired galaxies (RGs) and passive regions using both emission-line ratios and H$\alpha$ equivalent width. NGC~89 spans a broad range in the WHAN plane, with a substantial fraction of spaxels occupying the sAGN and wAGN regions, while others lie in the RG regime. The spatially resolved WHAN map (Fig.~\ref{fig:BPT_WHAN_NGC89}, bottom right) broadly mirrors the BPT classification, revealing a prominent sAGN nucleus surrounded by wAGN- and RG-like regions, along with a central star-forming ring. Extended sAGN emission toward the NE direction suggests the presence of a galaxy-scale outflow, consistent with the extraplanar emission reported by \citet{Presotto10}. However, no associated X-ray emission from the extended narrow-line region has been detected \citep{Trinchieri08}, despite the commonly observed correlation between H$\alpha$ outflows and X-ray features \citep{Bomans97, Strickland04, Zhang24}.

In addition to the Seyfert nucleus, NGC~89 hosts a young and metal-rich circumnuclear star-forming ring. The origin of both the nuclear ring and the outflow may be linked either to bar-driven gas inflow or to interaction-driven fuelling of the central black hole. Nuclear rings and discs are commonly observed in barred galaxies \citep{Gadotti19, Gadotti20, Keshri25a}, although merger-driven origins have also been proposed for unbarred systems \citep{Mayer08, Corsini12}. In merger-driven scenarios, nuclear discs are generally not expected to show strong radial gradients in stellar age and metallicity, unlike the well-defined gradients observed in barred galaxies \citep{Bittner2020}. Although the disturbed morphology of NGC~89 makes it difficult to constrain the presence of a bar photometrically, the clear stellar population gradients across the nuclear disc favour a bar-driven origin. These results suggest that bar-driven gas inflow is the most likely mechanism responsible for fuelling the AGN, driving the galactic-scale outflow, and sustaining the circumnuclear star-forming ring.

\begin{figure}[h]
    \centering
    \includegraphics[width=9cm, height = 8cm]{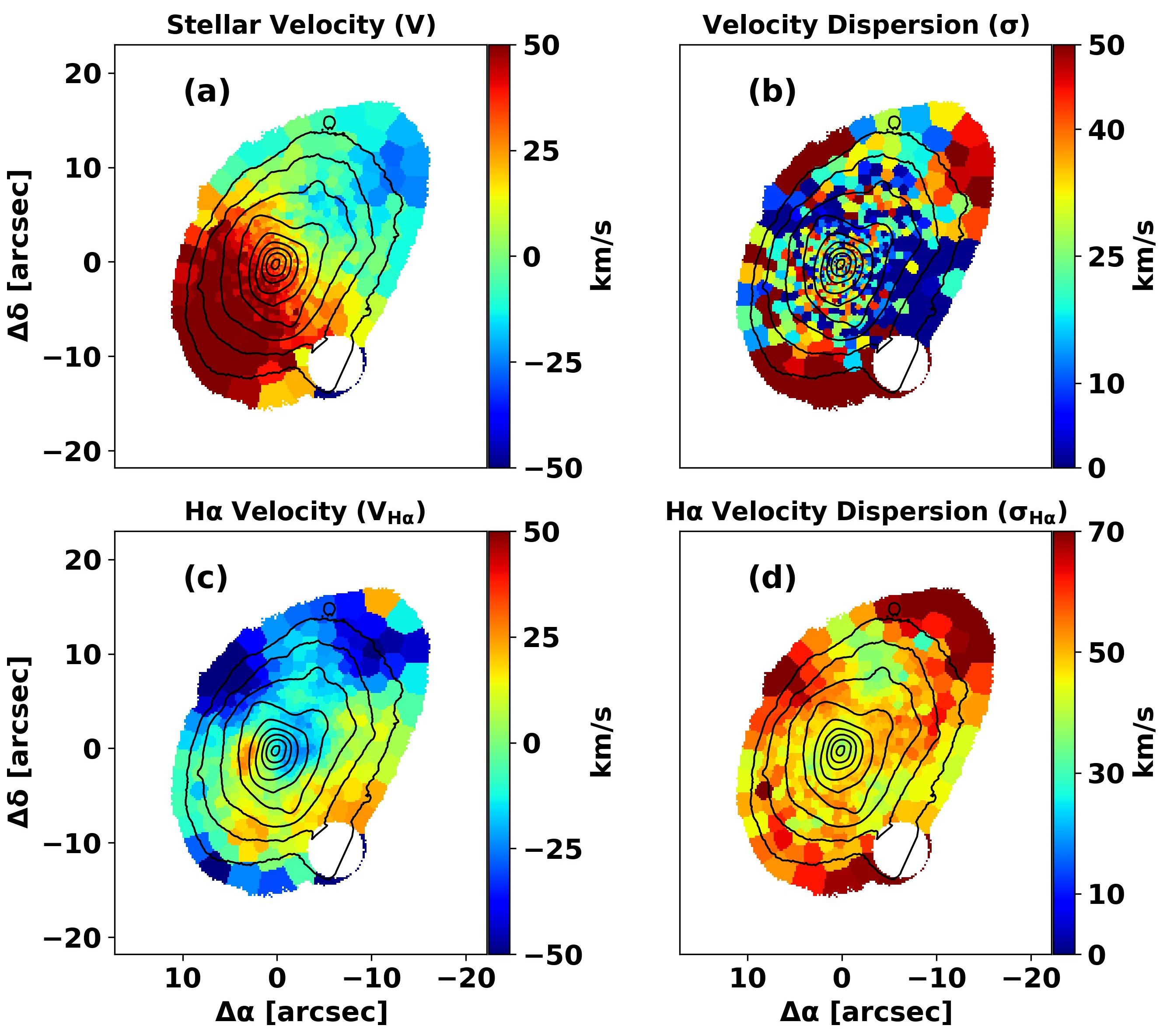}
    \caption{Spatially resolved maps of stellar and gas kinematics of NGC 88 derived from MUSE observations. All the panels are same as in Figure\ref{fig:NGC92_kin}.}
    \label{fig:NGC88_kin}
\end{figure}

\subsection{NGC 88}
NGC~88 is a peculiar S0 galaxy within RQ, exhibiting ongoing star formation, tidal features, and kinematically disturbed structures. SED modelling with \texttt{CIGALE} yields a stellar mass of $\sim6.26\times10^{9}M_{\odot}$, placing it among the low- to intermediate-mass systems in the group. The total SFR derived from the SED fit is ($\sim$0.4~$M_{\odot}~{\rm yr^{-1}}$), and in the sSFR–$M_{*}$ plane (Fig.~\ref{fig:ssfr-mass}), NGC 88 lies within the star-forming sequence. The FUV-based SFR is $\sim$0.44 $M_{\odot}~{\rm yr^{-1}}$, and the UV-derived age is $\sim$128 Myr. \citet{Pompei07} reported a common HI envelope between NGC~88 and NGC~92, suggesting that recent star formation may be triggered or sustained by ongoing gravitational interactions within the group.

The stellar velocity map (Fig.~\ref{fig:NGC88_kin}a) reveals a rotating disc with velocities reaching $\pm$50 km/s. The stellar velocity dispersion ($\sigma_{*}$, Fig.~\ref{fig:NGC88_kin}b) is generally low ($\sigma_{*} < 30$km/s), with localised enhancements up to $\sim$50 km/s in the outer regions. In contrast, the H$\alpha$ velocity field (Fig.~\ref{fig:NGC88_kin}c) is more irregular, with the core and outer disc exhibiting distinct rotation patterns. The outer gas disc is misaligned with respect to the stellar rotation axis, indicating kinematic decoupling and supporting a recent gas accretion scenario, while the central gas kinematics are broadly consistent with the stellar component. Such kinematic misalignments are commonly observed in S0 galaxies undergoing morphological transformation and are often attributed to minor mergers or external gas accretion \citep{McDermid07, Jin16, Bryant19, Zhou22, Ristea22, Zinchenko23}. Simulations show that misalignments between stellar and gas discs can persist for $\sim$2 Gyr, with the central regions realigning more rapidly than the outskirts \citep{van-de-Voort15}. The gas velocity dispersion ($\sigma_{H\alpha}$, Fig.~\ref{fig:NGC88_kin}d) is elevated in the outer regions and exhibits significant spatial variation, indicating turbulent gas dynamics, possibly driven by tidal heating.

\begin{figure}[h]
    \centering
    \includegraphics[width=9cm, height = 9cm]{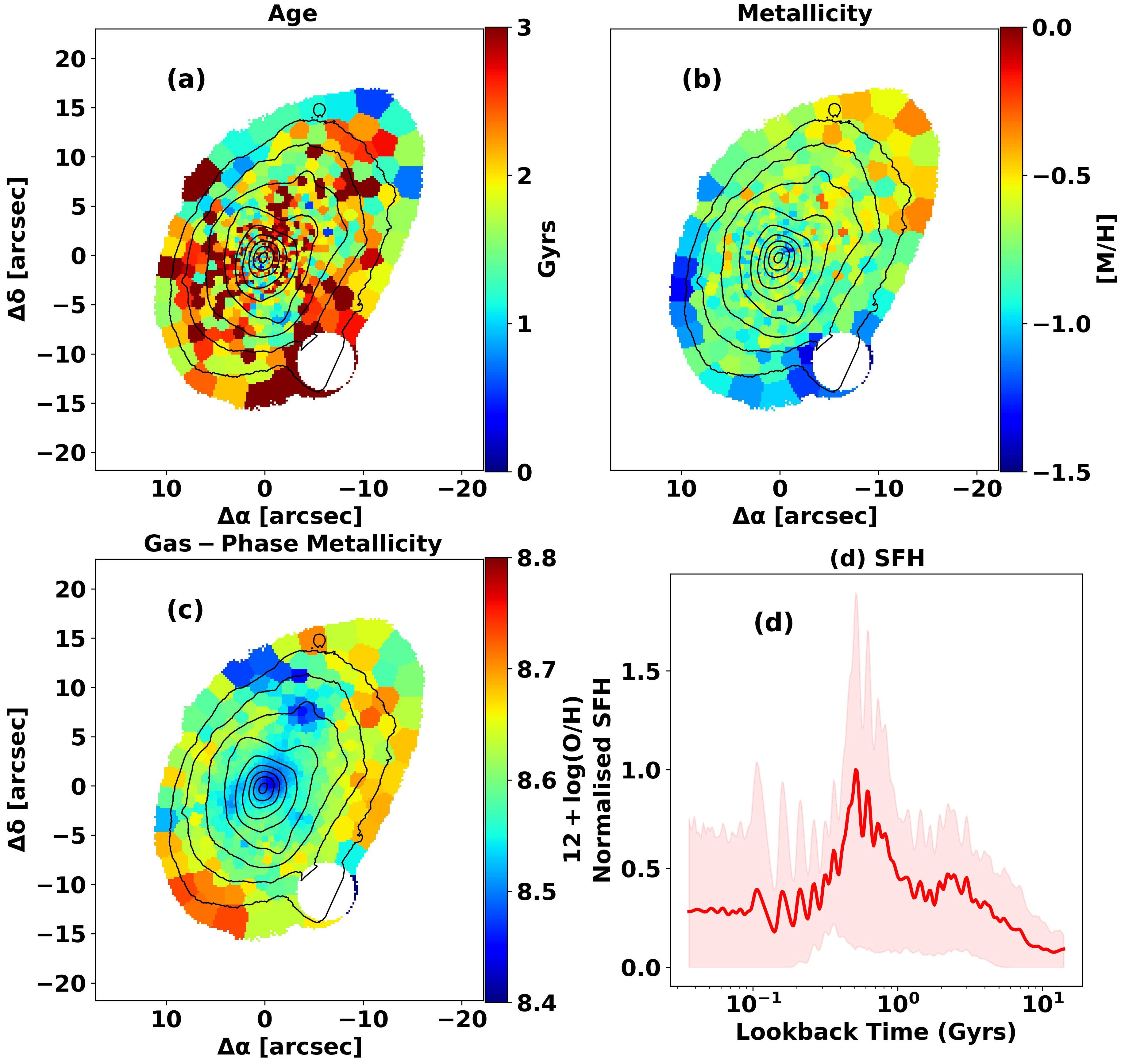}
    \caption{Spatially resolved maps of stellar population properties of NGC 88 derived from MUSE observations. All the panels are same as in Figure\ref{fig:NGC92_sfh}.}
    \label{fig:NGC88_sfh}
\end{figure}

The mass-weighted stellar age (Fig.~\ref{fig:NGC88_sfh}a) is dominated by an intermediate-age population ($\sim$1-3 Gyr). The stellar metallicity distribution ([M/H]; Fig.~\ref{fig:NGC88_sfh}b) spans $\sim$ -1.5 to -0.3 dex across the disc, with the highest values in the NW region. In contrast, the SE disc is comparatively metal-poor, indicating younger, less-evolved stellar populations, possibly associated with the inflow of metal-poor gas from NGC 92. The gas-phase metallicity (Fig.~\ref{fig:NGC88_sfh}c) ranges from $\sim$8.4 to 8.8 dex, with slightly lower values in the central regions than in the outskirts, and is broadly consistent with that of NGC~92. The average SFH distribution (Fig.~\ref{fig:NGC88_sfh}d) shows a prominent peak at $\sim$1 Gyr, indicating that a significant fraction of the stellar mass was assembled during this epoch.

The combined kinematic and stellar population properties of NGC 88 indicate ongoing gas transfer between NGC 88 and NGC 92, as evidenced by the observed kinematic misalignment between the stellar and gaseous discs. The stellar and gas-phase metallicity distributions further support this scenario.

\subsection{NGC 87}
NGC 87 is morphologically classified as an irregular and peculiar system. SED modelling with CIGALE yields a stellar mass of $M_{*}\sim 1.87\times 10^{9} M_{\odot}$ (Table~\ref{tab:sed_out}), placing it in the dwarf-galaxy regime. Dwarf galaxies typically have lower stellar masses ($M_{*}\le 10^{10} M_{\odot}$) and smaller physical sizes than galaxies along the Hubble sequence \citep{Lelli22}. \textit{GALEX} UV imaging reveals an extended UV disc, with an FUV-derived star formation rate (SFR) of 1.14 $M_{\odot} {\rm yr^{-1}}$ and a corresponding UV age of $\sim$32 Myr (Table~\ref{tab:sfr_age}). The combination of low stellar mass and elevated SFR classifies NGC 87 as a star-forming dwarf galaxy (SFDG). Approximately 70\% of dwarf galaxies are actively forming stars and are classified as SFDGs \citep{Karachentsev04}. Their compact sizes, low rotational velocities ($V_{rot} \le 100$ km/s), high gas fractions, and subsolar metallicities lead to star formation processes that differ from those in massive disc galaxies \citep{Bergvall12}. Stellar feedback and environmental effects further regulate their star formation activity \citep{Meurer95}. Although SFDGs are most commonly found in low-density environments with few massive neighbours \citep{Weisz14a, Weisz14b}, the presence of NGC 87 within RQ shows that such systems can also sustain active star formation in dense environments.

\begin{figure}
    \centering
    \includegraphics[width=9cm, height = 8cm]{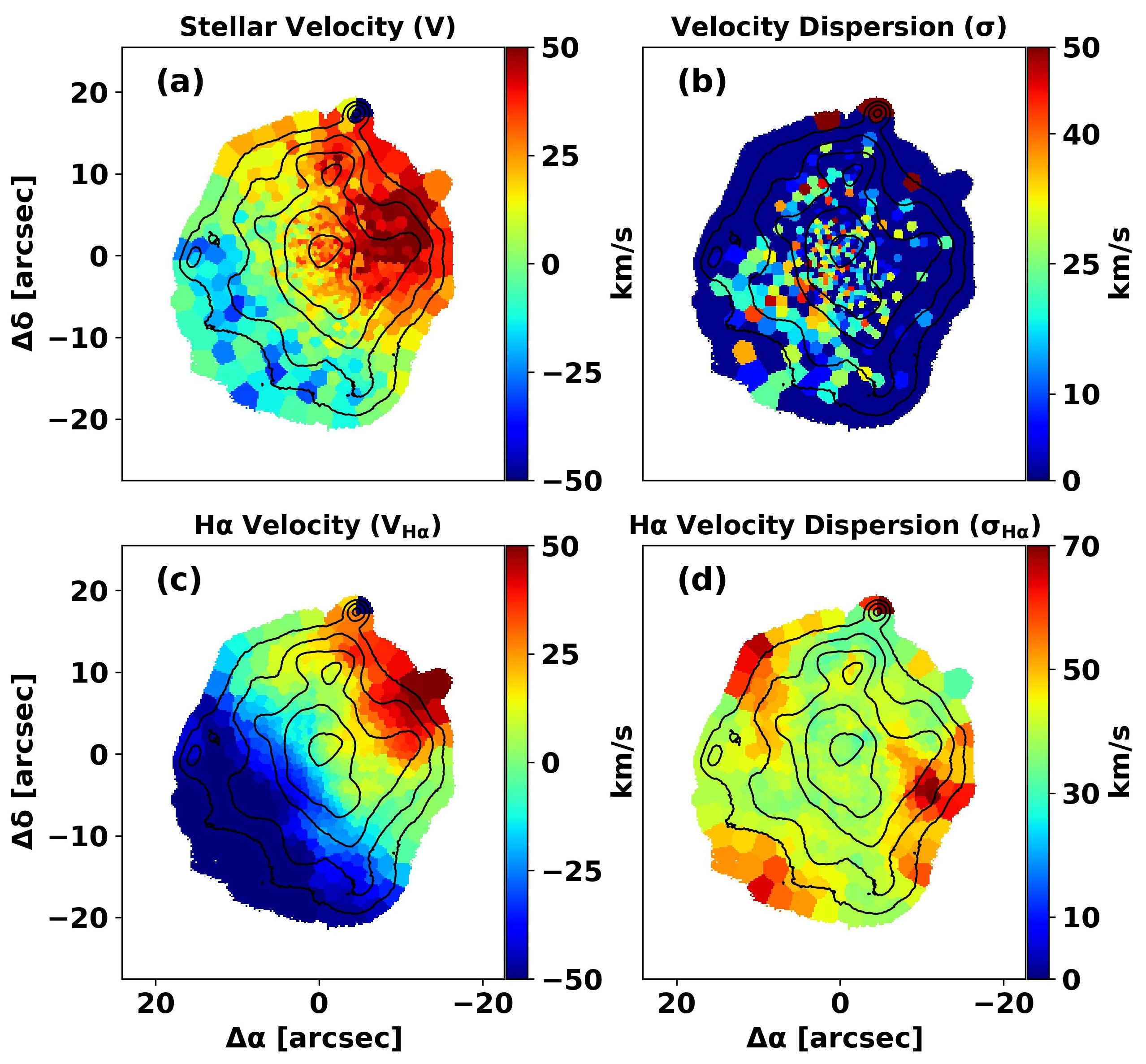}
    \caption{Spatially resolved maps of stellar and gas kinematics of NGC 87 derived from MUSE observations. All the panels are the same as in Figure\ref{fig:NGC92_kin}.}
    \label{fig:NGC87_kin}
\end{figure}

\begin{figure}
    \centering
    \includegraphics[width=9cm, height = 8cm]{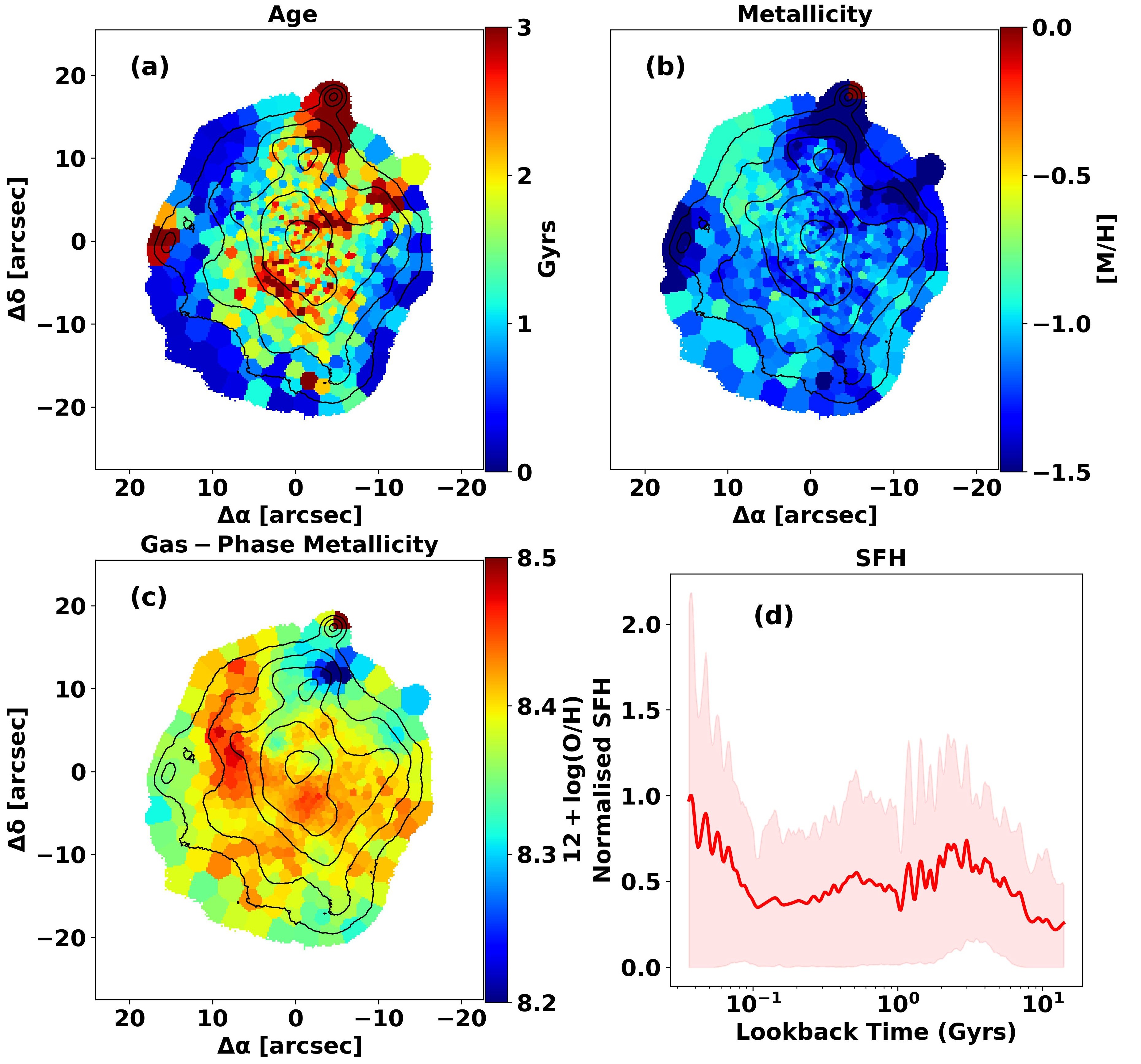}
    \caption{Spatially resolved maps of stellar population properties of NGC 87 derived from MUSE observations. All the panels are the same as in Figure\ref{fig:NGC92_sfh}.}
    \label{fig:NGC87_sfh}
\end{figure}

MUSE kinematic and stellar population maps of NGC 87 are shown in Fig.~\ref{fig:NGC87_kin} and Fig.~\ref{fig:NGC87_sfh}, respectively. Its stellar velocity field (Fig.~\ref{fig:NGC87_kin}a) exhibits a well-ordered rotation spanning $-50$ to $+50$ km/s, indicating the presence of a rotating disc despite the galaxy’s irregular morphology. The H$\alpha$ velocity field (Fig.~\ref{fig:NGC87_kin}c) closely follows the stellar kinematics, suggesting co-rotation of stars and gas. The velocity dispersion provides further insight into the dynamical state of the group. The stellar velocity dispersion ($\sigma_{*}$, Fig.~\ref{fig:NGC87_kin}b), remains low ($\sim$10-20 km/s) across most of the disc, consistent with a dynamically cold stellar component typical of dwarf galaxies \citep{Walker07}. In contrast, the gas velocity dispersion ($\sigma_{H\alpha}$, Fig.~\ref{fig:NGC87_kin}d), reaches values up to $\sim$70 km/s in the outer regions, indicating enhanced turbulence in the ionised gas component.

The mass-weighted stellar age map (Fig.~\ref{fig:NGC87_sfh}a) shows younger stellar populations ($\sim$0.5 Gyr) dominating the outer disc, while relatively older populations ($\sim$2--3 Gyr) are concentrated toward the centre. The stellar metallicity, [M/H] (Fig.~\ref{fig:NGC87_sfh}b), is subsolar across the galaxy, ranging from $-0.5$ to $-1.0$ dex, consistent with the low chemical enrichment expected in low-mass systems. The gas-phase metallicity (Fig.~\ref{fig:NGC87_sfh}c) is relatively uniform across the disc, with $12+\log({\rm O/H}) \sim$ 8.3-8.5. Low-mass galaxies with elevated star formation rates are known to exhibit lower metallicities \citep{Ellison08, Mannucci10, Yates12}, likely due to the removal of enriched material through stellar winds and supernova-driven outflows \citep{Mac99}. The reconstructed average SFH (Fig.~\ref{fig:NGC87_sfh}d) shows enhanced activity in recent epochs (100 Myr-1 Gyr), indicative of a bursty SFH.

NGC 87, despite its irregular morphology, exhibits ordered rotation. The low stellar and gas velocity dispersions ($\sigma_{*}$ and $\sigma_{H\alpha}$), together with a recent SFH and extended UV emission relative to the optical light, indicate the presence of young, massive stellar populations distributed across the galaxy. The subsolar stellar metallicity and relatively flat gas-phase metallicity gradient further suggest efficient mixing of metals throughout the disc. These results show that low-mass galaxies can sustain significant star formation even within a compact, interaction-dominated environment.

\section{Discussions}\label{sec:Disc}

The compact group Roberts Quartet (RQ) consists of four morphologically distinct galaxies in close proximity, spanning a stellar mass range of $9.2 < \log_{10}(M_{*}/M_{\odot}) < 10.8$. Despite sharing a common environment, the group exhibits a striking diversity in evolutionary states: three members are actively forming stars, while one galaxy is relatively quenched and appears to be transitioning toward a passive phase, likely driven by AGN activity and gas depletion within the group environment. Such coexistence of star-forming and quenched systems is a characteristic feature of compact groups, reflecting asynchronous galaxy evolution driven by repeated gravitational interactions, and RQ does represent a laboratory for studying CG evolution.

The interaction history of the RQ further supports this picture. Previous estimates suggest a recent encounter between NGC 92 and NGC 89 on timescales of $\sim$200–700 Myr \citep{Presotto10}, while our UV-based age estimates and non-parametric star formation histories indicate a similarly recent interaction epoch ($\lesssim$500 Myr) across all group members. This implies that the system is currently in an active phase of dynamical evolution, in which tidal interactions and gas exchange simultaneously trigger star formation and fuel nuclear activity. In this context, RQ can be regarded as a southern analogue of HCG~16, a well-studied compact group exhibiting strong kinematic disturbances and enhanced star formation. The presence of disturbed morphologies, asymmetric velocity fields, and widespread star formation in RQ reinforces the interpretation that tidal forces and repeated close encounters are efficiently redistributing angular momentum and funnelling gas toward galaxy centres, thereby sustaining both starburst and AGN phases. More broadly, this suggests that RQ represents a dynamically young compact group in which interaction-driven transformation is currently ongoing.

To further assess the dynamical state of RQ, we estimated the group crossing time and velocity dispersion (Sect.~\ref{sec:dynamics}). We derive a crossing time of $\sim$424 Myr ($log(H_{0}t_{c}) \sim -1.52$) and velocity dispersion of $\sim$74 km/s, placing the system in the early (stage 1) phase of compact group evolution as defined by \citet{Montaguth25}. Our estimate of crossing time differs from previous estimates \citep[e.g.,][]{Pompei07}, who treated RQ as a quintet by including an additional member. Their larger group membership led to a higher estimate of velocity dispersion. Our analysis considers only the four spectroscopically confirmed members of RQ, consistent with the current group definition. The presence of only one passive galaxy in the group further supports this early-stage classification, where most members are still actively forming stars.

The combination of short crossing time and low velocity dispersion indicates that RQ is a dynamically young system that has not yet reached virial equilibrium. Such conditions imply that the member galaxies are undergoing repeated close encounters on relatively short timescales, consistent with the interaction epoch inferred from UV ages and star formation histories ($\lesssim$500 Myr). The close agreement between the crossing time and the recent interaction timescale suggests that the current phase of enhanced star formation and AGN activity is directly linked to ongoing group assembly. Moreover, the observed kinematic disturbances and localised star-forming regions are characteristic of systems in which gravitational interactions are still actively redistributing mass and angular momentum. Taken together, these results support a scenario in which RQ is in the early stages of hierarchical assembly, where dynamical evolution, gas inflow, and star formation are tightly coupled.

The metallicity distribution of the group members provides additional insight into the assembly history of RQ. The predominantly sub-solar metallicities and relatively weak radial gradients are consistent with a scenario in which the system has been assembled through successive interactions and mergers involving low-mass, metal-poor progenitors. Such properties are commonly observed in dynamically young compact groups, where repeated encounters and gas mixing act to homogenise the chemical composition. In this context, RQ can be regarded as a nearby example of compact group formation driven by hierarchical assembly.

An important implication of our analysis emerges from the dynamical and structural complexity of NGC 92. The presence of kinematically distinct components, together with prominent tidal features and enhanced star formation, suggests that this system may be the remnant of a recent merger event rather than a single undisturbed spiral galaxy. If confirmed, this interpretation implies that the RQ may not represent a strictly four-member system, but rather the outcome of interactions involving additional progenitor galaxies, effectively making it a dynamically evolving group. This highlights the transient nature of compact groups, where ongoing mergers and accretion can modify both the internal structure of galaxies and the apparent group membership.

A similar picture is emerging from recent high-redshift observations with the \textit{James Webb Space Telescope} (JWST). Compact group-like systems identified at $z \sim 5$ \citep{Jin23} and $z \sim 7$ \citep{Wei26} exhibit dynamically young, non-virialised structures characterised by strong interactions, elevated star formation rates, and extreme ionisation conditions. These systems show clear signatures of rapid gas accretion, mergers, and efficient feedback operating on short dynamical timescales. The close resemblance between these properties and those observed in RQ suggests that the latter may represent a low-redshift analogue of such pre-virialised systems. In addition, recent JWST observations of Stephan’s Quintet have revealed the complex interplay between galaxy interactions, shocks, and the multiphase interstellar medium in compact environments, further reinforcing the connection between local and distant systems. As such, RQ provides a valuable nearby laboratory in which the physical processes governing galaxy assembly, gas accretion, and feedback have been studied in spatial detail, thereby offering a practical framework for interpreting JWST observations of galaxy evolution in the early Universe.

\section{Conclusions}\label{sec:Conc}
We have presented a multiwavelength analysis of the compact galaxy group Roberts Quartet (SCG0018-4854) using UV, optical, near-IR imaging, and spatially resolved MUSE IFU data. Our results indicate that the system is dynamically young and currently undergoing active interaction-driven evolution. The relatively low crossing time ($\sim$424 Myr), low velocity dispersion, and consistent interaction timescales ($\lesssim$500 Myr) derived from UV ages and star formation histories suggest that the group is in an early stage of assembly.

The member galaxies exhibit a diverse range of evolutionary states, including ongoing star formation, recent gas accretion, and AGN-driven quenching, highlighting the coexistence of multiple transformation pathways within a single galaxy group. The disturbed kinematics, enhanced star formation, and relatively uniform metallicity distributions further support a scenario in which repeated interactions and gas exchange are actively redistributing mass and angular momentum across the system.

A key result of this study is the dynamical and structural complexity of NGC~92, which shows evidence for a recent merger event. This suggests that RQ may not represent a strictly four-member system, but rather a dynamically evolving configuration involving additional progenitor galaxies. This highlights the transient nature of compact groups and the importance of hierarchical assembly in shaping their observed properties.

Finally, the physical conditions observed in RQ, characterised by strong interactions, enhanced star formation, and short dynamical timescales, closely resemble those seen in high-redshift galaxies. In this context, the system provides a valuable nearby analogue for compact, rapidly assembling structures now being probed with the \textit{JWST}. Our results demonstrate that detailed studies of local compact groups can offer important insights into the processes governing galaxy assembly, gas accretion, and feedback in the early Universe. Also, observations with ALMA could trace the molecular gas reservoirs and resolve the interplay between dense gas, star formation, and feedback processes, particularly in the ring-like structure of NGC 92. Complementary hydrodynamical simulations constrained by the specific mass distribution, orbital configuration, and gas content of RQ would allow a quantitative reconstruction of its interaction history and future evolution. Such models would be essential for testing the proposed origins of the ring structure, AGN fuelling, and gas mixing and for predicting the eventual fate of the group.

\begin{acknowledgement}
We are grateful to the anonymous referee for the thorough review and constructive comments, which have helped us improve the clarity and quality of the manuscript. MD and SB gratefully acknowledge the support of the Department of Science and Technology (DST) grant DST/WIDUSHI-A/PM/2023/25(G) for this research, and the support of the Science and Engineering Research Board (SERB) Core Research Grant CRG/2022/004531 for this study. SK thanks Dr Ramya Sethuram for long discussions. This paper has used the observations collected at the European Southern Observatory under ESO program 0101.C-0329. This research has also used UV data from GALEX, optical images from the Dark Energy Camera (DECam), which was constructed by the Dark Energy Survey (DES) collaboration. This work is based on observations made with VIRCAM/VISTA and WISE. This research has made use of NASA/IPAC Extragalactic Database (NED), which is operated by the Jet Propulsion Laboratory, California Institute of Technology (Caltech) under contract with NASA. 
\end{acknowledgement}
\bibliographystyle{aa}
\bibliography{Reference}

@article{Bacon2010,
  author  = {Bacon, R. and Accardo, M. and Adjali, L. and Anwand, H. and Bauer, S. M. and Biswas, I. and Blaizot, J. and Boudon, D. and Brau-Nogue, S. and Caillier, P. and others},
  title   = {The MUSE second-generation VLT instrument},
  journal = {Proc. SPIE},
  year    = {2010},
  volume  = {7735},
  pages   = {773508},
  doi     = {10.1117/12.856027}
}

@article{Hickson1982,
  author  = {Hickson, P.},
  title   = {Systematic properties of compact groups of galaxies},
  journal = {ApJ},
  year    = {1982},
  volume  = {255},
  pages   = {382--391},
  doi     = {10.1086/159838}
}

@article{Hickson1992,
  author  = {Hickson, P. and Mendes de Oliveira, C. and Huchra, J. P. and Palumbo, G. G.},
  title   = {A photometric catalog of compact groups of galaxies},
  journal = {ApJ},
  year    = {1992},
  volume  = {399},
  pages   = {353--377},
  doi     = {10.1086/171934}
}

@article{Barnes1992,
  author  = {Barnes, J. E. and Hernquist, L.},
  title   = {Dynamics of interacting galaxies},
  journal = {Nature},
  year    = {1992},
  volume  = {360},
  pages   = {715--718},
  doi     = {10.1038/360715a0}
}

@article{Springel2005,
  author  = {Springel, V. and Di Matteo, T. and Hernquist, L.},
  title   = {Modelling feedback from stars and black holes in galaxy mergers},
  journal = {MNRAS},
  year    = {2005},
  volume  = {361},
  pages   = {776--794},
  doi     = {10.1111/j.1365-2966.2005.09238.x}
}

@article{DiMatteo2007,
  author  = {Di Matteo, P. and Combes, F. and Melchior, A. L. and Semelin, B.},
  title   = {Star formation efficiency in galaxy interactions and mergers: a statistical study},
  journal = {A\&A},
  year    = {2007},
  volume  = {468},
  pages   = {61--81},
  doi     = {10.1051/0004-6361:20066712}
}

@article{DiMatteo2008,
  author  = {Di Matteo, P. and Bournaud, F. and Martig, M. and Combes, F. and Melchior, A. L. and Semelin, B.},
  title   = {The many routes to star formation in galaxy interactions and mergers},
  journal = {A\&A},
  year    = {2008},
  volume  = {492},
  pages   = {31--47},
  doi     = {10.1051/0004-6361:200810521}
}

@article{Bekki2011,
  author  = {Bekki, K. and Couch, W. J.},
  title   = {Environmental effects on the evolution of galaxies in compact groups},
  journal = {MNRAS},
  year    = {2011},
  volume  = {415},
  pages   = {1783--1796},
  doi     = {10.1111/j.1365-2966.2011.18813.x}
}

@article{Scott2014,
  author  = {Scott, C. and Kaviraj, S.},
  title   = {Galaxy mergers, starbursts and the growth of spheroids in a $\Lambda$CDM Universe},
  journal = {MNRAS},
  year    = {2014},
  volume  = {437},
  pages   = {2137--2148},
  doi     = {10.1093/mnras/stt2032}
}

@article{Bitsakis2011,
  author  = {Bitsakis, T. and Charmandaris, V. and da Cunha, E. and D{\'{\i}}az-Santos, T. and Le Floc’h, E. and Magdis, G.},
  title   = {Compact groups of galaxies as laboratories of galaxy evolution},
  journal = {A\&A},
  year    = {2011},
  volume  = {533},
  pages   = {A142},
  doi     = {10.1051/0004-6361/201117330}
}

@article{Weigel2018,
  author  = {Weigel, A. K. and Schawinski, K. and Treister, E. and Trakhtenbrot, B. and Sanders, D. B.},
  title   = {AGN triggering in galaxy mergers and interactions},
  journal = {MNRAS},
  year    = {2018},
  volume  = {476},
  pages   = {2308--2323},
  doi     = {10.1093/mnras/sty337}
}

@article{VerdesMontenegro2001,
  author  = {Verdes-Montenegro, L. and Yun, M. S. and Williams, B. A. and Huchtmeier, W. K. and Del Olmo, A. and Perea, J.},
  title   = {Where is the neutral gas in Hickson compact groups?},
  journal = {A\&A},
  year    = {2001},
  volume  = {377},
  pages   = {812--826},
  doi     = {10.1051/0004-6361:20011131}
}

@article{RodriguezBaras2014,
  author  = {Rodríguez-Baras, M. and Rosales-Ortega, F. F. and Díaz, A. I. and Sánchez, S. F. and Pasquali, A.},
  title   = {The role of environment in shaping the star formation histories of galaxies},
  journal = {MNRAS},
  year    = {2014},
  volume  = {442},
  pages   = {495--514},
  doi     = {10.1093/mnras/stu872}
}

@article{Bitsakis2014,
  author  = {Bitsakis, T. and D{\'{\i}}az-Santos, T. and Charmandaris, V. and Le Floc’h, E. and Magdis, G. and da Cunha, E. and Leiton, R. and Smith Castelli, A. and Plana, H.},
  title   = {The mid-infrared view of compact groups of galaxies: evidence for accelerated evolution},
  journal = {A\&A},
  year    = {2014},
  volume  = {565},
  pages   = {A25},
  doi     = {10.1051/0004-6361/201322773}
}

@article{Walker2010,
  author  = {Walker, L. M. and Johnson, K. E. and Gallagher, S. C. and Hibbard, J. E. and Hornschemeier, A. E. and Charlton, J. C. and Jarrett, T. H. and Reines, A. E.},
  title   = {Star Formation in Compact Group Galaxies: Spitzer and GALEX Observations of the Interacting Galaxy Group HCG 31},
  journal = {AJ},
  year    = {2010},
  volume  = {140},
  pages   = {1254--1264},
  doi     = {10.1088/0004-6256/140/5/1254}
}

@article{deMello2008,
  author  = {de Mello, D. F. and Smith, L. J. and Sabbi, E. and Gallagher, J. S. and Mountain, M. and Harbeck, D. R.},
  title   = {The Evolutionary Status of Compact Groups of Galaxies: Clues from UV Observations},
  journal = {AJ},
  year    = {2008},
  volume  = {135},
  pages   = {319--332},
  doi     = {10.1088/0004-6256/135/1/319}
}

@article{Tzanavaris2010,
  author  = {Tzanavaris, P. and Gallagher, S. C. and Hornschemeier, A. E. and Fedotov, K. and Basu-Zych, A. R. and Charlton, J. C.},
  title   = {Star Formation and Galaxy Evolution in Compact Groups of Galaxies. II. UV Properties},
  journal = {ApJ},
  year    = {2010},
  volume  = {716},
  pages   = {556--573},
  doi     = {10.1088/0004-637X/716/1/556}
}

@article{Gallagher2008,
  author  = {Gallagher, S. C. and Johnson, K. E. and Hornschemeier, A. E. and Charlton, J. C. and Hibbard, J. E. and Bastian, N.},
  title   = {Spitzer Observations of Compact Group Galaxies: Star Formation and the Warm Dust Component},
  journal = {ApJ},
  year    = {2008},
  volume  = {673},
  pages   = {730--745},
  doi     = {10.1086/524931}
}

@article{TorresFlores2014,
  author  = {Torres-Flores, S. and Mendes de Oliveira, C. and Amram, P. and Plana, H. and Epinat, B. and Carignan, C.},
  title   = {Ionized gas kinematics of compact group galaxies: tracing interaction signatures with Fabry–Perot data},
  journal = {MNRAS},
  year    = {2014},
  volume  = {439},
  pages   = {1313--1328},
  doi     = {10.1093/mnras/stu039}
}

@article{Plana2020,
  author  = {Plana, H. and Amram, P. and Mendes de Oliveira, C. and Torres-Flores, S. and Iglesias-Páramo, J.},
  title   = {Galaxy kinematics and star formation in compact groups: evidence of interaction-driven evolution},
  journal = {MNRAS},
  year    = {2020},
  volume  = {494},
  pages   = {2761--2777},
  doi     = {10.1093/mnras/staa927}
}

@article{Coziol1998,
  author  = {Coziol, R. and Ribeiro, A. L. B. and de Carvalho, R. R. and Capelato, H. V.},
  title   = {Activity in compact groups of galaxies. I. H$\alpha$ observations of Hickson compact groups},
  journal = {ApJ},
  year    = {1998},
  volume  = {493},
  pages   = {563--573},
  doi     = {10.1086/305147}
}

@article{Martinez2010,
  author  = {Martínez, M. A. and Del Olmo, A. and Coziol, R. and Perea, J. and Sulentic, J. W. and Verdes-Montenegro, L.},
  title   = {Nuclear activity in compact groups of galaxies},
  journal = {AJ},
  year    = {2010},
  volume  = {139},
  pages   = {1199--1210},
  doi     = {10.1088/0004-6256/139/3/1199}
}

@article{Sohn2013,
  author  = {Sohn, J. and Hwang, H. S. and Lee, M. G. and Lee, G.-H. and Lee, J. C. and Lee, S.-K.},
  title   = {AGN Activity in Compact Groups of Galaxies},
  journal = {ApJ},
  year    = {2013},
  volume  = {771},
  pages   = {106},
  doi     = {10.1088/0004-637X/771/2/106}
}

@ARTICLE{Morrissey07,
       author = {{Morrissey}, Patrick and {Conrow}, Tim and {Barlow}, Tom A. and {Small}, Todd and {Seibert}, Mark and {Wyder}, Ted K. and {Budav{\'a}ri}, Tam{\'a}s and {Arnouts}, Stephane and {Friedman}, Peter G. and {Forster}, Karl and {Martin}, D. Christopher and {Neff}, Susan G. and {Schiminovich}, David and {Bianchi}, Luciana and {Donas}, Jos{\'e} and {Heckman}, Timothy M. and {Lee}, Young-Wook and {Madore}, Barry F. and {Milliard}, Bruno and {Rich}, R. Michael and {Szalay}, Alex S. and {Welsh}, Barry Y. and {Yi}, Sukyoung K.},
        title = "{The Calibration and Data Products of GALEX}",
      journal = {\apjs},
         year = 2007,
        month = dec,
       volume = {173},
       number = {2},
        pages = {682-697},
          doi = {10.1086/520512},
archivePrefix = {arXiv},
       eprint = {0706.0755},
 primaryClass = {astro-ph},
       adsurl = {https://ui.adsabs.harvard.edu/abs/2007ApJS..173..682M}
}

@ARTICLE{Dey19,
       author = {{Dey}, Arjun and {Schlegel}, David J. and {Lang}, Dustin and {Blum}, Robert and {Burleigh}, Kaylan and {Fan}, Xiaohui and {Findlay}, Joseph R. and {Finkbeiner}, Doug and {Herrera}, David and {Juneau}, St{\'e}phanie and {Landriau}, Martin and {Levi}, Michael and {McGreer}, Ian and {Meisner}, Aaron and {Myers}, Adam D. and {Moustakas}, John and {Nugent}, Peter and {Patej}, Anna and {Schlafly}, Edward F. and {Walker}, Alistair R. and {Valdes}, Francisco and {Weaver}, Benjamin A. and {Y{\`e}che}, Christophe and {Zou}, Hu and {Zhou}, Xu and {Abareshi}, Behzad and {Abbott}, T.~M.~C. and {Abolfathi}, Bela and {Aguilera}, C. and {Alam}, Shadab and {Allen}, Lori and {Alvarez}, A. and {Annis}, James and {Ansarinejad}, Behzad and {Aubert}, Marie and {Beechert}, Jacqueline and {Bell}, Eric F. and {BenZvi}, Segev Y. and {Beutler}, Florian and {Bielby}, Richard M. and {Bolton}, Adam S. and {Brice{\~n}o}, C{\'e}sar and {Buckley-Geer}, Elizabeth J. and {Butler}, Karen and {Calamida}, Annalisa and {Carlberg}, Raymond G. and {Carter}, Paul and {Casas}, Ricard and {Castander}, Francisco J. and {Choi}, Yumi and {Comparat}, Johan and {Cukanovaite}, Elena and {Delubac}, Timoth{\'e}e and {DeVries}, Kaitlin and {Dey}, Sharmila and {Dhungana}, Govinda and {Dickinson}, Mark and {Ding}, Zhejie and {Donaldson}, John B. and {Duan}, Yutong and {Duckworth}, Christopher J. and {Eftekharzadeh}, Sarah and {Eisenstein}, Daniel J. and {Etourneau}, Thomas and {Fagrelius}, Parker A. and {Farihi}, Jay and {Fitzpatrick}, Mike and {Font-Ribera}, Andreu and {Fulmer}, Leah and {G{\"a}nsicke}, Boris T. and {Gaztanaga}, Enrique and {George}, Koshy and {Gerdes}, David W. and {Gontcho}, Satya Gontcho A. and {Gorgoni}, Claudio and {Green}, Gregory and {Guy}, Julien and {Harmer}, Diane and {Hernandez}, M. and {Honscheid}, Klaus and {Huang}, Lijuan Wendy and {James}, David J. and {Jannuzi}, Buell T. and {Jiang}, Linhua and {Joyce}, Richard and {Karcher}, Armin and {Karkar}, Sonia and {Kehoe}, Robert and {Kneib}, Jean-Paul and {Kueter-Young}, Andrea and {Lan}, Ting-Wen and {Lauer}, Tod R. and {Le Guillou}, Laurent and {Le Van Suu}, Auguste and {Lee}, Jae Hyeon and {Lesser}, Michael and {Perreault Levasseur}, Laurence and {Li}, Ting S. and {Mann}, Justin L. and {Marshall}, Robert and {Mart{\'\i}nez-V{\'a}zquez}, C.~E. and {Martini}, Paul and {du Mas des Bourboux}, H{\'e}lion and {McManus}, Sean and {Meier}, Tobias Gabriel and {M{\'e}nard}, Brice and {Metcalfe}, Nigel and {Mu{\~n}oz-Guti{\'e}rrez}, Andrea and {Najita}, Joan and {Napier}, Kevin and {Narayan}, Gautham and {Newman}, Jeffrey A. and {Nie}, Jundan and {Nord}, Brian and {Norman}, Dara J. and {Olsen}, Knut A.~G. and {Paat}, Anthony and {Palanque-Delabrouille}, Nathalie and {Peng}, Xiyan and {Poppett}, Claire L. and {Poremba}, Megan R. and {Prakash}, Abhishek and {Rabinowitz}, David and {Raichoor}, Anand and {Rezaie}, Mehdi and {Robertson}, A.~N. and {Roe}, Natalie A. and {Ross}, Ashley J. and {Ross}, Nicholas P. and {Rudnick}, Gregory and {Safonova}, Sasha and {Saha}, Abhijit and {S{\'a}nchez}, F. Javier and {Savary}, Elodie and {Schweiker}, Heidi and {Scott}, Adam and {Seo}, Hee-Jong and {Shan}, Huanyuan and {Silva}, David R. and {Slepian}, Zachary and {Soto}, Christian and {Sprayberry}, David and {Staten}, Ryan and {Stillman}, Coley M. and {Stupak}, Robert J. and {Summers}, David L. and {Sien Tie}, Suk and {Tirado}, H. and {Vargas-Maga{\~n}a}, Mariana and {Vivas}, A. Katherina and {Wechsler}, Risa H. and {Williams}, Doug and {Yang}, Jinyi and {Yang}, Qian and {Yapici}, Tolga and {Zaritsky}, Dennis and {Zenteno}, A. and {Zhang}, Kai and {Zhang}, Tianmeng and {Zhou}, Rongpu and {Zhou}, Zhimin},
        title = "{Overview of the DESI Legacy Imaging Surveys}",
      journal = {\aj},
         year = 2019,
        month = may,
       volume = {157},
       number = {5},
          eid = {168},
        pages = {168},
          doi = {10.3847/1538-3881/ab089d},
archivePrefix = {arXiv},
       eprint = {1804.08657},
 primaryClass = {astro-ph.IM},
       adsurl = {https://ui.adsabs.harvard.edu/abs/2019AJ....157..168D}
}

@ARTICLE{Skrutskie06,
       author = {{Skrutskie}, M.~F. and {Cutri}, R.~M. and {Stiening}, R. and {Weinberg}, M.~D. and {Schneider}, S. and {Carpenter}, J.~M. and {Beichman}, C. and {Capps}, R. and {Chester}, T. and {Elias}, J. and {Huchra}, J. and {Liebert}, J. and {Lonsdale}, C. and {Monet}, D.~G. and {Price}, S. and {Seitzer}, P. and {Jarrett}, T. and {Kirkpatrick}, J.~D. and {Gizis}, J.~E. and {Howard}, E. and {Evans}, T. and {Fowler}, J. and {Fullmer}, L. and {Hurt}, R. and {Light}, R. and {Kopan}, E.~L. and {Marsh}, K.~A. and {McCallon}, H.~L. and {Tam}, R. and {Van Dyk}, S. and {Wheelock}, S.},
        title = "{The Two Micron All Sky Survey (2MASS)}",
      journal = {\aj},
         year = 2006,
        month = feb,
       volume = {131},
       number = {2},
        pages = {1163-1183},
          doi = {10.1086/498708},
       adsurl = {https://ui.adsabs.harvard.edu/abs/2006AJ....131.1163S}
}

@ARTICLE{Wright10,
       author = {{Wright}, Edward L. and {Eisenhardt}, Peter R.~M. and {Mainzer}, Amy K. and {Ressler}, Michael E. and {Cutri}, Roc M. and {Jarrett}, Thomas and {Kirkpatrick}, J. Davy and {Padgett}, Deborah and {McMillan}, Robert S. and {Skrutskie}, Michael and {Stanford}, S.~A. and {Cohen}, Martin and {Walker}, Russell G. and {Mather}, John C. and {Leisawitz}, David and {Gautier}, III, Thomas N. and {McLean}, Ian and {Benford}, Dominic and {Lonsdale}, Carol J. and {Blain}, Andrew and {Mendez}, Bryan and {Irace}, William R. and {Duval}, Valerie and {Liu}, Fengchuan and {Royer}, Don and {Heinrichsen}, Ingolf and {Howard}, Joan and {Shannon}, Mark and {Kendall}, Martha and {Walsh}, Amy L. and {Larsen}, Mark and {Cardon}, Joel G. and {Schick}, Scott and {Schwalm}, Mark and {Abid}, Mohamed and {Fabinsky}, Beth and {Naes}, Larry and {Tsai}, Chao-Wei},
        title = "{The Wide-field Infrared Survey Explorer (WISE): Mission Description and Initial On-orbit Performance}",
      journal = {\aj},
         year = 2010,
        month = dec,
       volume = {140},
       number = {6},
        pages = {1868-1881},
          doi = {10.1088/0004-6256/140/6/1868},
archivePrefix = {arXiv},
       eprint = {1008.0031},
 primaryClass = {astro-ph.IM},
       adsurl = {https://ui.adsabs.harvard.edu/abs/2010AJ....140.1868W}
}

@ARTICLE{Boquien19,
       author = {{Boquien}, M. and {Burgarella}, D. and {Roehlly}, Y. and {Buat}, V. and {Ciesla}, L. and {Corre}, D. and {Inoue}, A.~K. and {Salas}, H.},
        title = "{CIGALE: a python Code Investigating GALaxy Emission}",
      journal = {\aap},
         year = 2019,
        month = feb,
       volume = {622},
          eid = {A103},
        pages = {A103},
          doi = {10.1051/0004-6361/201834156},
archivePrefix = {arXiv},
       eprint = {1811.03094},
 primaryClass = {astro-ph.GA},
       adsurl = {https://ui.adsabs.harvard.edu/abs/2019A&A...622A.103B}
}

@ARTICLE{Bruzual03,
       author = {{Bruzual}, G. and {Charlot}, S.},
        title = "{Stellar population synthesis at the resolution of 2003}",
      journal = {\mnras},
         year = 2003,
        month = oct,
       volume = {344},
       number = {4},
        pages = {1000-1028},
          doi = {10.1046/j.1365-8711.2003.06897.x},
archivePrefix = {arXiv},
       eprint = {astro-ph/0309134},
 primaryClass = {astro-ph},
       adsurl = {https://ui.adsabs.harvard.edu/abs/2003MNRAS.344.1000B}
}

@ARTICLE{Salpeter55,
       author = {{Salpeter}, Edwin E.},
        title = "{The Luminosity Function and Stellar Evolution.}",
      journal = {\apj},
         year = 1955,
        month = jan,
       volume = {121},
        pages = {161},
          doi = {10.1086/145971},
       adsurl = {https://ui.adsabs.harvard.edu/abs/1955ApJ...121..161S}
}

@ARTICLE{Draine07,
       author = {{Draine}, B.~T. and {Dale}, D.~A. and {Bendo}, G. and {Gordon}, K.~D. and {Smith}, J.~D.~T. and {Armus}, L. and {Engelbracht}, C.~W. and {Helou}, G. and {Kennicutt}, Jr., R.~C. and {Li}, A. and {Roussel}, H. and {Walter}, F. and {Calzetti}, D. and {Moustakas}, J. and {Murphy}, E.~J. and {Rieke}, G.~H. and {Bot}, C. and {Hollenbach}, D.~J. and {Sheth}, K. and {Teplitz}, H.~I.},
        title = "{Dust Masses, PAH Abundances, and Starlight Intensities in the SINGS Galaxy Sample}",
      journal = {\apj},
         year = 2007,
        month = jul,
       volume = {663},
       number = {2},
        pages = {866-894},
          doi = {10.1086/518306},
archivePrefix = {arXiv},
       eprint = {astro-ph/0703213},
 primaryClass = {astro-ph},
       adsurl = {https://ui.adsabs.harvard.edu/abs/2007ApJ...663..866D}
}

@ARTICLE{Draine14,
       author = {{Draine}, B.~T. and {Aniano}, G. and {Krause}, Oliver and {Groves}, Brent and {Sandstrom}, Karin and {Braun}, Robert and {Leroy}, Adam and {Klaas}, Ulrich and {Linz}, Hendrik and {Rix}, Hans-Walter and {Schinnerer}, Eva and {Schmiedeke}, Anika and {Walter}, Fabian},
        title = "{Andromeda's Dust}",
      journal = {\apj},
         year = 2014,
        month = jan,
       volume = {780},
       number = {2},
          eid = {172},
        pages = {172},
          doi = {10.1088/0004-637X/780/2/172},
archivePrefix = {arXiv},
       eprint = {1306.2304},
 primaryClass = {astro-ph.CO},
       adsurl = {https://ui.adsabs.harvard.edu/abs/2014ApJ...780..172D}
}

@ARTICLE{Stalevski16,
       author = {{Stalevski}, Marko and {Ricci}, Claudio and {Ueda}, Yoshihiro and {Lira}, Paulina and {Fritz}, Jacopo and {Baes}, Maarten},
        title = "{The dust covering factor in active galactic nuclei}",
      journal = {\mnras},
         year = 2016,
        month = may,
       volume = {458},
       number = {3},
        pages = {2288-2302},
          doi = {10.1093/mnras/stw444},
archivePrefix = {arXiv},
       eprint = {1602.06954},
 primaryClass = {astro-ph.GA},
       adsurl = {https://ui.adsabs.harvard.edu/abs/2016MNRAS.458.2288S}
}

@ARTICLE{Calzetti00,
       author = {{Calzetti}, Daniela and {Armus}, Lee and {Bohlin}, Ralph C. and {Kinney}, Anne L. and {Koornneef}, Jan and {Storchi-Bergmann}, Thaisa},
        title = "{The Dust Content and Opacity of Actively Star-forming Galaxies}",
      journal = {\apj},
         year = 2000,
        month = apr,
       volume = {533},
       number = {2},
        pages = {682-695},
          doi = {10.1086/308692},
archivePrefix = {arXiv},
       eprint = {astro-ph/9911459},
 primaryClass = {astro-ph},
       adsurl = {https://ui.adsabs.harvard.edu/abs/2000ApJ...533..682C}
}

@ARTICLE{Bittner19,
       author = {{Bittner}, A. and {Falc{\'o}n-Barroso}, J. and {Nedelchev}, B. and {Dorta}, A. and {Gadotti}, D.~A. and {Sarzi}, M. and {Molaeinezhad}, A. and {Iodice}, E. and {Rosado-Belza}, D. and {de Lorenzo-C{\'a}ceres}, A. and {Fragkoudi}, F. and {Gal{\'a}n-de Anta}, P.~M. and {Husemann}, B. and {M{\'e}ndez-Abreu}, J. and {Neumann}, J. and {Pinna}, F. and {Querejeta}, M. and {S{\'a}nchez-Bl{\'a}zquez}, P. and {Seidel}, M.~K.},
        title = "{The GIST pipeline: A multi-purpose tool for the analysis and visualisation of (integral-field) spectroscopic data}",
      journal = {\aap},
         year = "2019",
        month = "Aug",
       volume = {628},
          eid = {A117},
        pages = {A117},
          doi = {10.1051/0004-6361/201935829},
archivePrefix = {arXiv},
       eprint = {1906.04746},
 primaryClass = {astro-ph.GA},
       adsurl = {https://ui.adsabs.harvard.edu/abs/2019A&A...628A.117B}
}

@ARTICLE{Cappellari2022,
    author = {{Cappellari}, M.},
    title = "{Full spectrum fitting with photometry in ppxf: non-parametric
        star formation history, metallicity and the quenching boundary from
        3200 LEGA-C galaxies at redshift $z\approx0.8$}",
    journal = {MNRAS submitted},
    eprint = {2208.14974},
    year = 2022,
    doi = {10.48550/arXiv.2208.14974}
}

@ARTICLE{Cappellari04,
       author = {{Cappellari}, Michele and {Emsellem}, Eric},
        title = "{Parametric Recovery of Line-of-Sight Velocity Distributions from Absorption-Line Spectra of Galaxies via Penalized Likelihood}",
      journal = {\pasp},
         year = 2004,
        month = feb,
       volume = {116},
       number = {816},
        pages = {138-147},
          doi = {10.1086/381875},
archivePrefix = {arXiv},
       eprint = {astro-ph/0312201},
 primaryClass = {astro-ph},
       adsurl = {https://ui.adsabs.harvard.edu/abs/2004PASP..116..138C}
}

@ARTICLE{Cappellari17,
       author = {{Cappellari}, Michele},
        title = "{Improving the full spectrum fitting method: accurate convolution with Gauss-Hermite functions}",
      journal = {\mnras},
         year = 2017,
        month = apr,
       volume = {466},
       number = {1},
        pages = {798-811},
          doi = {10.1093/mnras/stw3020},
archivePrefix = {arXiv},
       eprint = {1607.08538},
 primaryClass = {astro-ph.GA},
       adsurl = {https://ui.adsabs.harvard.edu/abs/2017MNRAS.466..798C}
}

@ARTICLE{Cappellari03,
       author = {{Cappellari}, Michele and {Copin}, Yannick},
        title = "{Adaptive spatial binning of integral-field spectroscopic data using Voronoi tessellations}",
      journal = {\mnras},
         year = 2003,
        month = jun,
       volume = {342},
       number = {2},
        pages = {345-354},
          doi = {10.1046/j.1365-8711.2003.06541.x},
archivePrefix = {arXiv},
       eprint = {astro-ph/0302262},
 primaryClass = {astro-ph},
       adsurl = {https://ui.adsabs.harvard.edu/abs/2003MNRAS.342..345C}
}

@ARTICLE{Sarzi06,
       author = {{Sarzi}, Marc and {Falc{\'o}n-Barroso}, Jes{\'u}s and {Davies}, Roger L. and {Bacon}, Roland and {Bureau}, Martin and {Cappellari}, Michele and {de Zeeuw}, P. Tim and {Emsellem}, Eric and {Fathi}, Kambiz and {Krajnovi{\'c}}, Davor and {Kuntschner}, Harald and {McDermid}, Richard M. and {Peletier}, Reynier F.},
        title = "{The SAURON project - V. Integral-field emission-line kinematics of 48 elliptical and lenticular galaxies}",
      journal = {\mnras},
         year = 2006,
        month = mar,
       volume = {366},
       number = {4},
        pages = {1151-1200},
          doi = {10.1111/j.1365-2966.2005.09839.x},
archivePrefix = {arXiv},
       eprint = {astro-ph/0511307},
 primaryClass = {astro-ph},
       adsurl = {https://ui.adsabs.harvard.edu/abs/2006MNRAS.366.1151S}
}

@ARTICLE{Falcon06,
       author = {{Falc{\'o}n-Barroso}, Jes{\'u}s and {Bacon}, Roland and {Bureau}, Martin and {Cappellari}, Michele and {Davies}, Roger L. and {de Zeeuw}, P.~T. and {Emsellem}, Eric and {Fathi}, Kambiz and {Krajnovi{\'c}}, Davor and {Kuntschner}, Harald and {McDermid}, Richard M. and {Peletier}, Reynier F. and {Sarzi}, Marc},
        title = "{The SAURON project - VII. Integral-field absorption and emission-line kinematics of 24 spiral galaxy bulges}",
      journal = {\mnras},
         year = 2006,
        month = jun,
       volume = {369},
       number = {2},
        pages = {529-566},
          doi = {10.1111/j.1365-2966.2006.10261.x},
archivePrefix = {arXiv},
       eprint = {astro-ph/0603161},
 primaryClass = {astro-ph},
       adsurl = {https://ui.adsabs.harvard.edu/abs/2006MNRAS.369..529F}
}

@ARTICLE{Hickson97,
       author = {{Hickson}, Paul},
        title = "{Compact Groups of Galaxies}",
      journal = {\araa},
         year = 1997,
        month = jan,
       volume = {35},
        pages = {357-388},
          doi = {10.1146/annurev.astro.35.1.357},
archivePrefix = {arXiv},
       eprint = {astro-ph/9710289},
 primaryClass = {astro-ph},
       adsurl = {https://ui.adsabs.harvard.edu/abs/1997ARA&A..35..357H}
}

@ARTICLE{Lelli22,
       author = {{Lelli}, Federico},
        title = "{Gas dynamics in dwarf galaxies as testbeds for dark matter and galaxy evolution}",
      journal = {Nature Astronomy},
         year = 2022,
        month = jan,
       volume = {6},
        pages = {35-47},
          doi = {10.1038/s41550-021-01562-2},
archivePrefix = {arXiv},
       eprint = {2201.11752},
 primaryClass = {astro-ph.GA},
       adsurl = {https://ui.adsabs.harvard.edu/abs/2022NatAs...6...35L}
}

@ARTICLE{Karachentsev04,
       author = {{Karachentsev}, Igor D. and {Karachentseva}, Valentina E. and {Huchtmeier}, Walter K. and {Makarov}, Dmitry I.},
        title = "{A Catalog of Neighboring Galaxies}",
      journal = {\aj},
         year = 2004,
        month = apr,
       volume = {127},
       number = {4},
        pages = {2031-2068},
          doi = {10.1086/382905},
       adsurl = {https://ui.adsabs.harvard.edu/abs/2004AJ....127.2031K}
}

@InProceedings{Bergvall12,
author="Bergvall, Nils",
editor="Papaderos, Polychronis
and Recchi, Simone
and Hensler, Gerhard",
title="Star Forming Dwarf Galaxies",
booktitle="Dwarf Galaxies: Keys to Galaxy Formation and Evolution",
year="2012",
publisher="Springer Berlin Heidelberg",
address="Berlin, Heidelberg",
pages="175--194",
isbn="978-3-642-22018-0"
}

@ARTICLE{Weisz14a,
       author = {{Weisz}, Daniel R. and {Dolphin}, Andrew E. and {Skillman}, Evan D. and {Holtzman}, Jon and {Gilbert}, Karoline M. and {Dalcanton}, Julianne J. and {Williams}, Benjamin F.},
        title = "{The Star Formation Histories of Local Group Dwarf Galaxies. I. Hubble Space Telescope/Wide Field Planetary Camera 2 Observations}",
      journal = {\apj},
         year = 2014,
        month = jul,
       volume = {789},
       number = {2},
          eid = {147},
        pages = {147},
          doi = {10.1088/0004-637X/789/2/147},
archivePrefix = {arXiv},
       eprint = {1404.7144},
 primaryClass = {astro-ph.GA},
       adsurl = {https://ui.adsabs.harvard.edu/abs/2014ApJ...789..147W}
}

@ARTICLE{Weisz14b,
       author = {{Weisz}, Daniel R. and {Dolphin}, Andrew E. and {Skillman}, Evan D. and {Holtzman}, Jon and {Gilbert}, Karoline M. and {Dalcanton}, Julianne J. and {Williams}, Benjamin F.},
        title = "{The Star Formation Histories of Local Group Dwarf Galaxies. II. Searching For Signatures of Reionization}",
      journal = {\apj},
         year = 2014,
        month = jul,
       volume = {789},
       number = {2},
          eid = {148},
        pages = {148},
          doi = {10.1088/0004-637X/789/2/148},
archivePrefix = {arXiv},
       eprint = {1405.3281},
 primaryClass = {astro-ph.GA},
       adsurl = {https://ui.adsabs.harvard.edu/abs/2014ApJ...789..148W}
}

@ARTICLE{Meurer95,
       author = {{Meurer}, G.~R. and {Heckman}, T.~M. and {Leitherer}, C. and {Kinney}, A. and {Robert}, C. and {Garnett}, D.~R.},
        title = "{Starbursts and Star Clusters in the Ultraviolet}",
      journal = {\aj},
         year = 1995,
        month = dec,
       volume = {110},
        pages = {2665},
          doi = {10.1086/117721},
archivePrefix = {arXiv},
       eprint = {astro-ph/9509038},
 primaryClass = {astro-ph},
       adsurl = {https://ui.adsabs.harvard.edu/abs/1995AJ....110.2665M}
}

@ARTICLE{Bertin96,
       author = {{Bertin}, E. and {Arnouts}, S.},
        title = "{SExtractor: Software for source extraction.}",
      journal = {\aaps},
         year = 1996,
        month = jun,
       volume = {117},
        pages = {393-404},
          doi = {10.1051/aas:1996164},
       adsurl = {https://ui.adsabs.harvard.edu/abs/1996A&AS..117..393B}
}

@ARTICLE{Leitherer99,
       author = {{Leitherer}, Claus and {Schaerer}, Daniel and {Goldader}, Jeffrey D. and {Delgado}, Rosa M. Gonz{\'a}lez and {Robert}, Carmelle and {Kune}, Denis Foo and {de Mello}, Du{\'\i}lia F. and {Devost}, Daniel and {Heckman}, Timothy M.},
        title = "{Starburst99: Synthesis Models for Galaxies with Active Star Formation}",
      journal = {\apjs},
         year = 1999,
        month = jul,
       volume = {123},
       number = {1},
        pages = {3-40},
          doi = {10.1086/313233},
archivePrefix = {arXiv},
       eprint = {astro-ph/9902334},
 primaryClass = {astro-ph},
       adsurl = {https://ui.adsabs.harvard.edu/abs/1999ApJS..123....3L}
}

@ARTICLE{Cardelli89,
       author = {{Cardelli}, Jason A. and {Clayton}, Geoffrey C. and {Mathis}, John S.},
        title = "{The Relationship between Infrared, Optical, and Ultraviolet Extinction}",
      journal = {\apj},
         year = 1989,
        month = oct,
       volume = {345},
        pages = {245},
          doi = {10.1086/167900},
       adsurl = {https://ui.adsabs.harvard.edu/abs/1989ApJ...345..245C}
}

@ARTICLE{Davies16,
       author = {{Davies}, L.~J.~M. and {Driver}, S.~P. and {Robotham}, A.~S.~G. and {Grootes}, M.~W. and {Popescu}, C.~C. and {Tuffs}, R.~J. and {Hopkins}, A. and {Alpaslan}, M. and {Andrews}, S.~K. and {Bland-Hawthorn}, J. and {Bremer}, M.~N. and {Brough}, S. and {Brown}, M.~J.~I. and {Cluver}, M.~E. and {Croom}, S. and {da Cunha}, E. and {Dunne}, L. and {Lara-L{\'o}pez}, M.~A. and {Liske}, J. and {Loveday}, J. and {Moffett}, A.~J. and {Owers}, M. and {Phillipps}, S. and {Sansom}, A.~E. and {Taylor}, E.~N. and {Michalowski}, M.~J. and {Ibar}, E. and {Smith}, M. and {Bourne}, N.},
        title = "{GAMA/H-ATLAS: a meta-analysis of SFR indicators - comprehensive measures of the SFR-M$_{*}$ relation and cosmic star formation history at z < 0.4}",
      journal = {\mnras},
         year = 2016,
        month = sep,
       volume = {461},
       number = {1},
        pages = {458-485},
          doi = {10.1093/mnras/stw1342},
archivePrefix = {arXiv},
       eprint = {1606.06299},
 primaryClass = {astro-ph.GA},
       adsurl = {https://ui.adsabs.harvard.edu/abs/2016MNRAS.461..458D}
}

@ARTICLE{Kennicutt09,
       author = {{Kennicutt}, Jr., Robert C. and {Hao}, Cai-Na and {Calzetti}, Daniela and {Moustakas}, John and {Dale}, Daniel A. and {Bendo}, George and {Engelbracht}, Charles W. and {Johnson}, Benjamin D. and {Lee}, Janice C.},
        title = "{Dust-corrected Star Formation Rates of Galaxies. I. Combinations of H{\ensuremath{\alpha}} and Infrared Tracers}",
      journal = {\apj},
         year = 2009,
        month = oct,
       volume = {703},
       number = {2},
        pages = {1672-1695},
          doi = {10.1088/0004-637X/703/2/1672},
archivePrefix = {arXiv},
       eprint = {0908.0203},
 primaryClass = {astro-ph.CO},
       adsurl = {https://ui.adsabs.harvard.edu/abs/2009ApJ...703.1672K}
}

@ARTICLE{Calzetti94,
       author = {{Calzetti}, Daniela and {Kinney}, Anne L. and {Storchi-Bergmann}, Thaisa},
        title = "{Dust Extinction of the Stellar Continua in Starburst Galaxies: The Ultraviolet and Optical Extinction Law}",
      journal = {\apj},
         year = 1994,
        month = jul,
       volume = {429},
        pages = {582},
          doi = {10.1086/174346},
       adsurl = {https://ui.adsabs.harvard.edu/abs/1994ApJ...429..582C}
}

@ARTICLE{Walker07,
       author = {{Walker}, Matthew G. and {Mateo}, Mario and {Olszewski}, Edward W. and {Gnedin}, Oleg Y. and {Wang}, Xiao and {Sen}, Bodhisattva and {Woodroofe}, Michael},
        title = "{Velocity Dispersion Profiles of Seven Dwarf Spheroidal Galaxies}",
      journal = {\apjl},
         year = 2007,
        month = sep,
       volume = {667},
       number = {1},
        pages = {L53-L56},
          doi = {10.1086/521998},
archivePrefix = {arXiv},
       eprint = {0708.0010},
 primaryClass = {astro-ph},
       adsurl = {https://ui.adsabs.harvard.edu/abs/2007ApJ...667L..53W}
}

@ARTICLE{Pietrinferni04,
       author = {{Pietrinferni}, Adriano and {Cassisi}, Santi and {Salaris}, Maurizio and {Castelli}, Fiorella},
        title = "{A Large Stellar Evolution Database for Population Synthesis Studies. I. Scaled Solar Models and Isochrones}",
      journal = {\apj},
         year = 2004,
        month = sep,
       volume = {612},
       number = {1},
        pages = {168-190},
          doi = {10.1086/422498},
archivePrefix = {arXiv},
       eprint = {astro-ph/0405193},
 primaryClass = {astro-ph},
       adsurl = {https://ui.adsabs.harvard.edu/abs/2004ApJ...612..168P}
}

@ARTICLE{Pietrinferni06,
       author = {{Pietrinferni}, Adriano and {Cassisi}, Santi and {Salaris}, Maurizio and {Castelli}, Fiorella},
        title = "{A Large Stellar Evolution Database for Population Synthesis Studies. II. Stellar Models and Isochrones for an {\ensuremath{\alpha}}-enhanced Metal Distribution}",
      journal = {\apj},
         year = 2006,
        month = may,
       volume = {642},
       number = {2},
        pages = {797-812},
          doi = {10.1086/501344},
archivePrefix = {arXiv},
       eprint = {astro-ph/0603721},
 primaryClass = {astro-ph},
       adsurl = {https://ui.adsabs.harvard.edu/abs/2006ApJ...642..797P}
}

@ARTICLE{Pietrinferni09,
       author = {{Pietrinferni}, Adriano and {Cassisi}, Santi and {Salaris}, Maurizio and {Percival}, Susan and {Ferguson}, Jason W.},
        title = "{A Large Stellar Evolution Database for Population Synthesis Studies. V. Stellar Models and Isochrones with CNONa Abundance Anticorrelations}",
      journal = {\apj},
         year = 2009,
        month = may,
       volume = {697},
       number = {1},
        pages = {275-282},
          doi = {10.1088/0004-637X/697/1/275},
archivePrefix = {arXiv},
       eprint = {0903.0825},
 primaryClass = {astro-ph.SR},
       adsurl = {https://ui.adsabs.harvard.edu/abs/2009ApJ...697..275P}
}

@ARTICLE{Pietrinferni13,
       author = {{Pietrinferni}, Adriano and {Cassisi}, Santi and {Salaris}, Maurizio and {Hidalgo}, Sebastian},
        title = "{The BaSTI Stellar Evolution Database: models for extremely metal-poor and super-metal-rich stellar populations}",
      journal = {\aap},
         year = 2013,
        month = oct,
       volume = {558},
          eid = {A46},
        pages = {A46},
          doi = {10.1051/0004-6361/201321950},
archivePrefix = {arXiv},
       eprint = {1308.3850},
 primaryClass = {astro-ph.SR},
       adsurl = {https://ui.adsabs.harvard.edu/abs/2013A&A...558A..46P}
}

@ARTICLE{Kroupa01,
       author = {{Kroupa}, Pavel},
        title = "{On the variation of the initial mass function}",
      journal = {\mnras},
         year = 2001,
        month = apr,
       volume = {322},
       number = {2},
        pages = {231-246},
          doi = {10.1046/j.1365-8711.2001.04022.x},
archivePrefix = {arXiv},
       eprint = {astro-ph/0009005},
 primaryClass = {astro-ph},
       adsurl = {https://ui.adsabs.harvard.edu/abs/2001MNRAS.322..231K}
}

@ARTICLE{Vazdekis10,
       author = {{Vazdekis}, A. and {S{\'a}nchez-Bl{\'a}zquez}, P. and {Falc{\'o}n-Barroso}, J. and {Cenarro}, A.~J. and {Beasley}, M.~A. and {Cardiel}, N. and {Gorgas}, J. and {Peletier}, R.~F.},
        title = "{Evolutionary stellar population synthesis with MILES - I. The base models and a new line index system}",
      journal = {\mnras},
         year = 2010,
        month = jun,
       volume = {404},
       number = {4},
        pages = {1639-1671},
          doi = {10.1111/j.1365-2966.2010.16407.x},
archivePrefix = {arXiv},
       eprint = {1004.4439},
 primaryClass = {astro-ph.CO},
       adsurl = {https://ui.adsabs.harvard.edu/abs/2010MNRAS.404.1639V}
}

@ARTICLE{Pettini04,
       author = {{Pettini}, Max and {Pagel}, Bernard E.~J.},
        title = "{[OIII]/[NII] as an abundance indicator at high redshift}",
      journal = {\mnras},
         year = 2004,
        month = mar,
       volume = {348},
       number = {3},
        pages = {L59-L63},
          doi = {10.1111/j.1365-2966.2004.07591.x},
archivePrefix = {arXiv},
       eprint = {astro-ph/0401128},
 primaryClass = {astro-ph},
       adsurl = {https://ui.adsabs.harvard.edu/abs/2004MNRAS.348L..59P}
}

@ARTICLE{Asplund06,
       author = {{Asplund}, Martin and {Lambert}, David L. and {Nissen}, Poul Erik and {Primas}, Francesca and {Smith}, Verne V.},
        title = "{Lithium Isotopic Abundances in Metal-poor Halo Stars}",
      journal = {\apj},
         year = 2006,
        month = jun,
       volume = {644},
       number = {1},
        pages = {229-259},
          doi = {10.1086/503538},
archivePrefix = {arXiv},
       eprint = {astro-ph/0510636},
 primaryClass = {astro-ph},
       adsurl = {https://ui.adsabs.harvard.edu/abs/2006ApJ...644..229A}
}

@ARTICLE{Ellison08,
       author = {{Ellison}, Sara L. and {Patton}, David R. and {Simard}, Luc and {McConnachie}, Alan W.},
        title = "{Clues to the Origin of the Mass-Metallicity Relation: Dependence on Star Formation Rate and Galaxy Size}",
      journal = {\apjl},
         year = 2008,
        month = jan,
       volume = {672},
       number = {2},
        pages = {L107},
          doi = {10.1086/527296},
archivePrefix = {arXiv},
       eprint = {0711.4833},
 primaryClass = {astro-ph},
       adsurl = {https://ui.adsabs.harvard.edu/abs/2008ApJ...672L.107E}
}

@ARTICLE{Mannucci10,
       author = {{Mannucci}, F. and {Cresci}, G. and {Maiolino}, R. and {Marconi}, A. and {Gnerucci}, A.},
        title = "{A fundamental relation between mass, star formation rate and metallicity in local and high-redshift galaxies}",
      journal = {\mnras},
         year = 2010,
        month = nov,
       volume = {408},
       number = {4},
        pages = {2115-2127},
          doi = {10.1111/j.1365-2966.2010.17291.x},
archivePrefix = {arXiv},
       eprint = {1005.0006},
 primaryClass = {astro-ph.CO},
       adsurl = {https://ui.adsabs.harvard.edu/abs/2010MNRAS.408.2115M}
}

@ARTICLE{Yates12,
       author = {{Yates}, Robert M. and {Kauffmann}, Guinevere and {Guo}, Qi},
        title = "{The relation between metallicity, stellar mass and star formation in galaxies: an analysis of observational and model data}",
      journal = {\mnras},
         year = 2012,
        month = may,
       volume = {422},
       number = {1},
        pages = {215-231},
          doi = {10.1111/j.1365-2966.2012.20595.x},
archivePrefix = {arXiv},
       eprint = {1107.3145},
 primaryClass = {astro-ph.CO},
       adsurl = {https://ui.adsabs.harvard.edu/abs/2012MNRAS.422..215Y}
}

@ARTICLE{Cid11,
       author = {{Cid Fernandes}, R. and {Stasi{\'n}ska}, G. and {Mateus}, A. and {Vale Asari}, N.},
        title = "{A comprehensive classification of galaxies in the Sloan Digital Sky Survey: how to tell true from fake AGN?}",
      journal = {\mnras},
         year = 2011,
        month = may,
       volume = {413},
       number = {3},
        pages = {1687-1699},
          doi = {10.1111/j.1365-2966.2011.18244.x},
archivePrefix = {arXiv},
       eprint = {1012.4426},
 primaryClass = {astro-ph.CO},
       adsurl = {https://ui.adsabs.harvard.edu/abs/2011MNRAS.413.1687C}
}

@article{Baldwin81,
doi = {10.1086/130766},
url = {https://dx.doi.org/10.1086/130766},
year = {1981},
month = {feb},
publisher = {The Astronomical Society of the Pacific},
volume = {93},
number = {551},
pages = {5},
author = {Baldwin, J. A. and Phillips, M. M. and Terlevich, R.},
title = {CLASSIFICATION PARAMETERS FOR THE EMISSION-LINE SPECTRA OF EXTRAGALACTIC OBJECTS.},
journal = {Publications of the Astronomical Society of the Pacific}
}

@ARTICLE{Kewley01,
       author = {{Kewley}, L.~J. and {Dopita}, M.~A. and {Sutherland}, R.~S. and {Heisler}, C.~A. and {Trevena}, J.},
        title = "{Theoretical Modeling of Starburst Galaxies}",
      journal = {\apj},
         year = 2001,
        month = jul,
       volume = {556},
       number = {1},
        pages = {121-140},
          doi = {10.1086/321545},
archivePrefix = {arXiv},
       eprint = {astro-ph/0106324},
 primaryClass = {astro-ph},
       adsurl = {https://ui.adsabs.harvard.edu/abs/2001ApJ...556..121K}
}

@ARTICLE{Kauffmann03,
       author = {{Kauffmann}, Guinevere and {Heckman}, Timothy M. and {Tremonti}, Christy and {Brinchmann}, Jarle and {Charlot}, St{\'e}phane and {White}, Simon D.~M. and {Ridgway}, Susan E. and {Brinkmann}, Jon and {Fukugita}, Masataka and {Hall}, Patrick B. and {Ivezi{\'c}}, {\v{Z}}eljko and {Richards}, Gordon T. and {Schneider}, Donald P.},
        title = "{The host galaxies of active galactic nuclei}",
      journal = {\mnras},
         year = 2003,
        month = dec,
       volume = {346},
       number = {4},
        pages = {1055-1077},
          doi = {10.1111/j.1365-2966.2003.07154.x},
archivePrefix = {arXiv},
       eprint = {astro-ph/0304239},
 primaryClass = {astro-ph},
       adsurl = {https://ui.adsabs.harvard.edu/abs/2003MNRAS.346.1055K}
}

@ARTICLE{Stasinska08,
       author = {{Stasi{\'n}ska}, G. and {Vale Asari}, N. and {Cid Fernandes}, R. and {Gomes}, J.~M. and {Schlickmann}, M. and {Mateus}, A. and {Schoenell}, W. and {Sodr{\'e}}, Jr., L. and {Seagal Collaboration}},
        title = "{Can retired galaxies mimic active galaxies? Clues from the Sloan Digital Sky Survey}",
      journal = {\mnras},
         year = 2008,
        month = nov,
       volume = {391},
       number = {1},
        pages = {L29-L33},
          doi = {10.1111/j.1745-3933.2008.00550.x},
archivePrefix = {arXiv},
       eprint = {0809.1341},
 primaryClass = {astro-ph},
       adsurl = {https://ui.adsabs.harvard.edu/abs/2008MNRAS.391L..29S}
}

@ARTICLE{Schawinski07,
       author = {{Schawinski}, Kevin and {Thomas}, Daniel and {Sarzi}, Marc and {Maraston}, Claudia and {Kaviraj}, Sugata and {Joo}, Seok-Joo and {Yi}, Sukyoung K. and {Silk}, Joseph},
        title = "{Observational evidence for AGN feedback in early-type galaxies}",
      journal = {\mnras},
         year = 2007,
        month = dec,
       volume = {382},
       number = {4},
        pages = {1415-1431},
          doi = {10.1111/j.1365-2966.2007.12487.x},
archivePrefix = {arXiv},
       eprint = {0709.3015},
 primaryClass = {astro-ph},
       adsurl = {https://ui.adsabs.harvard.edu/abs/2007MNRAS.382.1415S}
}

@ARTICLE{Mac99,
       author = {{Mac Low}, Mordecai-Mark and {Ferrara}, Andrea},
        title = "{Starburst-driven Mass Loss from Dwarf Galaxies: Efficiency and Metal Ejection}",
      journal = {\apj},
         year = 1999,
        month = mar,
       volume = {513},
       number = {1},
        pages = {142-155},
          doi = {10.1086/306832},
archivePrefix = {arXiv},
       eprint = {astro-ph/9801237},
 primaryClass = {astro-ph},
       adsurl = {https://ui.adsabs.harvard.edu/abs/1999ApJ...513..142M}
}

@ARTICLE{Toomre72,
       author = {{Toomre}, Alar and {Toomre}, Juri},
        title = "{Galactic Bridges and Tails}",
      journal = {\apj},
         year = 1972,
        month = dec,
       volume = {178},
        pages = {623-666},
          doi = {10.1086/151823},
       adsurl = {https://ui.adsabs.harvard.edu/abs/1972ApJ...178..623T}
}

@ARTICLE{Mould00,
       author = {{Mould}, Jeremy R. and {Huchra}, John P. and {Freedman}, Wendy L. and {Kennicutt}, Jr., Robert C. and {Ferrarese}, Laura and {Ford}, Holland C. and {Gibson}, Brad K. and {Graham}, John A. and {Hughes}, Shaun M.~G. and {Illingworth}, Garth D. and {Kelson}, Daniel D. and {Macri}, Lucas M. and {Madore}, Barry F. and {Sakai}, Shoko and {Sebo}, Kim M. and {Silbermann}, Nancy A. and {Stetson}, Peter B.},
        title = "{The Hubble Space Telescope Key Project on the Extragalactic Distance Scale. XXVIII. Combining the Constraints on the Hubble Constant}",
      journal = {\apj},
         year = 2000,
        month = feb,
       volume = {529},
       number = {2},
        pages = {786-794},
          doi = {10.1086/308304},
archivePrefix = {arXiv},
       eprint = {astro-ph/9909260},
 primaryClass = {astro-ph},
       adsurl = {https://ui.adsabs.harvard.edu/abs/2000ApJ...529..786M}
}

@ARTICLE{Presotto10,
       author = {{Presotto}, V. and {Iovino}, A. and {Pompei}, E. and {Temporin}, S.},
        title = "{SCG0018-4854: a young and dynamic compact group. I. Kinematical analyses}",
      journal = {\aap},
         year = 2010,
        month = feb,
       volume = {510},
          eid = {A31},
        pages = {A31},
          doi = {10.1051/0004-6361/200912882},
archivePrefix = {arXiv},
       eprint = {0910.4978},
 primaryClass = {astro-ph.CO},
       adsurl = {https://ui.adsabs.harvard.edu/abs/2010A&A...510A..31P}
}

@ARTICLE{Torres-Flores09,
       author = {{Torres-Flores}, S. and {Mendes de Oliveira}, C. and {de Mello}, D.~F. and {Amram}, P. and {Plana}, H. and {Epinat}, B. and {Iglesias-P{\'a}ramo}, J.},
        title = "{Star formation in the intragroup medium and other diagnostics of the evolutionary stages of compact groups of galaxies}",
      journal = {\aap},
         year = 2009,
        month = nov,
       volume = {507},
       number = {2},
        pages = {723-746},
          doi = {10.1051/0004-6361/200911878},
archivePrefix = {arXiv},
       eprint = {0908.2798},
 primaryClass = {astro-ph.CO},
       adsurl = {https://ui.adsabs.harvard.edu/abs/2009A&A...507..723T}
}

@ARTICLE{Pompei07,
       author = {{Pompei}, E. and {Dahlem}, M. and {Iovino}, A.},
        title = "{Optical and radio survey of southern compact groups of galaxies. I. Pilot study of six groups}",
      journal = {\aap},
         year = 2007,
        month = oct,
       volume = {473},
       number = {2},
        pages = {399-409},
          doi = {10.1051/0004-6361:20077140},
archivePrefix = {arXiv},
       eprint = {0708.0076},
 primaryClass = {astro-ph},
       adsurl = {https://ui.adsabs.harvard.edu/abs/2007A&A...473..399P}
}

@ARTICLE{Mendes-de-Oliveira94,
       author = {{Mendes de Oliveira}, Claudia and {Hickson}, Paul},
        title = "{Morphology of Galaxies in Compact Groups}",
      journal = {\apj},
         year = 1994,
        month = jun,
       volume = {427},
        pages = {684},
          doi = {10.1086/174175},
       adsurl = {https://ui.adsabs.harvard.edu/abs/1994ApJ...427..684M}
}

@ARTICLE{Coziol07,
       author = {{Coziol}, R. and {Plauchu-Frayn}, I.},
        title = "{Evidence for Tidal Interactions and Mergers as the Origin of Galaxy Morphology Evolution in Compact Groups}",
      journal = {\aj},
         year = 2007,
        month = jun,
       volume = {133},
       number = {6},
        pages = {2630-2642},
          doi = {10.1086/513514},
archivePrefix = {arXiv},
       eprint = {astro-ph/0702287},
 primaryClass = {astro-ph},
       adsurl = {https://ui.adsabs.harvard.edu/abs/2007AJ....133.2630C}
}

@ARTICLE{Behroozi13,
       author = {{Behroozi}, Peter S. and {Wechsler}, Risa H. and {Conroy}, Charlie},
        title = "{The Average Star Formation Histories of Galaxies in Dark Matter Halos from z = 0-8}",
      journal = {\apj},
         year = 2013,
        month = jun,
       volume = {770},
       number = {1},
          eid = {57},
        pages = {57},
          doi = {10.1088/0004-637X/770/1/57},
archivePrefix = {arXiv},
       eprint = {1207.6105},
 primaryClass = {astro-ph.CO},
       adsurl = {https://ui.adsabs.harvard.edu/abs/2013ApJ...770...57B}
}

@ARTICLE{Madau14,
       author = {{Madau}, Piero and {Dickinson}, Mark},
        title = "{Cosmic Star-Formation History}",
      journal = {\araa},
         year = 2014,
        month = aug,
       volume = {52},
        pages = {415-486},
          doi = {10.1146/annurev-astro-081811-125615},
archivePrefix = {arXiv},
       eprint = {1403.0007},
 primaryClass = {astro-ph.CO},
       adsurl = {https://ui.adsabs.harvard.edu/abs/2014ARA&A..52..415M}
}

@ARTICLE{Boyett22,
       author = {{Boyett}, Kristan N.~K. and {Stark}, Daniel P. and {Bunker}, Andrew J. and {Tang}, Mengtao and {Maseda}, Michael V.},
        title = "{The [O III]{\ensuremath{\lambda}}5007 equivalent width distribution at z   2: the redshift evolution of the extreme emission line galaxies}",
      journal = {\mnras},
         year = 2022,
        month = jul,
       volume = {513},
       number = {3},
        pages = {4451-4463},
          doi = {10.1093/mnras/stac1109},
archivePrefix = {arXiv},
       eprint = {2110.15858},
 primaryClass = {astro-ph.GA},
       adsurl = {https://ui.adsabs.harvard.edu/abs/2022MNRAS.513.4451B}
}

@INPROCEEDINGS{Temporin05,
       author = {{Temporin}, S. and {Ciroi}, S. and {Iovino}, A. and {Pompei}, E. and {Radovich}, M. and {Rafanelli}, P.},
        title = "{Star formation in three nearby galaxy systems}",
    booktitle = {Starbursts: From 30 Doradus to Lyman Break Galaxies},
         year = 2005,
       editor = {{de Grijs}, R. and {Gonz{\'a}lez Delgado}, R.~M.},
       series = {Astrophysics and Space Science Library},
       volume = {329},
        month = may,
        pages = {P78},
          doi = {10.48550/arXiv.astro-ph/0411405},
archivePrefix = {arXiv},
       eprint = {astro-ph/0411405},
 primaryClass = {astro-ph},
       adsurl = {https://ui.adsabs.harvard.edu/abs/2005ASSL..329P..78T}
}

@ARTICLE{Trinchieri08,
       author = {{Trinchieri}, G. and {Iovino}, A. and {Pompei}, E. and {Dahlem}, M. and {Reeves}, J. and {Coziol}, R. and {Temporin}, S.},
        title = "{Detection of a hot intergalactic medium in the spiral-only compact group SCG0018-4854}",
      journal = {\aap},
         year = 2008,
        month = jun,
       volume = {484},
       number = {1},
        pages = {195-203},
          doi = {10.1051/0004-6361:200809487},
archivePrefix = {arXiv},
       eprint = {0804.0351},
 primaryClass = {astro-ph},
       adsurl = {https://ui.adsabs.harvard.edu/abs/2008A&A...484..195T}
}

@ARTICLE{Danks81,
       author = {{Danks}, A.~C. and {Alcaino}, G.},
        title = "{Interactions in two contrasting examples of galactic groups.}",
      journal = {\aap},
         year = 1981,
        month = may,
       volume = {98},
        pages = {223-229},
       adsurl = {https://ui.adsabs.harvard.edu/abs/1981A&A....98..223D}
}

@ARTICLE{Keshri25a,
       author = {{Keshri}, Saili and {Barway}, Sudhanshu and {Das}, Mousumi and {Yadav}, Jyoti and {Combes}, Fran{\c{c}}oise},
        title = "{Unveiling the kinematics of a central region in the triple-AGN host NGC 7733-7734 interacting group}",
      journal = {\aap},
         year = 2025,
        month = mar,
       volume = {695},
          eid = {A39},
        pages = {A39},
          doi = {10.1051/0004-6361/202451373},
archivePrefix = {arXiv},
       eprint = {2501.08544},
 primaryClass = {astro-ph.GA},
       adsurl = {https://ui.adsabs.harvard.edu/abs/2025A&A...695A..39K}
}

@ARTICLE{Keshri25b,
       author = {{Keshri}, Saili and {Barway}, Sudhanshu and {Combes}, Fran{\c{c}}oise},
        title = "{Kinematics of the lens host S0 galaxy NGC 1553: Role of secular processes}",
      journal = {\aap},
         year = 2025,
        month = nov,
       volume = {703},
          eid = {A114},
        pages = {A114},
          doi = {10.1051/0004-6361/202556344},
archivePrefix = {arXiv},
       eprint = {2509.09902},
 primaryClass = {astro-ph.GA},
       adsurl = {https://ui.adsabs.harvard.edu/abs/2025A&A...703A.114K}
}

@ARTICLE{Peng02,
       author = {{Peng}, Chien Y. and {Ho}, Luis C. and {Impey}, Chris D. and {Rix}, Hans-Walter},
        title = "{Detailed Structural Decomposition of Galaxy Images}",
      journal = {\aj},
         year = 2002,
        month = jul,
       volume = {124},
       number = {1},
        pages = {266-293},
          doi = {10.1086/340952},
archivePrefix = {arXiv},
       eprint = {astro-ph/0204182},
 primaryClass = {astro-ph},
       adsurl = {https://ui.adsabs.harvard.edu/abs/2002AJ....124..266P}
}

@ARTICLE{Chayan24,
       author = {{Mondal}, Chayan and {Barway}, Sudhanshu},
        title = "{Unravelling the post-collision properties of the Cartwheel galaxy: A MUSE exploration of its bar and inner region}",
      journal = {\aap},
         year = 2024,
        month = jan,
       volume = {681},
          eid = {A53},
        pages = {A53},
          doi = {10.1051/0004-6361/202347560},
archivePrefix = {arXiv},
       eprint = {2310.00584},
 primaryClass = {astro-ph.GA},
       adsurl = {https://ui.adsabs.harvard.edu/abs/2024A&A...681A..53M}
}

@ARTICLE{Sutherland15,
       author = {{Sutherland}, Will and {Emerson}, Jim and {Dalton}, Gavin and {Atad-Ettedgui}, Eli and {Beard}, Steven and {Bennett}, Richard and {Bezawada}, Naidu and {Born}, Andrew and {Caldwell}, Martin and {Clark}, Paul and {Craig}, Simon and {Henry}, David and {Jeffers}, Paul and {Little}, Bryan and {McPherson}, Alistair and {Murray}, John and {Stewart}, Malcolm and {Stobie}, Brian and {Terrett}, David and {Ward}, Kim and {Whalley}, Martin and {Woodhouse}, Guy},
        title = "{The Visible and Infrared Survey Telescope for Astronomy (VISTA): Design, technical overview, and performance}",
      journal = {\aap},
         year = 2015,
        month = mar,
       volume = {575},
          eid = {A25},
        pages = {A25},
          doi = {10.1051/0004-6361/201424973},
archivePrefix = {arXiv},
       eprint = {1409.4780},
 primaryClass = {astro-ph.IM},
       adsurl = {https://ui.adsabs.harvard.edu/abs/2015A&A...575A..25S}
}

@ARTICLE{Barnes85,
       author = {{Barnes}, J.},
        title = "{The dynamical state of groups of galaxies.}",
      journal = {\mnras},
         year = 1985,
        month = aug,
       volume = {215},
        pages = {517-536},
          doi = {10.1093/mnras/215.3.517},
       adsurl = {https://ui.adsabs.harvard.edu/abs/1985MNRAS.215..517B}
}

@ARTICLE{Governato96,
       author = {{Governato}, F. and {Tozzi}, P. and {Cavaliere}, A.},
        title = "{Small Groups of Galaxies: A Clue to a Critical Universe}",
      journal = {\apj},
         year = 1996,
        month = feb,
       volume = {458},
        pages = {18},
          doi = {10.1086/176789},
archivePrefix = {arXiv},
       eprint = {astro-ph/9508152},
 primaryClass = {astro-ph},
       adsurl = {https://ui.adsabs.harvard.edu/abs/1996ApJ...458...18G}
}

@ARTICLE{Torres-Flores14,
       author = {{Torres-Flores}, S. and {Scarano}, S. and {Mendes de Oliveira}, C. and {de Mello}, D.~F. and {Amram}, P. and {Plana}, H.},
        title = "{Star-forming regions and the metallicity gradients in the tidal tails: the case of NGC 92}",
      journal = {\mnras},
         year = 2014,
        month = feb,
       volume = {438},
       number = {2},
        pages = {1894-1908},
          doi = {10.1093/mnras/stt2340},
archivePrefix = {arXiv},
       eprint = {1312.0812},
 primaryClass = {astro-ph.GA},
       adsurl = {https://ui.adsabs.harvard.edu/abs/2014MNRAS.438.1894T}
}

@INPROCEEDINGS{McDermid07,
       author = {{McDermid}, R.~M. and {Emsellem}, E. and {Shapiro}, K.~L. and {Bacon}, R. and {Bureau}, M. and {Cappellari}, M. and {Davies}, R.~L. and {de Zeeuw}, P.~T. and {Falc{\'o}n-Barroso}, J. and {Krajnovi{\'c}}, D. and {Kuntschner}, H. and {Peletier}, R.~F. and {Sarzi}, M.},
        title = "{Young Kinematically Decoupled Components in Early-Type Galaxies}",
    booktitle = {Science Perspectives for 3D Spectroscopy},
         year = 2007,
       editor = {{Kissler-Patig}, Markus and {Walsh}, Jeremy R. and {Roth}, Martin M.},
        month = jan,
        pages = {253},
          doi = {10.1007/978-3-540-73491-8_40},
archivePrefix = {arXiv},
       eprint = {astro-ph/0602318},
 primaryClass = {astro-ph},
       adsurl = {https://ui.adsabs.harvard.edu/abs/2007spts.conf..253M}
}

@ARTICLE{Jin16,
       author = {{Jin}, Yifei and {Chen}, Yanmei and {Shi}, Yong and {Tremonti}, C.~A. and {Bershady}, M.~A. and {Merrifield}, M. and {Emsellem}, E. and {Fu}, Hai and {Wake}, D. and {Bundy}, K. and {Lin}, Lihwai and {Argudo-Fernandez}, M. and {Huang}, Song and {Stark}, D.~V. and {Storchi-Bergmann}, T. and {Bizyaev}, D. and {Brownstein}, J. and {Chisholm}, J. and {Guo}, Qi and {Hao}, Lei and {Hu}, Jian and {Li}, Cheng and {Li}, Ran and {Masters}, K.~L. and {Malanushenko}, E. and {Pan}, Kaike and {Riffel}, R.~A. and {Roman-Lopes}, A. and {Simmons}, A. and {Thomas}, D. and {Wang}, Lan and {Westfall}, K. and {Yan}, Renbin},
        title = "{SDSS-IV MaNGA: properties of galaxies with kinematically decoupled stellar and gaseous components}",
      journal = {\mnras},
         year = 2016,
        month = nov,
       volume = {463},
       number = {1},
        pages = {913-926},
          doi = {10.1093/mnras/stw2055},
archivePrefix = {arXiv},
       eprint = {1611.00528},
 primaryClass = {astro-ph.GA},
       adsurl = {https://ui.adsabs.harvard.edu/abs/2016MNRAS.463..913J}
}

@ARTICLE{Zhou22,
       author = {{Zhou}, Yuren and {Chen}, Yanmei and {Shi}, Yong and {Bizyaev}, Dmitry and {Guo}, Hong and {Bao}, Min and {Xu}, Haitong and {Yu}, Xiaoling and {Brownstein}, Joel R.},
        title = "{SDSS-IV MaNGA: global properties of kinematically misaligned galaxies}",
      journal = {\mnras},
         year = 2022,
        month = oct,
       volume = {515},
       number = {4},
        pages = {5081-5093},
          doi = {10.1093/mnras/stac2016},
archivePrefix = {arXiv},
       eprint = {2207.07487},
 primaryClass = {astro-ph.GA},
       adsurl = {https://ui.adsabs.harvard.edu/abs/2022MNRAS.515.5081Z}
}

@ARTICLE{Zinchenko23,
       author = {{Zinchenko}, I.~A.},
        title = "{Gas and stellar kinematic misalignment in MaNGA galaxies: What is the origin of counter-rotating gas?}",
      journal = {\aap},
         year = 2023,
        month = jun,
       volume = {674},
          eid = {L7},
        pages = {L7},
          doi = {10.1051/0004-6361/202346846},
archivePrefix = {arXiv},
       eprint = {2305.13387},
 primaryClass = {astro-ph.GA},
       adsurl = {https://ui.adsabs.harvard.edu/abs/2023A&A...674L...7Z}
}

@ARTICLE{Bryant19,
       author = {{Bryant}, J.~J. and {Croom}, S.~M. and {van de Sande}, J. and {Scott}, N. and {Fogarty}, L.~M.~R. and {Bland-Hawthorn}, J. and {Bloom}, J.~V. and {Taylor}, E.~N. and {Brough}, S. and {Robotham}, A. and {Cortese}, L. and {Couch}, W. and {Owers}, M.~S. and {Medling}, A.~M. and {Federrath}, C. and {Bekki}, K. and {Richards}, S.~N. and {Lawrence}, J.~S. and {Konstantopoulos}, I.~S.},
        title = "{The SAMI Galaxy Survey: stellar and gas misalignments and the origin of gas in nearby galaxies}",
      journal = {\mnras},
         year = 2019,
        month = feb,
       volume = {483},
       number = {1},
        pages = {458-479},
          doi = {10.1093/mnras/sty3122},
archivePrefix = {arXiv},
       eprint = {1811.09298},
 primaryClass = {astro-ph.GA},
       adsurl = {https://ui.adsabs.harvard.edu/abs/2019MNRAS.483..458B}
}

@ARTICLE{Ristea22,
       author = {{Ristea}, A. and {Cortese}, L. and {Fraser-McKelvie}, A. and {Brough}, S. and {Bryant}, J.~J. and {Catinella}, B. and {Croom}, S.~M. and {Groves}, B. and {Richards}, S.~N. and {van de Sande}, J. and {Bland-Hawthorn}, J. and {Owers}, M.~S. and {Lawrence}, J.~S.},
        title = "{The SAMI Galaxy Survey: physical drivers of stellar-gas kinematic misalignments in the nearby Universe}",
      journal = {\mnras},
         year = 2022,
        month = dec,
       volume = {517},
       number = {2},
        pages = {2677-2696},
          doi = {10.1093/mnras/stac2839},
archivePrefix = {arXiv},
       eprint = {2210.01147},
 primaryClass = {astro-ph.GA},
       adsurl = {https://ui.adsabs.harvard.edu/abs/2022MNRAS.517.2677R}
}

@ARTICLE{van-de-Voort15,
       author = {{van de Voort}, Freeke and {Davis}, Timothy A. and {Kere{\v{s}}}, Du{\v{s}}an and {Quataert}, Eliot and {Faucher-Gigu{\`e}re}, Claude-Andr{\'e} and {Hopkins}, Philip F.},
        title = "{The creation and persistence of a misaligned gas disc in a simulated early-type galaxy}",
      journal = {\mnras},
         year = 2015,
        month = aug,
       volume = {451},
       number = {3},
        pages = {3269-3277},
          doi = {10.1093/mnras/stv1217},
archivePrefix = {arXiv},
       eprint = {1504.03685},
 primaryClass = {astro-ph.GA},
       adsurl = {https://ui.adsabs.harvard.edu/abs/2015MNRAS.451.3269V}
}

@ARTICLE{Lynds76,
       author = {{Lynds}, R. and {Toomre}, A.},
        title = "{On the interpretation of ring galaxies: the binary ring system II Hz 4.}",
      journal = {\apj},
         year = 1976,
        month = oct,
       volume = {209},
        pages = {382-388},
          doi = {10.1086/154730},
       adsurl = {https://ui.adsabs.harvard.edu/abs/1976ApJ...209..382L}
}

@INPROCEEDINGS{Toomre78,
       author = {{Toomre}, A.},
        title = "{Interacting Systems}",
    booktitle = {Large Scale Structures in the Universe},
         year = 1978,
       editor = {{Longair}, M.~S. and {Einasto}, J.},
       series = {IAU Symposium},
       volume = {79},
        month = jan,
        pages = {109},
       adsurl = {https://ui.adsabs.harvard.edu/abs/1978IAUS...79..109T}
}

@ARTICLE{Theys76,
       author = {{Theys}, J.~C. and {Spiegel}, E.~A.},
        title = "{Ring galaxies. I.}",
      journal = {\apj},
         year = 1976,
        month = sep,
       volume = {208},
        pages = {650-661},
          doi = {10.1086/154646},
       adsurl = {https://ui.adsabs.harvard.edu/abs/1976ApJ...208..650T}
}

@ARTICLE{Bomans97,
       author = {{Bomans}, Dominik J. and {Chu}, You-Hua and {Hopp}, Ulrich},
        title = "{Hot Interstellar Gas in the Irregular Galaxy NGC 4449}",
      journal = {\aj},
         year = 1997,
        month = may,
       volume = {113},
        pages = {1678-1690},
          doi = {10.1086/118384},
archivePrefix = {arXiv},
       eprint = {astro-ph/9702121},
 primaryClass = {astro-ph},
       adsurl = {https://ui.adsabs.harvard.edu/abs/1997AJ....113.1678B}
}

@ARTICLE{Strickland04,
       author = {{Strickland}, David K. and {Heckman}, Timothy M. and {Colbert}, Edward J.~M. and {Hoopes}, Charles G. and {Weaver}, Kimberly A.},
        title = "{A High Spatial Resolution X-Ray and H{\ensuremath{\alpha}} Study of Hot Gas in the Halos of Star-forming Disk Galaxies. I. Spatial and Spectral Properties of the Diffuse X-Ray Emission}",
      journal = {\apjs},
         year = 2004,
        month = apr,
       volume = {151},
       number = {2},
        pages = {193-236},
          doi = {10.1086/382214},
archivePrefix = {arXiv},
       eprint = {astro-ph/0306592},
 primaryClass = {astro-ph},
       adsurl = {https://ui.adsabs.harvard.edu/abs/2004ApJS..151..193S}
}

@ARTICLE{Zhang24,
       author = {{Zhang}, He-Shou and {Ponti}, Gabriele and {Carretti}, Ettore and {Liu}, Ruo-Yu and {Morris}, Mark R. and {Haverkorn}, Marijke and {Locatelli}, Nicola and {Zheng}, Xueying and {Aharonian}, Felix and {Zhang}, Hai-Ming and {Zhang}, Yi and {Stel}, Giovanni and {Strong}, Andrew and {Yeung}, Michael C.~H. and {Merloni}, Andrea},
        title = "{A magnetized Galactic halo from inner Galaxy outflows}",
      journal = {Nature Astronomy},
         year = 2024,
        month = nov,
       volume = {8},
        pages = {1416-1428},
          doi = {10.1038/s41550-024-02362-0},
archivePrefix = {arXiv},
       eprint = {2408.06312},
 primaryClass = {astro-ph.GA},
       adsurl = {https://ui.adsabs.harvard.edu/abs/2024NatAs...8.1416Z}
}

@ARTICLE{Bittner2020,
       author = {{Bittner}, Adrian and {S{\'a}nchez-Bl{\'a}zquez}, Patricia and {Gadotti}, Dimitri A. and {Neumann}, Justus and {Fragkoudi}, Francesca and {Coelho}, Paula and {de Lorenzo-C{\'a}ceres}, Adriana and {Falc{\'o}n-Barroso}, Jes{\'u}s and {Kim}, Taehyun and {Leaman}, Ryan and {Mart{\'\i}n-Navarro}, Ignacio and {M{\'e}ndez-Abreu}, Jairo and {P{\'e}rez}, Isabel and {Querejeta}, Miguel and {Seidel}, Marja K. and {van de Ven}, Glenn},
        title = "{Inside-out formation of nuclear discs and the absence of old central spheroids in barred galaxies of the TIMER survey}",
      journal = {\aap},
         year = 2020,
        month = nov,
       volume = {643},
          eid = {A65},
        pages = {A65},
          doi = {10.1051/0004-6361/202038450},
archivePrefix = {arXiv},
       eprint = {2009.01856},
 primaryClass = {astro-ph.GA},
       adsurl = {https://ui.adsabs.harvard.edu/abs/2020A&A...643A..65B}
}

@ARTICLE{Gadotti19,
       author = {{Gadotti}, Dimitri A. and {S{\'a}nchez-Bl{\'a}zquez}, Patricia and {Falc{\'o}n-Barroso}, Jes{\'u}s and {Husemann}, Bernd and {Seidel}, Marja K. and {P{\'e}rez}, Isabel and {de Lorenzo-C{\'a}ceres}, Adriana and {Martinez-Valpuesta}, Inma and {Fragkoudi}, Francesca and {Leung}, Gigi and {van de Ven}, Glenn and {Leaman}, Ryan and {Coelho}, Paula and {Martig}, Marie and {Kim}, Taehyun and {Neumann}, Justus and {Querejeta}, Miguel},
        title = "{Time Inference with MUSE in Extragalactic Rings (TIMER): properties of the survey and high-level data products}",
      journal = {\mnras},
         year = 2019,
        month = jan,
       volume = {482},
       number = {1},
        pages = {506-529},
          doi = {10.1093/mnras/sty2666},
archivePrefix = {arXiv},
       eprint = {1810.01425},
 primaryClass = {astro-ph.GA},
       adsurl = {https://ui.adsabs.harvard.edu/abs/2019MNRAS.482..506G}
}

@ARTICLE{Gadotti20,
       author = {{Gadotti}, Dimitri A. and {Bittner}, Adrian and {Falc{\'o}n-Barroso}, Jes{\'u}s and {M{\'e}ndez-Abreu}, Jairo and {Kim}, Taehyun and {Fragkoudi}, Francesca and {de Lorenzo-C{\'a}ceres}, Adriana and {Leaman}, Ryan and {Neumann}, Justus and {Querejeta}, Miguel and {S{\'a}nchez-Bl{\'a}zquez}, Patricia and {Martig}, Marie and {Mart{\'\i}n-Navarro}, Ignacio and {P{\'e}rez}, Isabel and {Seidel}, Marja K. and {van de Ven}, Glenn},
        title = "{Kinematic signatures of nuclear discs and bar-driven secular evolution in nearby galaxies of the MUSE TIMER project}",
      journal = {\aap},
         year = 2020,
        month = nov,
       volume = {643},
          eid = {A14},
        pages = {A14},
          doi = {10.1051/0004-6361/202038448},
archivePrefix = {arXiv},
       eprint = {2009.01852},
 primaryClass = {astro-ph.GA},
       adsurl = {https://ui.adsabs.harvard.edu/abs/2020A&A...643A..14G}
}

@ARTICLE{Corsini12,
       author = {{Corsini}, E.~M. and {M{\'e}ndez-Abreu}, J. and {Pastorello}, N. and {Dalla Bont{\`a}}, E. and {Morelli}, L. and {Beifiori}, A. and {Pizzella}, A. and {Bertola}, F.},
        title = "{Polar bulges and polar nuclear discs: the case of NGC 4698}",
      journal = {\mnras},
         year = 2012,
        month = jun,
       volume = {423},
       number = {1},
        pages = {L79-L83},
          doi = {10.1111/j.1745-3933.2012.01261.x},
archivePrefix = {arXiv},
       eprint = {1204.2265},
 primaryClass = {astro-ph.CO},
       adsurl = {https://ui.adsabs.harvard.edu/abs/2012MNRAS.423L..79C}
}

@ARTICLE{Mayer08,
       author = {{Mayer}, L. and {Kazantzidis}, S. and {Escala}, A.},
        title = "{Formation of Nuclear Disks and Supermassive Black Hole Binaries in Galaxy Mergers}",
      journal = {\memsai},
         year = 2008,
        month = jan,
       volume = {79},
        pages = {1284},
          doi = {10.48550/arXiv.0807.3329},
archivePrefix = {arXiv},
       eprint = {0807.3329},
 primaryClass = {astro-ph},
       adsurl = {https://ui.adsabs.harvard.edu/abs/2008MmSAI..79.1284M}
}

@ARTICLE{Salim18,
       author = {{Salim}, Samir and {Boquien}, M{\'e}d{\'e}ric and {Lee}, Janice C.},
        title = "{Dust Attenuation Curves in the Local Universe: Demographics and New Laws for Star-forming Galaxies and High-redshift Analogs}",
      journal = {\apj},
         year = 2018,
        month = may,
       volume = {859},
       number = {1},
          eid = {11},
        pages = {11},
          doi = {10.3847/1538-4357/aabf3c},
archivePrefix = {arXiv},
       eprint = {1804.05850},
 primaryClass = {astro-ph.GA},
       adsurl = {https://ui.adsabs.harvard.edu/abs/2018ApJ...859...11S}
}

@ARTICLE{Salim16,
       author = {{Salim}, Samir and {Lee}, Janice C. and {Janowiecki}, Steven and {da Cunha}, Elisabete and {Dickinson}, Mark and {Boquien}, M{\'e}d{\'e}ric and {Burgarella}, Denis and {Salzer}, John J. and {Charlot}, St{\'e}phane},
        title = "{GALEX-SDSS-WISE Legacy Catalog (GSWLC): Star Formation Rates, Stellar Masses, and Dust Attenuations of 700,000 Low-redshift Galaxies}",
      journal = {\apjs},
         year = 2016,
        month = nov,
       volume = {227},
       number = {1},
          eid = {2},
        pages = {2},
          doi = {10.3847/0067-0049/227/1/2},
archivePrefix = {arXiv},
       eprint = {1610.00712},
 primaryClass = {astro-ph.GA},
       adsurl = {https://ui.adsabs.harvard.edu/abs/2016ApJS..227....2S}
}

@ARTICLE{Iovino02,
       author = {{Iovino}, Angela},
        title = "{Detecting Fainter Compact Groups: Results from a New Automated Algorithm}",
      journal = {\aj},
         year = 2002,
        month = nov,
       volume = {124},
       number = {5},
        pages = {2471-2489},
          doi = {10.1086/343059},
       adsurl = {https://ui.adsabs.harvard.edu/abs/2002AJ....124.2471I}
}

@ARTICLE{Zheng21,
       author = {{Zheng}, Yun-Liang and {Shen}, Shi-Yin},
        title = "{Compact Groups of Galaxies in Sloan Digital Sky Survey and LAMOST Spectral Survey. II. Dynamical Properties of Isolated and Embedded Groups}",
      journal = {\apj},
         year = 2021,
        month = apr,
       volume = {911},
       number = {2},
          eid = {105},
        pages = {105},
          doi = {10.3847/1538-4357/abeaa2},
archivePrefix = {arXiv},
       eprint = {2102.12804},
 primaryClass = {astro-ph.GA},
       adsurl = {https://ui.adsabs.harvard.edu/abs/2021ApJ...911..105Z}
}

@ARTICLE{Beers90,
       author = {{Beers}, Timothy C. and {Flynn}, Kevin and {Gebhardt}, Karl},
        title = "{Measures of Location and Scale for Velocities in Clusters of Galaxies---A Robust Approach}",
      journal = {\aj},
         year = 1990,
        month = jul,
       volume = {100},
        pages = {32},
          doi = {10.1086/115487},
       adsurl = {https://ui.adsabs.harvard.edu/abs/1990AJ....100...32B}
}

@ARTICLE{Montaguth25,
       author = {{Montaguth}, Gissel P. and {Monachesi}, Antonela and {Torres-Flores}, Sergio and {G{\'o}mez}, Facundo A. and {Lima-Dias}, Ciria and {Cortesi}, Arianna and {Mendes de Oliveira}, Claudia and {Telles}, Eduardo and {Panda}, Swayamtrupta and {Grossi}, Marco and {Lopes}, Paulo A.~A. and {O'Mill}, Ana Laura and {Hernandez-Jimenez}, Jose A. and {Olave-Rojas}, Daniela E. and {Demarco}, Ricardo and {Kanaan}, Antonio and {Ribeiro}, Tiago and {Schoenell}, William},
        title = "{Galaxy evolution in compact groups: II. Witnessing the influence of major structures in their evolution}",
      journal = {\aap},
         year = 2025,
        month = apr,
       volume = {696},
          eid = {A240},
        pages = {A240},
          doi = {10.1051/0004-6361/202451198},
archivePrefix = {arXiv},
       eprint = {2406.14671},
 primaryClass = {astro-ph.GA},
       adsurl = {https://ui.adsabs.harvard.edu/abs/2025A&A...696A.240M}
}

@ARTICLE{Jin23,
       author = {{Jin}, Shuowen and {Sillassen}, Nikolaj B. and {Magdis}, Georgios E. and {Vijayan}, Aswin P. and {Brammer}, Gabriel B. and {Kokorev}, Vasily and {Weaver}, John R. and {Gobat}, Raphael and {Gim{\'e}nez-Arteaga}, Clara and {Valentino}, Francesco and {Brinch}, Malte and {G{\'o}mez-Guijarro}, Carlos and {Shuntov}, Marko and {Toft}, Sune and {Greve}, Thomas R. and {Blanquez Sese}, David},
        title = "{Massive galaxy formation caught in action at z {\ensuremath{\sim}} 5 with JWST}",
      journal = {\aap},
         year = 2023,
        month = feb,
       volume = {670},
          eid = {L11},
        pages = {L11},
          doi = {10.1051/0004-6361/202245724},
archivePrefix = {arXiv},
       eprint = {2212.09372},
 primaryClass = {astro-ph.GA},
       adsurl = {https://ui.adsabs.harvard.edu/abs/2023A&A...670L..11J}
}

@ARTICLE{Wei26,
       author = {{Wei}, Xiaoyang and {Cai}, Zheng and {Yu}, Fujiang and {Li}, Mingyu and {Wu}, Yunjing},
        title = "{Pre-Virialized Assembly at Cosmic Dawn: The Dynamics and Extreme Ionization of Compact Group CGG-z7 at $z\sim7.04$}",
      journal = {arXiv e-prints},
         year = 2026,
        month = mar,
          eid = {arXiv:2603.08066},
        pages = {arXiv:2603.08066},
          doi = {10.48550/arXiv.2603.08066},
archivePrefix = {arXiv},
       eprint = {2603.08066},
 primaryClass = {astro-ph.GA},
       adsurl = {https://ui.adsabs.harvard.edu/abs/2026arXiv260308066W}
}

@ARTICLE{Jyoti23,
       author = {{Yadav}, Jyoti and {Das}, Mousumi and {Barway}, Sudhanshu and {Combes}, Francoise},
        title = "{An FUV and optical study of star formation in closely interacting galaxies: star-forming rings, tidal arms, and nuclear outflows}",
      journal = {\mnras},
         year = 2023,
        month = nov,
       volume = {526},
       number = {1},
        pages = {198-216},
          doi = {10.1093/mnras/stad2672},
archivePrefix = {arXiv},
       eprint = {2308.16152},
 primaryClass = {astro-ph.GA},
       adsurl = {https://ui.adsabs.harvard.edu/abs/2023MNRAS.526..198Y}
}

@article{yang05,
    author = {Yang, Xiaohu and Mo, H. J. and van den Bosch, Frank C. and Jing, Y. P.},
    title = {A halo-based galaxy group finder: calibration and application to the 2dFGRS},
    journal = {Monthly Notices of the Royal Astronomical Society},
    volume = {356},
    number = {4},
    pages = {1293-1307},
    year = {2005},
    month = {02},
    issn = {0035-8711},
    doi = {10.1111/j.1365-2966.2005.08560.x},
    url = {https://doi.org/10.1111/j.1365-2966.2005.08560.x},
    eprint = {https://academic.oup.com/mnras/article-pdf/356/4/1293/3296536/356-4-1293.pdf},
}

@ARTICLE{Eke04,
       author = {{Eke}, V.~R. and {Baugh}, Carlton M. and {Cole}, Shaun and {Frenk}, Carlos S. and {Norberg}, Peder and {Peacock}, John A. and {Baldry}, Ivan K. and {Bland-Hawthorn}, Joss and {Bridges}, Terry and {Cannon}, Russell and {Colless}, Matthew and {Collins}, Chris and {Couch}, Warrick and {Dalton}, Gavin and {de Propris}, Roberto and {Driver}, Simon P. and {Efstathiou}, George and {Ellis}, Richard S. and {Glazebrook}, Karl and {Jackson}, Carole and {Lahav}, Ofer and {Lewis}, Ian and {Lumsden}, Stuart and {Maddox}, Steve and {Madgwick}, Darren and {Peterson}, Bruce A. and {Sutherland}, Will and {Taylor}, Keith},
        title = "{Galaxy groups in the 2dFGRS: the group-finding algorithm and the 2PIGG catalogue}",
      journal = {\mnras},
         year = 2004,
        month = mar,
       volume = {348},
       number = {3},
        pages = {866-878},
          doi = {10.1111/j.1365-2966.2004.07408.x},
archivePrefix = {arXiv},
       eprint = {astro-ph/0402567},
 primaryClass = {astro-ph},
       adsurl = {https://ui.adsabs.harvard.edu/abs/2004MNRAS.348..866E}
}

\begin{appendix}

\section{Detection of SFCs and SFR estimation}
Star-forming (SF) regions appear as bright clumps in the background-subtracted \textit{GALEX} FUV and NUV images. To systematically detect these regions across all galaxies of RQ, we used the \texttt{SExtractor} software \citep{Bertin96}. Source detection was performed on the \textit{GALEX} NUV images with the following configuration: \texttt{DETECT\_THRESH}=5$\sigma$ (where $\sigma$ is the global background noise), \texttt{DETECT\_MINAREA}=10 pixels, \texttt{DEBLEND\_NTHRESH}=16, and \texttt{DEBLEND\_MINCONT}=0.0005. For NGC 87, NGC 88, and NGC 89, one prominent SF region was detected in each galaxy, which covers the light from the entire galaxy, whereas we were able to detect six regions in NGC 92, consistent with the spatial resolution of the \textit{GALEX} data. Aperture photometry of the detected regions was performed using the \texttt{Photutils} package\footnote{\url{https://photutils.readthedocs.io/en/stable/aperture.html}}, yielding FUV and NUV fluxes for each source.

These UV-bright regions trace populations of young, massive stars, whose observed magnitudes are affected by both Galactic and internal extinction. Galactic extinction in the FUV and NUV bands was corrected using the extinction law of \citet{Cardelli89}, expressed as
\begin{equation}
    <\frac{A(\lambda)}{A(V)}> = a(x)+\frac{b(x)}{R_{v}}
    \label{eq:1}
\end{equation}
where $x$ is the wavenumber and the functions $a(x)$ and $b(x)$ are defined in \citet{Cardelli89}. We adopted $R_V = 3.1$ for the Milky Way and used $A(V) = 0.054$ for NGC 87 and $A(V) = 0.053$ for NGC 88, NGC 89, and NGC 92.

In addition to Galactic extinction, it is crucial to correct for internal extinction, which arises from dust within the star-forming regions themselves. It was corrected using the UV spectral slope ($\beta$), defined by $f_{\lambda} \propto \lambda^{\beta}$, which quantifies dust attenuation in the UV continuum after Galactic correction and is applicable over the wavelength range 1250–2600$\AA$ \citep{Calzetti94}. For the \textit{GALEX} filters, $\beta$ was computed as
\begin{equation}
    \beta = -0.4 \frac{m_{FUV} - m_{NUV}}{log(\lambda_{FUV}/\lambda_{NUV})} - 2.0
    \label{eq:2}
\end{equation}
where $m_{\mathrm{FUV}}$ and $m_{\mathrm{NUV}}$ are the Milky Way extinction-corrected magnitudes, and $\lambda_{\mathrm{FUV}}$ and $\lambda_{\mathrm{NUV}}$ are the central wavelengths of the respective filters. The colour excess was then derived following \citet{Calzetti00}:
\begin{equation}
    \beta = -2.616 + 4.684E(B-V)
    \label{eq:3}
\end{equation}
From E(B-V), the internal extinction at a given wavelength was calculated using the  \citep{Calzetti00} relation: 

\begin{equation}
    A_{\lambda} = k^{'}(\lambda)E_{s}(B-V)
    \label{eq:4}
\end{equation}
Where $E_{s}(B-V)$ is the colour excess applicable to the stellar continuum, and $k^{'}(\lambda)$ is the attenuation curve defined as:
\begin{equation}
    k^{'}(\lambda) = -2.659(-2.15 + \frac{1.509}{\lambda} - \frac{0.198}{\lambda^{2}} + \frac{0.011}{\lambda^{3}}) + R^{'}_{V}
    \label{eq:5}
\end{equation}
Here, $R^{'}_{V}$ = 4.05 $\pm$ 0.80 and $\lambda$ $\in$ [1200\AA, 6300\AA] \citep{Calzetti00}. 

The stellar colour excess is related to the total colour excess by:
\begin{equation}
    E_{s}(B-V) = (0.44\pm0.03)E(B-V)
    \label{eq:6}
\end{equation}
Using the estimated extinction, we calculate the extinction-corrected luminosity as:
\begin{equation}
    L_{corr} = L_{obs} \times 10^{0.4 E_{\lambda}(B-V)K^{'}(\lambda)}
    \label{eq:7}
\end{equation}
Here $L_{obs}$ denotes the luminosity measured before correcting for dust extinction.

To investigate star formation activity in galaxies, it is essential to examine SF regions across multiple wavelengths, as different populations of stars emit at different wavelengths. Ultraviolet (UV) emission serves as an effective tracer of recent star formation (up to $\sim$300 Myr), as it predominantly originates from young high mass O, B-type stars \citep{Davies16}.

We estimated the star formation rates (SFRs) using the GALEX FUV band, following the calibrations provided by \citet{Kennicutt09}. The extinction-corrected luminosities were used to derive the SFRs according to the following relation:
\begin{equation}
    SFR_{FUV}(M_\odot\ yr^{-1}) = 4.6\times10^{-44}L_{corr}(FUV).
	\label{eq:8}
\end{equation}
Here, $L_{corr}$ denotes the extinction-corrected luminosity in units of erg s$^{-1}$. The derived SFRs for each galaxy are presented in Table~\ref{tab:sfr_age}. \\

\section{Age of the SF regions}
\begin{table}
	\centering
        \caption{Starburst99 model parameters used in this work.}
	\begin{tabular}{lccccr} 
		\hline
		  Parameter & Value\\
            \hline
		Star Formation & Instantaneous \\
		Cluster mass range &  $10^{7} M_{\odot}- 10^{9} M_{\odot}$\\
            Stellar IMF & Kroupa (1.3,2.3)\\
            Mass Boundaries for IMF & $0.1M_{\odot}$, $0.5M_{\odot}$, $100M_{\odot}$ \\
		Stellar evolution track & Geneva (high mass loss)\\
  		Metallicity & Solar (Z =0.02)\\
            Age range & 1-900 Myr\\
            Supernova cut-off mass & 8 $M_{\odot}$ \\
            Black hole cut-off mass & 120 $M_{\odot}$ \\
            \hline
	\end{tabular}
	\label{tab:starburst}
\end{table}

To estimate the ages of the star-forming regions, we employed the \texttt{Starburst99} stellar population synthesis code \citep{Leitherer99}, widely used for modelling actively star-forming galaxies. \texttt{Starburst99} provides spectrophotometric predictions for simple stellar populations (SSPs) based on evolutionary tracks and various stellar atmosphere models. Synthetic stellar spectra were generated using the input parameters listed in Table~\ref{tab:starburst}. The model spectra were convolved with the effective area curves of the \textit{GALEX} FUV and NUV filters to compute the model-predicted fluxes in each band, using

 \begin{equation}
 F_{\mathrm{cal}}(\lambda) = \frac{\int F(\lambda) , EA(\lambda) , d\lambda}{\int EA(\lambda) , d\lambda},
 \label{eq:10}
 \end{equation}
 
\begin{figure*}
\centering
\includegraphics[width=\textwidth]{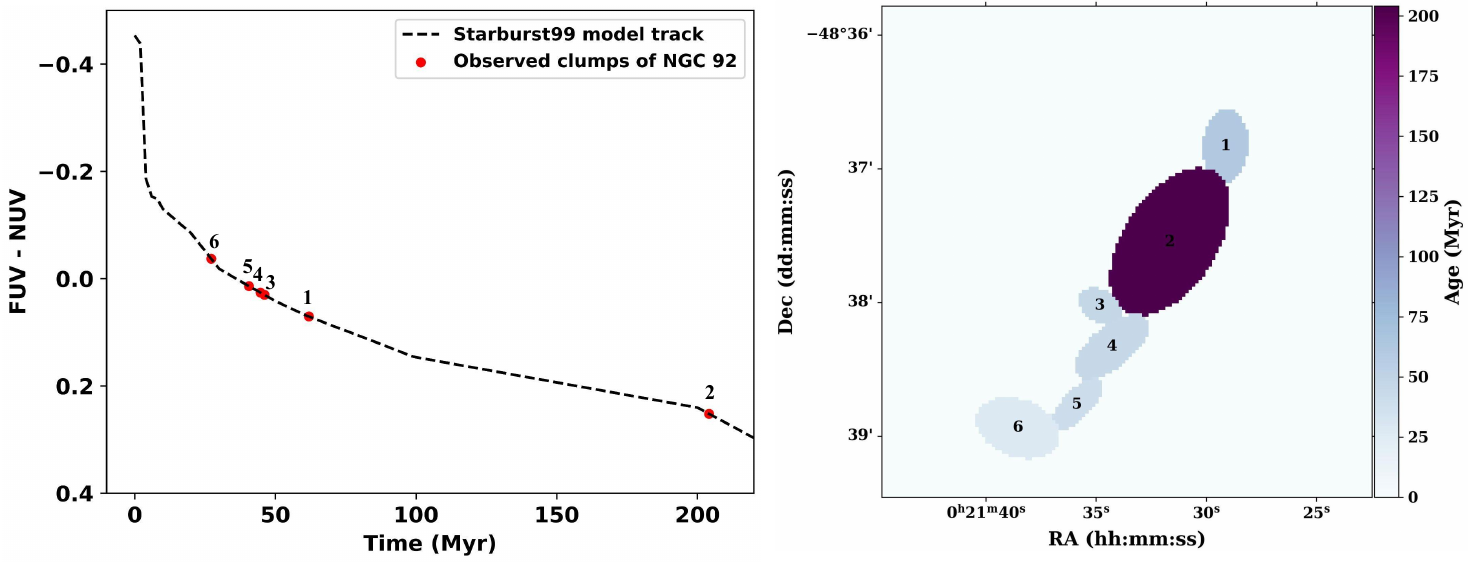}
\caption{\textbf{Left:} Extinction-corrected observed (FUV-NUV) colors of star-forming regions in NGC 92 (red circles) plotted as a function of age. The black dashed line represents the theoretical color–age relation obtained from a Starburst99 model. \textbf{Right:} Spatial distribution of star-forming regions in NGC 92, color-coded by their derived ages.}
\label{fig:age_NGC92}
\end{figure*}

Where $F_{\mathrm{cal}}(\lambda)$ is the calibrated flux, $F(\lambda)$ is the model flux and $EA(\lambda)$ denote the effective area of the corresponding \textit{GALEX} filter. All fluxes are expressed in units of erg s$^{-1}$ cm$^{-2}$ \AA$^{-1}$. The resulting fluxes were converted to AB magnitudes to compute theoretical FUV–NUV colors. By comparing these with the extinction-corrected observed FUV–NUV colors, we estimated the ages of the star-forming clumps in each galaxy. The derived ages are summarised in Table~\ref{tab:sfr_age}. Fig.~\ref{fig:age_NGC92} presents the age distribution of star-forming regions in NGC 92, based on their UV colours (left panel) and their corresponding spatial distribution within the galaxy (right panel).

\begin{table}
	\centering
        \caption{SFR and Age of the regions identified in each galaxy.}
	\begin{tabular}{lccccr}
		\hline
		  Name & $SFR_{FUV}$ & Age\\
                 & $(M_\odot\ yr^{-1})$ & (Myr)\\
            \hline
            \\
		NGC 87 & 1.142 $\pm$ 0.037  & 31.97 \\
        NGC 88 & 0.444 $\pm$ 0.007 & 127.57 \\
		NGC 89 & 0.586 $\pm$ 0.017 & 275.54 \\
        NGC 92 & 0.018 $\pm$ 0.002 & 61.90 (1)\\
               & 0.942 $\pm$ 0.029 & 204.13 (2)\\
               & 0.015 $\pm$ 0.002 & 46.10 (3)\\
               & 0.044 $\pm$ 0.006 & 44.70 (4)\\
               & 0.022 $\pm$ 0.003 & 40.63 (5)\\
               & 0.072 $\pm$ 0.008 & 27.18 (6)\\                  
            \hline
	\end{tabular}
    \begin{flushleft}
    \footnotesize
    \textbf{Description:} {For NGC 87, NGC 88, and NGC 89, one prominent SF region was detected, which covers the light from the entire galaxy, whereas we were able to detect six regions in NGC 92.}
    \end{flushleft}
	\label{tab:sfr_age}
\end{table}

\section{BPT classification of galaxies}

\begin{figure}
     \centering
     \includegraphics[width=\linewidth]{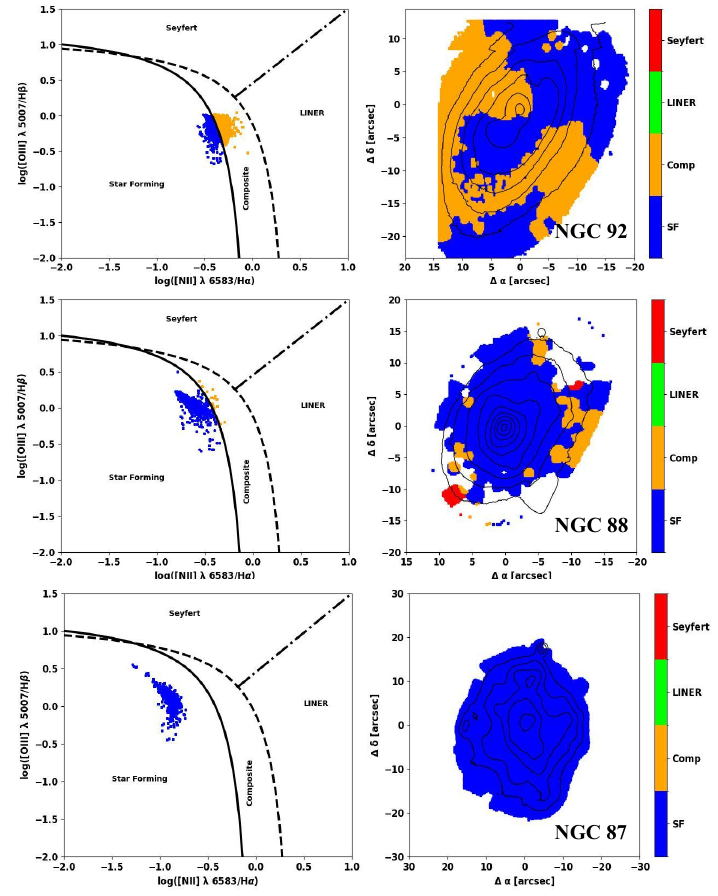}
     \caption{BPT classification of NGC 92 (top), NGC 88 (middle) and NGC 87(bottom).}
     \label{fig:BPT}
\end{figure}

\section{Input and output parameters of CIGALE modelling}

\begin{figure*}
\centering
\includegraphics[width=0.92\textwidth]{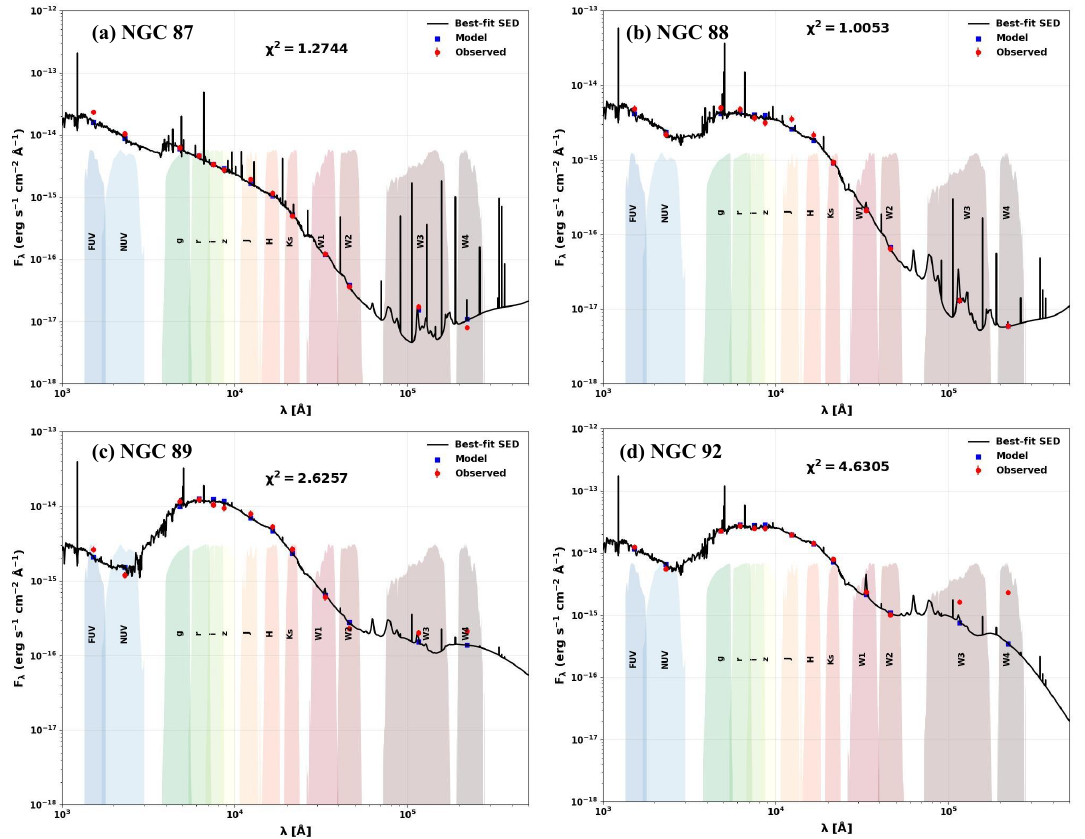}
\caption{CIGALE SED fits for the RQ quartet: NGC 87 (top-left), NGC 88 (top-right), NGC 89 (bottom-left), and NGC 92 (bottom-right). The best-fitting models are overlaid on observed photometric data in each panel.}
\label{fig:sed}
\end{figure*}

\begin{table*}
     \centering
     \caption{Input Parameters for SED Fitting with CIGALE.}
         \begin{tabularx}{\textwidth}{lX}
         \hline
         \textbf{Parameter} & \textbf{Value} \\
         \hline
         \textbf{Star formation history (sfhdelayedbq)} &  \\
         e-folding time of the main stellar population & 200.0, 500.0, 700.0, 1000.0, 2000.0, 3000.0, 4000.0, 5000.0 \\
          Age of the main stellar population & 1000, 2000, 3500, 5000, 6500, 8000, 10500  \\
          Age of the late burst/quench episode & 100.0, 200.0, 300.0, 400.0, 500.0 \\
          Ratio of the SFR after/before burst/quench & 0.0, 0.1, 0.2, 0.4, 0.7, 1.0, 5.0, 10.0, 100.0 \\
          Instance without any burst & 1.0 \\
          \hline
          \textbf{Stellar population synthesis (bc03)} \\
          \\
          IMF & Salpeter \\
          Metallicity & 0.004, 0.008, 0.02, 0.05 \\
          Separation between the young and the old stellar populations & 10 Myr \\
          \hline       
          \textbf{Nebular emission (nebular)} \\
          \\
         Ionization parameter & -2.0 \\
         Gas metallicity & 0.004, 0.008, 0.012, 0.02, 0.033 \\
         Electron density  & 100 $cm^{-3}$ \\
         \hline
         \textbf{Dust attenuation (dustatt \_ modified \_ starburst)} \\
         \\
         Colour excess of the nebular lines light for both the young and old population & 0.3 \\
         Reduction factor to apply on E\_BV\_lines to compute E(B-V)s & 0.44 \\
         Central wavelength of the UV bump in nm & 217.5 \\
         Width (FWHM) of the UV bump in nm & 35.0 \\
         UV bump amplitude & 0.0 \\
         Slope delta of the power law & 0.0 \\
         Filters for which the attenuation will be computed and added & FUV and NUV \\
         \hline        
         \textbf{Dust emission (dl2014)} \\
         \\
         Mass fraction of polycyclic aromatic hydrocarbon (PAH) & 0.5, 1.12, 2.5, 3.19, 5.26 \\
         Minimum radiation field  & 0.5, 1.0, 5, 10, 15, 25  \\
         Power-law slope & 1.5, 2.0 \\
         Fraction illuminated from Umin to Umax &  0.001, 0.01, 0.1, 0.5 \\
         \hline     
         \textbf{AGN module (skirtor2016)} \textbf{(NGC 89)}\\
         \\     
         Average edge-on optical depth at 9.7 $\mu m$ & 3, 7 \\
         Power-law exponent that sets the radial gradient with the polar angle & 1.0 \\
         Index that sets the dust density gradient with the polar angle & 1.0 \\
         Inclination, i.e., viewing angle, position of the instrument w.r.t. the AGN axis & 30, 70 \\
         Disk spectrum & Schartmann spectrum \\
         AGN fraction & 0.0, 0.05, 0.1, 0.2, 0.3, 0.4, 0.8 \\
         Wavelength range in microns used to compute the AGN fraction & 0/0 \\
         Extinction law of the polar dust & SMC \\
         E(B-V ) for the extinction in the polar direction in magnitudes & 0.0, 0.03, 0.2, 0.4 \\
         Temperature of the polar dust & 100.0 K \\
         Emissivity index of the polar dust  & 1.6 \\      
         \hline
         \end{tabularx}
     \label{tab:sed}
 \end{table*}

\begin{figure}
    \centering
    \includegraphics[width=\linewidth]{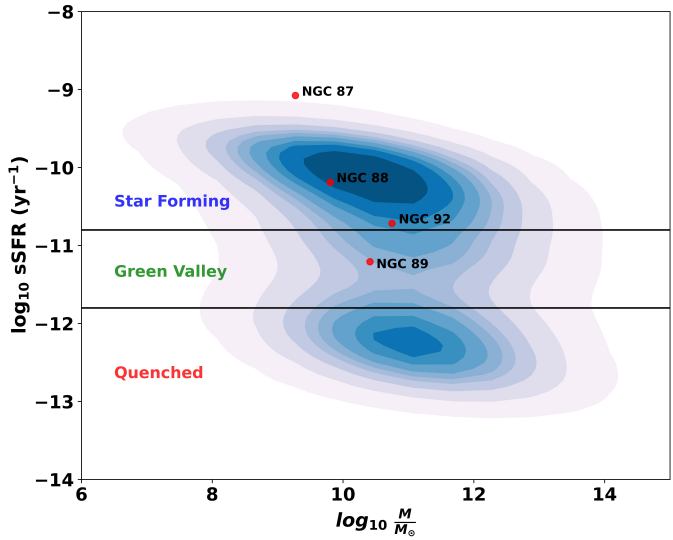}
    \caption{The star-forming main sequence of galaxies. Here, the red symbols represent the galaxies of RQ. The contour shows the density map of the data points from the GSLWC-X2 catalogue \citep{Salim16, Salim18}}.
    \label{fig:ssfr-mass}
\end{figure}

\end{appendix}
\end{document}